\documentclass[ twocolumn, times, astrosymb, numberedappendix, appendixfloats]{aastex631}
\usepackage[utf8]{inputenc}
\usepackage{amsmath}
\usepackage{booktabs} 
\usepackage{tabularx}
\usepackage{threeparttable}
\usepackage{float}
\usepackage{url}

\newcommand{\Av}{$A(V)$}

\newcommand{\msol}{M$_\odot$}
\newcommand{\mass}{M$_*$}

\newcommand{\zsol}{Z$_\odot$}

\newcommand{\fastpp}{{\tt FAST++}}
\newcommand{\prospector}{{\tt Prospector}}
\newcommand{\slinefit}{{\tt slinefit}}

\newcommand{\OVI}{[\hbox{{\rm O}\kern 0.1em{\sc vi}}]}

\newcommand{\NV}{\hbox{{\rm N}\kern 0.1em{\sc v}}}
\newcommand{\SiIV}{\hbox{{\rm Si}\kern 0.1em{\sc iv}}}
\newcommand{\OIV}{[\hbox{{\rm O}\kern 0.1em{\sc iv}}]}
\newcommand{\NIV}{[\hbox{{\rm N}\kern 0.1em{\sc iv}}]}
\newcommand{\CIV}{\hbox{{\rm C}\kern 0.1em{\sc iv}}}
\newcommand{\HeII}{\hbox{{\rm He}\kern 0.1em{\sc ii}\kern 0.1em{$\lambda1640$} }}
\newcommand{\OIII}{[\hbox{{\rm O}\kern 0.1em{\sc iii}}]{$\lambda5007$}}
\newcommand{\OIIId}{[\hbox{{\rm O}\kern 0.1em{\sc iii}}]{$\lambda4959\lambda5007$}}
\newcommand{\NIII}{[\hbox{{\rm N}\kern 0.1em{\sc iii}}]}
\newcommand{\AlIII}{\hbox{{\rm Al}\kern 0.1em{\sc iii}}}
\newcommand{\SiIII}{\hbox{{\rm Si}\kern 0.1em{\sc iii}}}
\newcommand{\CIII}{\hbox{{\rm C}\kern 0.1em{\sc iii}]}}
\newcommand{\NeIV}{[\hbox{{\rm Ne}\kern 0.1em{\sc iv}}]}
\newcommand{\MgII}{\hbox{{\rm Mg}\kern 0.1em{\sc ii}}}

\newcommand{\CII}{[\hbox{{\rm C}\kern 0.1em{\sc ii}]}}

\newcommand{\He}{\hbox{{\rm He}\kern 0.1em{\sc ii}\kern 0.1em{$\lambda1640\lambda4686$}}}

\newcommand{\Halpha}{H$\alpha$}
\newcommand{\Hbeta}{H$\beta$}
\newcommand{\SII}{[\hbox{{\rm S}\kern 0.1em{\sc ii}}]$\lambda6716\lambda6731$}
\newcommand{\NII}{[\hbox{{\rm N}\kern 0.1em{\sc ii}}]}
\newcommand{\NIId}{[\hbox{{\rm N}\kern 0.1em{\sc ii}}]$\lambda6548\lambda6584$}
\newcommand{\OII}{[\hbox{{\rm O}\kern 0.1em{\sc ii}}]}
\newcommand{\MgI}{\hbox{{\rm Mg}\kern 0.1em{\sc i}}}
\newcommand{\FeII}{\hbox{{\rm Fe}\kern 0.1em{\sc ii}}}

\newcommand{\OI}{\hbox{{\rm O}\kern 0.1em{\sc i}}}
\newcommand{\NeII}{[\hbox{{\rm Ne}\kern 0.1em{\sc ii}}] }
\newcommand{\NaI}{[\hbox{{\rm Na}\kern 0.1em{\sc i}}] }
\newcommand{\NeIII}{[\hbox{{\rm Ne}\kern 0.1em{\sc iii}}] }

\submitjournal{ApJ}

\shorttitle{Massive Quiescent Galaxies in the $3<z<4.5$ Universe}
\shortauthors{Nanayakkara et al.}
\graphicspath{{./}{figures/}}

\begin{document}

\title{The formation histories of massive and quiescent galaxies in the $3<z<4.5$ Universe.}

\author[0000-0003-2804-0648 ]{Themiya Nanayakkara}
\affiliation{Centre for Astrophysics and Supercomputing, Swinburne University of Technology, PO Box 218, Hawthorn, VIC 3122, Australia}

\author[0000-0002-3254-9044]{Karl Glazebrook}
\affiliation{Centre for Astrophysics and Supercomputing, Swinburne University of Technology, PO Box 218, Hawthorn, VIC 3122, Australia}

\author[0000-0003-0942-5198]{Corentin Schreiber}
\affiliation{IBEX Innovations, Sedgefield, Stockton-on-Tees, TS21 3FF, United Kingdom}

\author[0000-0001-5856-8713]{Harry Chittenden}
\affiliation{Centre for Astrophysics and Supercomputing, Swinburne University of Technology, PO Box 218, Hawthorn, VIC 3122, Australia}

\author[0000-0003-2680-005X]{Gabriel Brammer}
\affiliation{Cosmic DAWN Center, Niels Bohr Institute, University of Copenhagen,  Jagtvej 128, Copenhagen N, DK-2200, Denmark}

\author[0000-0003-2680-005X]{James Esdaile}
\affiliation{Centre for Astrophysics and Supercomputing, Swinburne University of Technology, PO Box 218, Hawthorn, VIC 3122, Australia}

\author[0000-0003-4239-4055]{Colin Jacobs}
\affiliation{Centre for Astrophysics and Supercomputing, Swinburne University of Technology, PO Box 218, Hawthorn, VIC 3122, Australia}

\author[0000-0003-1362-9302]{Glenn G. Kacprzak}
\affiliation{Centre for Astrophysics and Supercomputing, Swinburne University of Technology, PO Box 218, Hawthorn, VIC 3122, Australia}

\author[0000-0003-4032-2445]{Lalitwadee Kawinwanichakij}
\affiliation{Centre for Astrophysics and Supercomputing, Swinburne University of Technology, PO Box 218, Hawthorn, VIC 3122, Australia}

\author[0009-0006-8337-8712]{Lucas C. Kimmig}
\affiliation{University Observatory Munich, Faculty of Physics, Ludwig-Maximilians-University, Scheinerstrasse 1, 81679, Munich, Germany}

\author[0000-0002-2057-5376]{Ivo Labbe}
\affiliation{Centre for Astrophysics and Supercomputing, Swinburne University of Technology, PO Box 218, Hawthorn, VIC 3122, Australia}

\author[0000-0003-3021-8564]{Claudia Lagos}
\affiliation{Cosmic DAWN Center, Niels Bohr Institute, University of Copenhagen,  Jagtvej 128, Copenhagen N, DK-2200, Denmark}
\affiliation{ARC Centre for Excellence in All-Sky Astrophysics in 3D}
\affiliation{International Centre for Radio Astronomy Research, University of Western Australia, 7 Fairway, Crawley, 6009, WA, Australia}

\author[0000-0001-9002-3502]{Danilo Marchesini}
\affiliation{Tufts University, Physics and Astronomy Department, 574 Boston Avenue, Medford, MA, 02155, USA}

\author[0000-0002-6701-1666]{M. Mart\`inez-Mar\`in}
\affiliation{Centre for Astrophysics and Supercomputing, Swinburne University of Technology, PO Box 218, Hawthorn, VIC 3122, Australia}

\author[0000-0002-7248-1566]{Z.~Cemile Marsan}
\affiliation{Department of Physics and Astronomy, York University, 4700 Keele Street, Toronto, ON, M3J 1P3, Canada}

\author[0000-0001-5851-6649]{Pascal A. Oesch}
\affiliation{Cosmic DAWN Center, Niels Bohr Institute, University of Copenhagen,  Jagtvej 128, Copenhagen N, DK-2200, Denmark}
\affiliation{Department of Astronomy, University of Geneva,  Chemin Pegasi 51, Versoix, CH-1290, Switzerland}

\author[0000-0001-7503-8482]{Casey Papovich}
\affiliation{Department of Physics and Astronomy, and George P. and Cynthia Woods Mitchell Institute for Fundamental Physics and Astronomy, Texas A\&M University, College Station, TX 77843-4242, USA}

\author[0009-0008-9260-7278]{Rhea-Silvia Remus}
\affiliation{University Observatory Munich, Faculty of Physics, Ludwig-Maximilians-University, Scheinerstrasse 1, 81679, Munich, Germany}

\author[0000-0001-9208-2143]{Kim-Vy H. Tran}
\affiliation{Center for Astrophysics, Harvard \& Smithsonian, Cambridge, MA, USA}
\affiliation{ARC Centre for Excellence in All-Sky Astrophysics in 3D}

\begin{abstract}
We present the formation histories of 19 massive  ($\gtrsim3\times10^{10}$\msol) quiescent (sSFR$<0.15\ \text{Gyr}^{-1}$) galaxy candidates at $z\sim3.0-4.5$ observed using JWST/NIRSpec. 
This completes the spectroscopic confirmation of the 24 $K$-selected quiescent galaxy sample from the ZFOURGE and 3DHST surveys \citep{Schreiber2018}. 
Utilizing Prism $1-5\mu$m spectroscopy, we confirm that all 12 sources that eluded confirmation by ground-based spectroscopy lie at $z>3$, resulting in a spectroscopically confirmed number density of $\sim1.4\times10^{-5}\text{Mpc}^{-3}$ between $z\sim3-4$.
Rest-frame $U-V$ vs $V-J$ color selections show high effectiveness in identifying quiescent galaxies, with a purity of $\sim90\%$.
Our analysis shows that parametric star-formation histories (SFHs) from \fastpp\ and binned SFHs from \prospector\ on average yield consistent results, revealing diverse formation and quenching times. The oldest galaxy formed $\sim6\times10^{10}$ \msol\ by $z\sim10$ and has been quiescent for over 1 Gyr at $z\sim3.2$. We detect two galaxies with ongoing star formation and six with active galactic nuclei (AGN).
We demonstrate that the choice of stellar population models, stellar libraries, and nebular or AGN contributions does not significantly affect the derived average SFHs of the galaxies. We demonstrate that extending spectral fitting beyond the rest-frame optical regime reduces the inferred average star formation rates in the earliest time bins of the SFH reconstruction.
The assumed SFH prior influences the star formation rate at early times, where spectral diagnostic power is limited. Simulated $z\sim3$ quiescent galaxies from IllustrisTNG, SHARK, and Magneticum broadly match the average SFHs of the observed sample but struggle to capture the full diversity, particularly at early stages. Our results emphasize the need for mechanisms that rapidly build stellar mass and quench star formation within the first billion years of the Universe.
\end{abstract}


\section{Introduction} \label{sec:intro}

In the last 10 years, from wide and deep near infra-red (NIR) surveys on $>4 m$ telescopes, we have made significant advancements in identifying the first generation of massive ($\sim$\mass$>10^{10}$\msol) quiescent galaxies in the Universe \cite[e.g.][]{Gobat2012a,Straatman2014,Schreiber2018}. Ground based $J,H,K$ band imaging accompanied by deep HST and Spitzer imaging led the way through surveys such as ZFOURGE \citep{Straatman2016} and UltraVISTA \citep{McCracken2012} to detect possible $z>3$ massive quiescent candidates.

Ground based spectroscopic confirmations required long exposure deep $H$ and $K$ band spectroscopy from 8-10m class telescopes \citep[e.g.][]{Glazebrook2017,Schreiber2018,Valentino2020,Forrest2022a,Carnall2020a}. Due to atmospheric cutoffs in the NIR and high thermal backgrounds, detailed stellar population analysis exploring rest-frame optical spectral features was challenging \citep[e.g.][]{Nanayakkara2022b}. 
Furthermore, due to limitations in achieving sensitive continuum magnitudes, the majority of confirmed galaxies were $K$ bright \citep[$K_s\lesssim22.5$ AB, ][]{Antwi-Danso2025a}. 
This adds a bias towards recently quenched galaxies because the A type stars of a passively evolving galaxy will transition out of the main sequence making the galaxy $K$ faint (increasing the mass/K band luminosity ratio) at $z\sim3-4$. 
Thus, there is a potential bias for spectroscopically confirmed quiescent galaxies' quenching time scales to be within the last $\sim 500$ Myrs of the observation \citep[e.g.][]{Forrest2020b}

However, there were at least two known galaxies that hinted an older underlying population of quiescent galaxies with ages $\gtrsim 500$ Myr at $z\sim3-4$.
The SFH reconstruction of ZF-COS-20115 \citep[][also presented in this analysis]{Glazebrook2017,Schreiber2018} utilizing ZFOURGE multiband imaging and Keck/MOSFIRE spectroscopy pointed it to have quenched $\lesssim700$ Myr before the  observed redshift of $z=3.7$. 
Similarly, $HST$ grism spectroscopic sample presented by \citet{DEugenio2021a}  also hinted at the presence of an underlying older population for at least one $z=3.0$ galaxy. 
The $\sim10^{11}$\msol\ of these galaxies and the old ages meant that these galaxies had to rapidly buildup their stellar masses within the first billion years of the Universe and have mechanisms within them to abruptly cease the star-formation. 
Thus, at the peak of their star-formation histories, these galaxies likely required SFRs $>1000$\msol/yr with star-forming episodes likely limited to $\Delta t\lesssim250$ Myr \citep{Glazebrook2017}. 
Additionally, SFHs from multi-band photometric analysis of photometrically selected quiescent galaxies further hinted at the existence of $>1$Gyr old stellar populations, however, these galaxies were unable to be spectroscopically confirmed from ground, even with $\sim10$h of Keck/MOSFIRE spectroscopy \cite[e.g. ZF-UDS-7329][]{Schreiber2018}.

With the launch of JWST, the ability to gain deeper insights into the first massive quiescent galaxies in the Universe has significantly expanded. Imaging data from NIRCam has identified new \(z \sim 3{-}5\) massive quiescent candidates \citep[e.g.][]{Carnall2023a}. Photometric studies provide broad constraints on quiescent galaxy populations through multi-band imaging, often utilizing Bayesian spectral energy distribution fitting codes, while spectroscopic data is essential for tightening SFH constraints by analyzing detailed age, metallicity, and quenching timescales via rest-frame optical features such as the Balmer break and absorption lines \citep{Nanayakkara2024a}. SFHs reconstructed with spectroscopic data have revealed that some galaxies formed \(\sim 10^{11}\, M_\odot\) within the first billion years of the Universe \citep{Carnall2024a,Glazebrook2024a}. Furthermore, spectroscopic studies have uncovered evidence of temporarily quiescent low-mass galaxies up to \(z \sim 7\) \citep{Looser2024a}.

The $1-5\mu$m spectroscopy afforded by JWST NIRSpec provides detailed rest-frame optical spectral features that are instrumental to explore the stellar population properties of $z>3$ massive quiescent galaxies. When galaxies enter quiescence their continuum gets dominated by late-B type and main sequence A stars. This gives rise to a Balmer break which transitions to a D4000\AA\ feature with the passive evolution of stars after $>800$ Myr \citep[e.g. see Section 4.1 in][]{Bruzual2003}. 
Additionally, the rest-optical spectrum of these galaxies is dominated by absorption features from hydrogen and other stellar nucleosynthesis elements, such as Mg, Na, Ca, and Fe, which provide critical insights into the underlying stellar populations. These elements are produced through various nucleosynthesis channels that depend on the stellar masses of the contributing stars. Since the lifetimes of stars are linked to their masses, studying the abundance of different elements in galaxies allows us to infer whether their stellar populations grew quickly or gradually. For instance, if \(\alpha\)-elements produced by massive stars are more abundant in a galaxy, it indicates rapid star formation without sufficient time for subsequent generations of stars to form from \(\alpha\)-enhanced material. Thus, through detailed element abundance analyses, a forensic reconstruction of star formation histories can reveal how galaxies built up their stellar masses.

Ground based \citep{Martinez-Marin2024a} and JWST/NIRspec spectroscopy has discovered galaxies that are categorized as quiescent but with broad optical emission lines, indicative of an AGN \citep{Carnall2023b,DEugenio2024a,Nanayakkara2024a}. 
This suggests that AGN might be a mechanism that could quench star-formation in these galaxies.

Recent results from JWST reveal a variety of quenching timescales in galaxies, ranging from short and abrupt quenching episodes lasting less than 250 Myr \citep{Carnall2023b,Carnall2024a,deGraaff2024a,Glazebrook2024a,Nanayakkara2024a,Perez-Gonzalez2024a,Weibel2024a}, to more gradual processes extending over several hundred Myrs \citep{Carnall2023a,Carnall2024a,Nanayakkara2024a,Setton2024a}. These findings provide critical insights for hydrodynamical simulations, which aim to develop pathways for both rapid and extended quenching mechanisms of massive galaxies in the early Universe \citep[e.g.,][]{Hartley2023a,Lagos2025a}.
The low mass post-starburst like systems discovered at $z>5$ \citep[e.g.][]{Looser2024a,Strait2023a} point towards temporary quiescence due to the stochastic nature of star-formation in the early Universe. 
However, simulations suggest that for higher mass galaxies quiescence is expected to be a longer term phenomena driven by AGN activity \citep{Xie2024a}. 
Given their higher masses, space for further substantial mass increases is limited, thus, these galaxies are expected to grow passively via dry minor mergers to $z\sim0$ \citep{Oser2012a}.

Most analysis conducted in pre-JWST area were targeted towards interesting sources identified in deep imaging surveys over HST legacy fields. 
In \citet{Schreiber2018} a $K$-selected sample of galaxies at $z\sim3-4$ was used to obtain ground based spectroscopic confirmations of their quiescence. 
$K$-band covers the rest-frame optical at these redshifts, thus,  this translates approximately to a mass selected sample. 
Out of the 24 galaxies selected for spectroscopic followup,  \citet{Schreiber2018} was successful in obtaining redshifts for 12 sources. 
10 of these were confirmed to be at the correct redshifts ($z>3$), and a mass selected number density for $z>3$ massive quiescent galaxies was obtained based on the ZFOURGE survey. 
Both spectroscopically confirmed and unconfirmed sources spanned $K~21.5-24$. JWST observations showed that the limitation for ground based spectroscopic confirmations for \citet{Schreiber2018} sources was also driven by missing key spectral features in ground based spectra due to atmospheric cutoff \citep{Nanayakkara2024a}.

In this analysis we present a JWST NIRSpec spectroscopic followup and uniform SFH analysis of the remaining 12 galaxies that were beyond the reach of ground based spectroscopy. Given the multiplexing nature of NIRSpec, we also present improved spectroscopy of 7 galaxies that were previously spectroscopically confirmed by \citet{Schreiber2018}.
As outlined in Table \ref{tab:sample}, out of the 19 galaxies used in our analysis 12 galaxies were presented in \citet{Nanayakkara2024a}. 
This includes ZF-UDS-7329, whose formation history was looked into in detail by \cite{Glazebrook2024a} to find that it has formed $\sim10^{11}$\msol\ by $z\sim10$ and quenched for $1>$Gyr by the time it was observed at $z\sim3.2$.
In this paper, we build upon \citet{Nanayakkara2024a} by presenting a detailed analysis of the star-formation histories of the full massive quiescent galaxy candidate sample observed by our program.
In Section \ref{sec:sample} we present our observation and data reduction strategy.  
In Section \ref{sec:sfh} we present an analysis of the stellar populations and star formation histories of our galaxies using \fastpp\ and \prospector\ SED fitting codes, considering parametric and non-parametric (binned) SFHs and various stellar libraries and stellar population models.
In Section \ref{sec:discussion}, we explore our results in the broader context of galaxy evolution and cosmological models, and in Section \ref{sec:conclusions} we present the conclusions and present ideas for future directions of work. 
Unless otherwise stated, we assume a \citet{Chabrier2003} IMF and a cosmology with  H$_{0}= 70$ km/s/Mpc, $\Omega_\Lambda=0.7$ and $\Omega_m= 0.3$. 
All magnitudes are expressed using the AB system \citep{Oke1983}.

\section{Observed Sample} \label{sec:sample}

\begin{deluxetable}{lrll}
\tabletypesize{\scriptsize}
\tablecaption{ $z>3$ massive quiescent galaxy candidates observed by our program \label{tab:sample}}
\tablecolumns{4}
\tablewidth{0pt} 
\tablehead{ \colhead{Name}    &  \colhead{Observation ID}  &  \colhead{Comments }}
\startdata
S18 $z_{phot}$ only \\
ZF-COS-10559 & 301	& Observation 1 rescheduled by \\		 
			 &      & WOPR 88655 due to NIRSpec short.  \\
ZF-COS-14907 &   2	&  \\	
3D-EGS-27584 &	 6  & Partial spectral coverage  \\	
3D-EGS-34322$^{*}$ &	 6  & Presented in \citet{Nanayakkara2024a}.\\	
ZF-UDS-3651	 & 100	& Presented in \citet{Nanayakkara2024a}. \\  	 
ZF-UDS-4347	 & 100	& Presented in \citet{Nanayakkara2024a}. \\  	 
ZF-UDS-6496	 & 100	& Presented in \citet{Nanayakkara2024a}$^{**}$. \\  	 
ZF-UDS-7329	 & 200	& Presented in \citet{Glazebrook2024a} \& \\  
			 &      & \citet{Nanayakkara2024a}$^{**}$\\
ZF-UDS-7542	 & 200	& Presented in \citet{Nanayakkara2024a}. \\  	 
3D-UDS-35168 & 300	& Presented in \citet{Nanayakkara2024a}. \\ 	
3D-UDS-39102$^{*}$ & 300	& Presented in \citet{Nanayakkara2024a}. \\ 	
3D-UDS-41232 & 300	& Presented in \citet{Nanayakkara2024a}. \\	
\hline
S18 low confident spec-$z$	\\
ZF-COS-18842 & 7	& \\
ZF-COS-19589 & 7	& \\
3D-EGS-31322 & 6	& Presented in \citet{Nanayakkara2024a}.\\
\hline	
S18 robust spec-$z$ \\
ZF-COS-20115 & 7	& \\
ZF-COS-20133 & 2	& Partial spectral coverage \\
3D-EGS-18996 & 6	& Presented in \citet{Nanayakkara2024a}.\\
ZF-UDS-8197  & 200	& Presented in \citet{Nanayakkara2024a}.\\
\enddata
\tablecomments{S18 refers to \citet{Schreiber2018}.\\  $^{*}$ Categorized as star-forming based on sSFR and rest-frame $U$, $V$, and $J$ color selections (Figures \ref{fig:sSFRs} and \ref{fig:UVJ}). $^{**}R\sim1000$ spectroscopy presented by \citet{Carnall2024a}. }
\end{deluxetable}

In \cite{Schreiber2018}, we presented ground based spectroscopy for a sample of $z\sim3-4$ massive quiescent galaxy candidates from the ZFOURGE \citep{Straatman2016} and 3DHST \citep{Skelton2014} surveys. 
These galaxies were selected by applying a magnitude cut $K<24.5$, a mass cut of $M\geq 10^{10}$\msol, a photometric redshift cut of $z_{phot}>2.8$, and a $UVJ$ color selection \citep{Williams2009} to select galaxies that are classified as quiescent.  
We cross matched the photometrically selected galaxies with Keck archival data to find galaxies that were observed with MOSFIRE \citep{McLean2012} $H$ and/or $K$ bands. 
There were 24 galaxies that were selected based on the above selection criteria, however, we could only obtain spectroscopic redshift measurements for 12 of the galaxies. 

In JWST Cycle 1 observations, we were awarded 14.4 hours of NIRSpec prime time (GO-2565 ``How Many Quiescent Galaxies Are There at $3<z<4$ Really?" ) to spectroscopically followup the remaining 12 galaxies. 
This analysis presents the full massive quiescent galaxy sample observed by our program. 

Our galaxies are  spread over three HST legacy fields, All-Wavelength Extended Growth Strip International Survey \citep[EGS/AEGIS]{Skelton2014}, The Cosmological Evolution Survey \citep[COSMOS][]{Straatman2016}, and UKIDSS Ultra-Deep Survey \citep[UDS][]{Skelton2014,Straatman2016} fields. Due to the spatial distribution, 7 NIRSpec pointings were required to obtain prism spectroscopy of the 12 galaxies. Due to close clustering, we were also able to target 7 galaxies that were spectroscopically confirmed to be at $z>3$ by \citep{Schreiber2018}. 

Observations were carried out between August 2022 to May 2023. More details about the observations are provided in Table \ref{tab:sample}. Each observation utilized 5 slitlet shutters with 3 dither positions. Each dither position was observed for $657s$, resulting in $33$ min of exposure time per target. 
There was one object that was affected by a failed closed shutter in the MSA. 

All data was reduced using the publicly available STScI JWST pipeline {\tt jwst v1.12.5} using the latest calibration reference files available at the time.  For galaxies that had NIRCam coverage from PRIMER (GO-1837, PI Dunlop) or CEERS \citep[DD-ERS-1345, PI Finkelstein][]{Finkelstein2023a} surveys, NIRCam imaging data was used to calibrate the spectra to match with the total fluxes\footnote{galaxies were cross matched with the DAWN JWST Archive v7 data release \url{https:// dawn-cph.github.io/dja} \citep{Valentino2023a}}. There were three galaxies with no NIRCam imaging coverage (ZF-COS-19589, 3D-UDS-39102, and 3D-UDS-41232). The JWST/NIRCam footprint from the PRIMER and CEERS surveys differ from the HST footprints of the CANDELS/3DHST surveys. 
Consequently, certain regions within the Hubble Legacy fields are not covered by NIRCam. As a result, the three galaxies mentioned above lack corresponding NIRCam photometry for which we used multiband photometric data from 3DHST \citep{Skelton2014}  and ZFOURGE \citep{Straatman2016} surveys to calibrate the spectra. We did extensive tests to make sure that there are no calibration effects applied to the spectral shape of the spectrum based on artificial slit mask images overlaid on the NIRcam images as outlined in \citet{Nanayakkara2024a} and that the errors reported by the STScI JWST pipeline are reasonable as outlined in \citet{Glazebrook2024a}.

\begin{figure*}
\includegraphics[scale=0.75, trim= 0.1 0 0.1 0, clip]{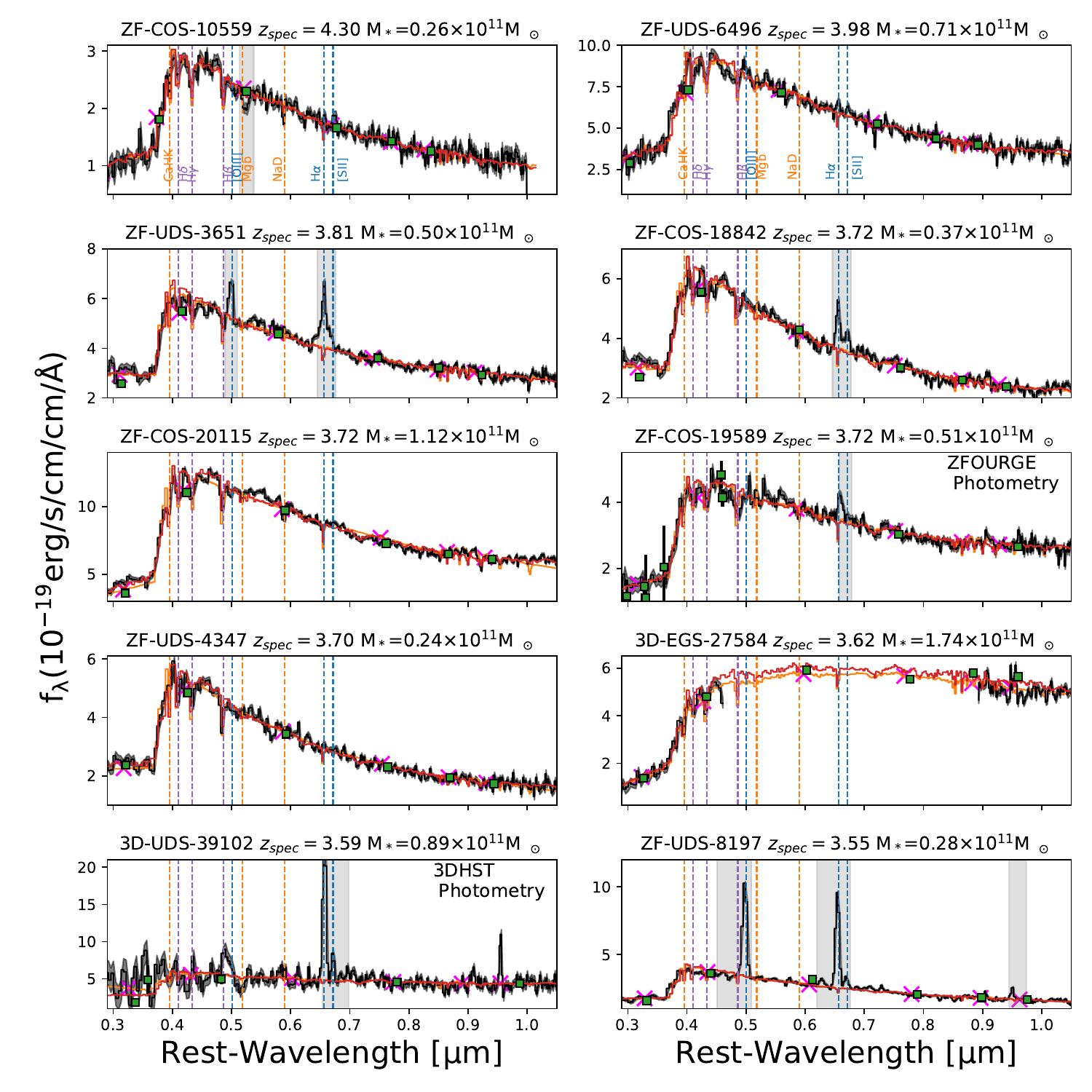}
\caption{The 19 massive quiescent galaxy candidates presented in our analysis.  The observed JWST/NIRSpec PRISM spectra are shown by black with its uncertainty shaded in dark grey. The best fit \fastpp\ utilizing \citet{Bruzual2003} stellar population models and \prospector\ models utilizing the C3K stellar library \citep{Conroy2009} are shown by red and orange respectively. The emission line regions masked for spectral fitting are shown by the vertical light-grey stripes. Multi-band photometric data are shown by the green squares. The \slinefit\ redshifts and \fastpp\ stellar masses are marked in each panel along with key rest-frame optical lines. \fastpp\ and \prospector\ masses estimates agree within $\sim0.4$ dex.
\label{fig:full_qu_sample_spectra_set_1}}
\end{figure*}

\begin{figure*}
\includegraphics[scale=0.75, trim= 0.1 0 0.1 0, clip]{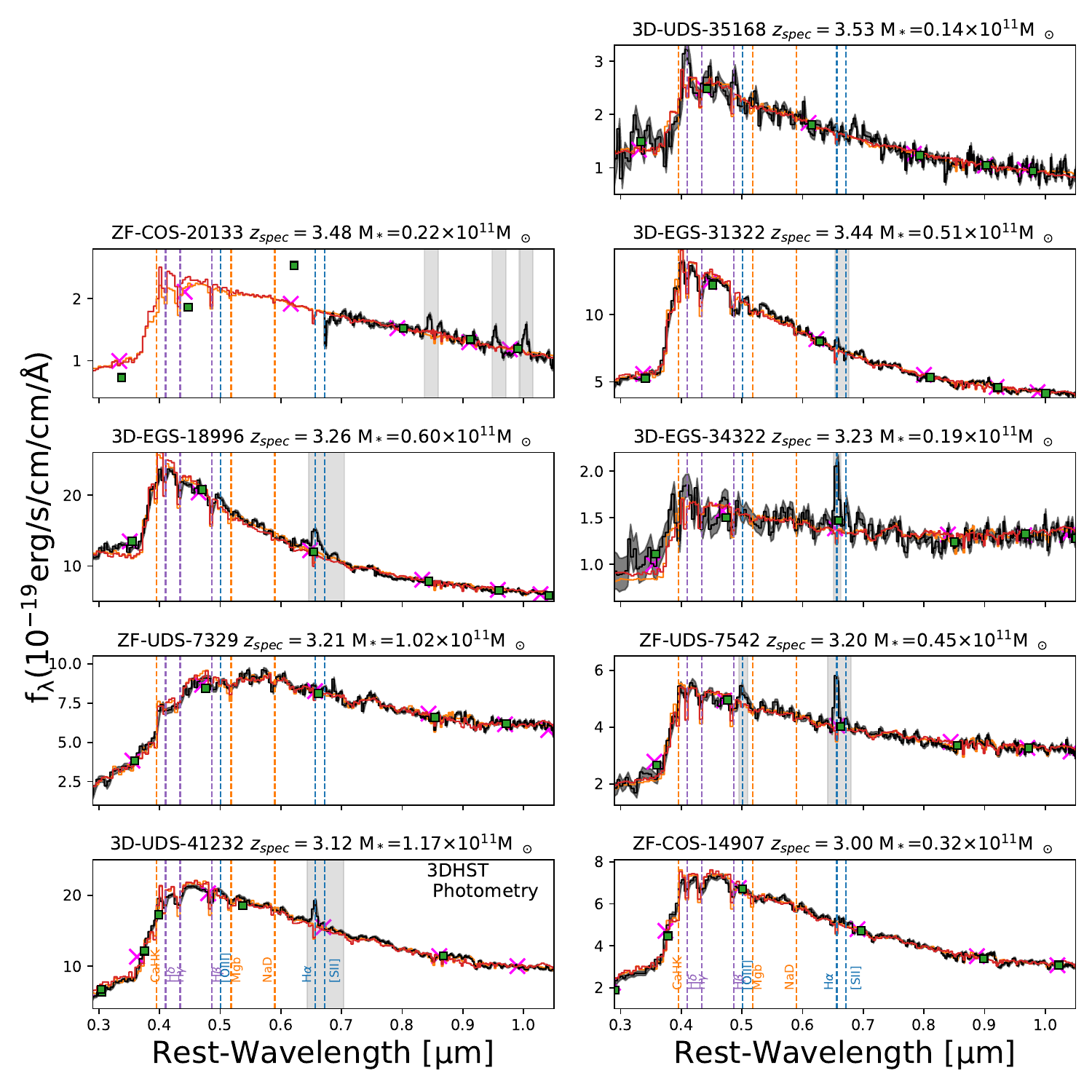}
\caption{Continuation of Figure \ref{fig:full_qu_sample_spectra_set_1}.
\label{fig:full_qu_sample_spectra_set_2}}
\end{figure*}

\begin{figure*}
\includegraphics[scale=0.5, trim= 10 10 10 0, clip]{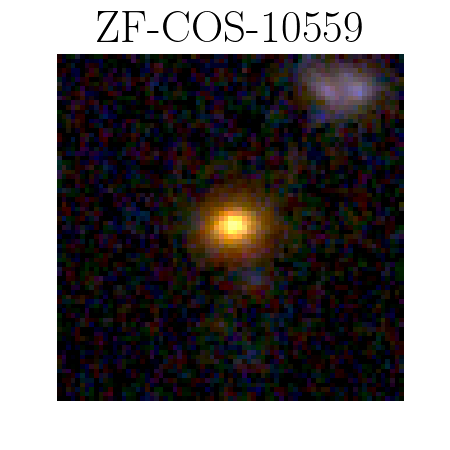}
\includegraphics[scale=0.5, trim= 10 10 10 0, clip]{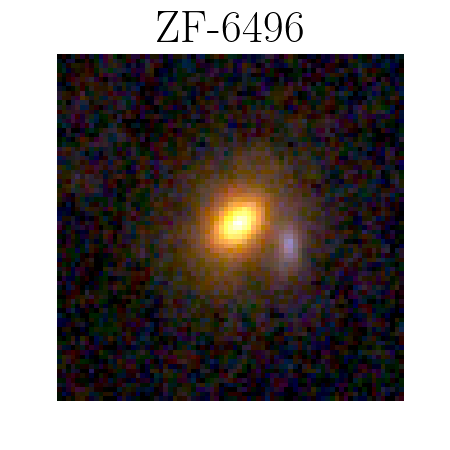}
\includegraphics[scale=0.5, trim= 10 10 10 0, clip]{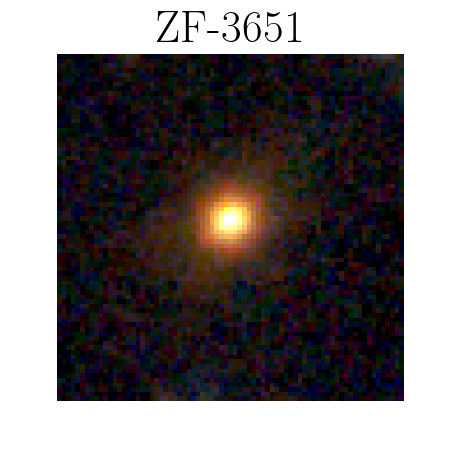}
\includegraphics[scale=0.5, trim= 10 10 10 0, clip]{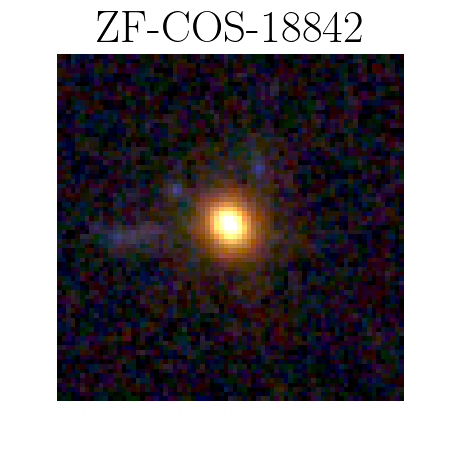}
\includegraphics[scale=0.5, trim= 10 10 10 0, clip]{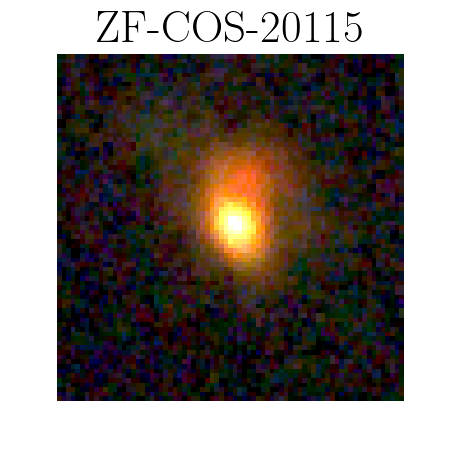}
\includegraphics[scale=0.5, trim= 10 10 10 0, clip]{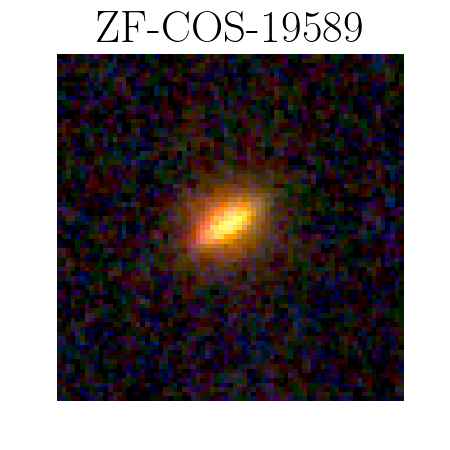}
\includegraphics[scale=0.5, trim= 10 10 10 0, clip]{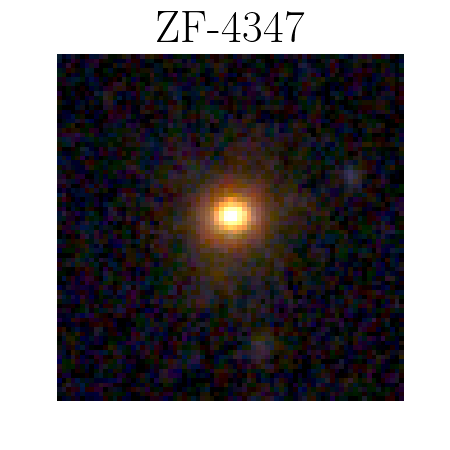}
\includegraphics[scale=0.5, trim= 10 10 10 0, clip]{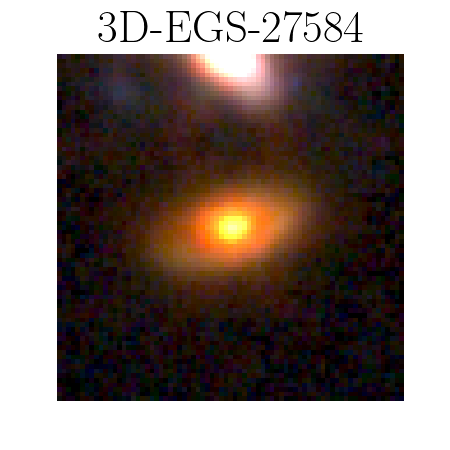}
\includegraphics[scale=0.5, trim= 10 10 10 0, clip]{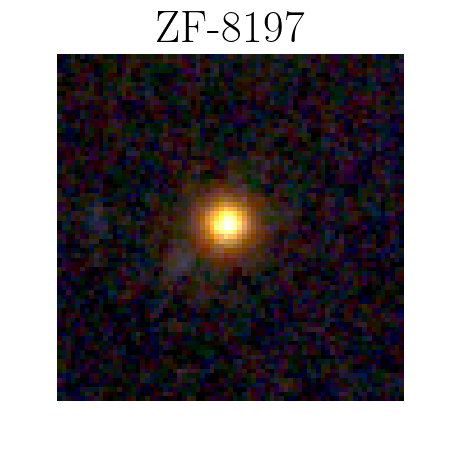}
\includegraphics[scale=0.5, trim= 10 10 10 0, clip]{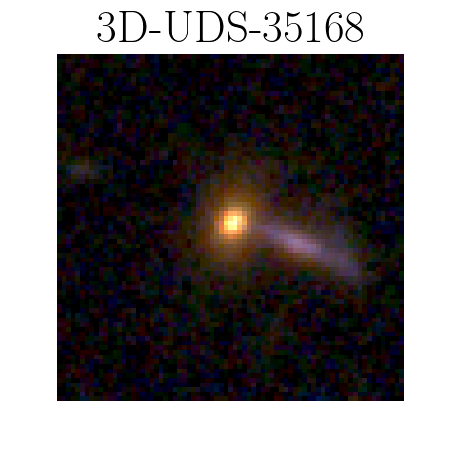}
\includegraphics[scale=0.5, trim= 10 10 10 0, clip]{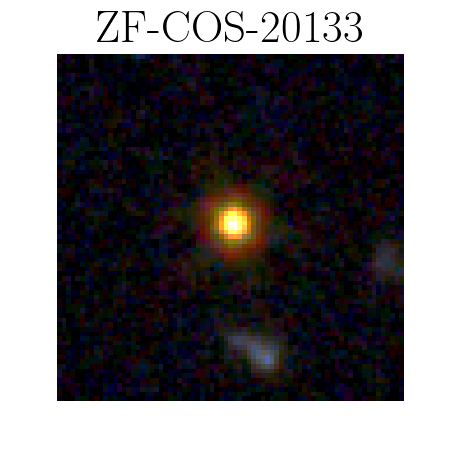}
\includegraphics[scale=0.5, trim= 10 10 10 0, clip]{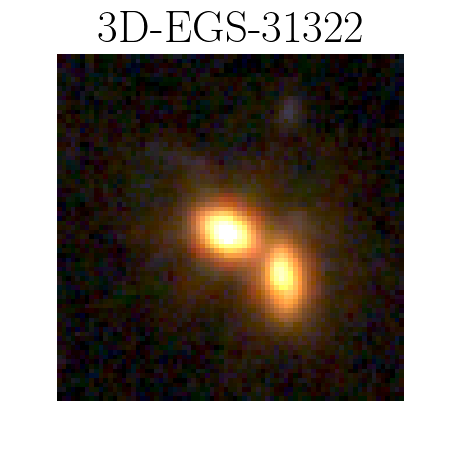}
\includegraphics[scale=0.5, trim= 10 10 10 0, clip]{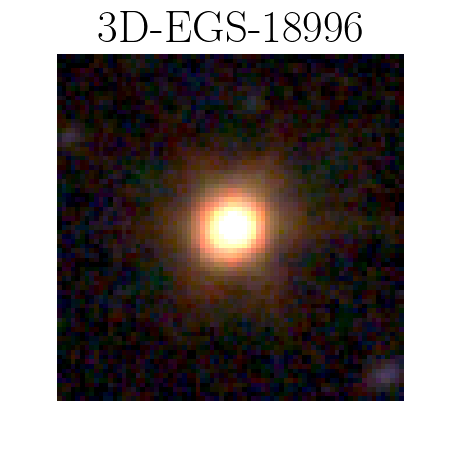}
\includegraphics[scale=0.5, trim= 10 10 10 0, clip]{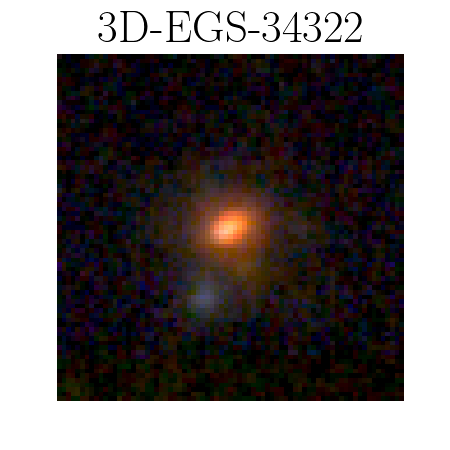}
\includegraphics[scale=0.5, trim= 10 10 10 0, clip]{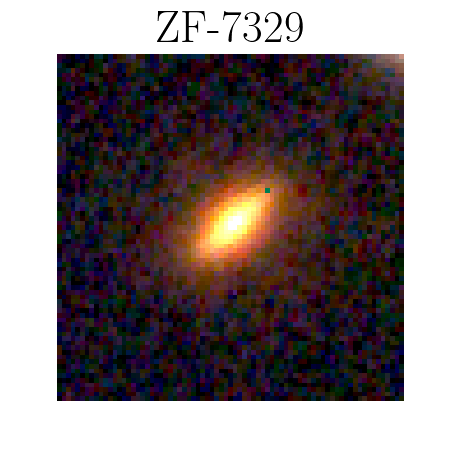}
\includegraphics[scale=0.2250, trim= 10 10 10 0, clip]{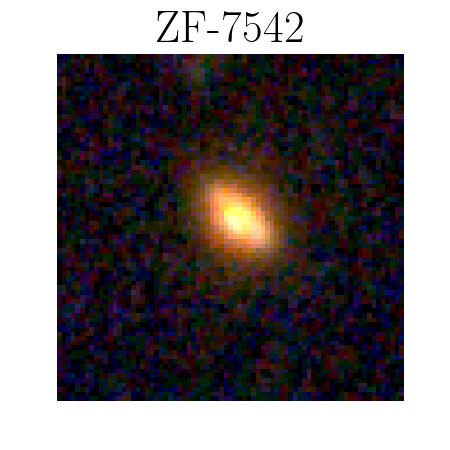}
\includegraphics[scale=0.5, trim= 10 10 10 0, clip]{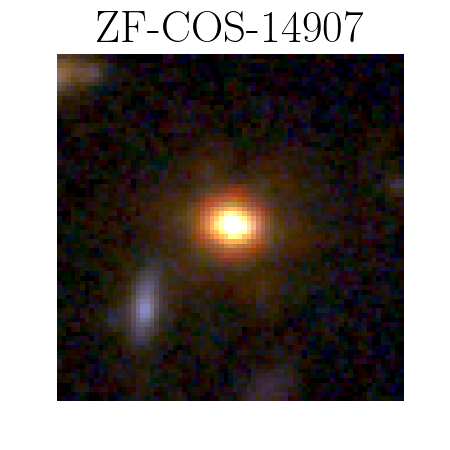}
\caption{Red (F444W), green (F277W), and blue (F150W) color $3\arcsec\times3\arcsec$images of our sample which are covered by JWST/NIRCam imaging.
\label{fig:images}}
\end{figure*}

In Figures \ref{fig:full_qu_sample_spectra_set_1} and \ref{fig:full_qu_sample_spectra_set_2} we show the \citet{Schreiber2018} massive quiescent galaxies observed by our program. The continuum of all galaxies with the exception of 3D-EGS-39102 is observed with a median S/N$>18$. With the exception of ZF-COS-20133, all galaxies have coverage of the Balmer break. The $\sim 2-4\mu$m observed wavelength range of 3D-EGS-27584 falls in the detector gap. We use \slinefit\footnote{\url{https://
github.com/cschreib/slinefit}} to measure redshifts using custom line spread functions as outlined in \citet{Nanayakkara2024a}.  
Redshift errors are determined using 200 Monte-Carlo simulations obtained by perturbing the observed spectrum within its error levels. 
The median redshift error is $\sim0.002$.
All sources have secure spectroscopic redshifts based on the Balmer break and/or emission line detections. All galaxies that were unable to be spectroscopically confirmed by \citet{Schreiber2018} are now confirmed to be at $z>3$. Thus, there are no new low redshift outliers. 
The best-fit redshifts for our sample is presented in  Figures \ref{fig:full_qu_sample_spectra_set_1} and \ref{fig:full_qu_sample_spectra_set_2} (and can also be found in Table \ref{table:fast_results}). 
Color images of our galaxies with JWST/NIRCam coverage are shown by Figure \ref{fig:images}.
The image of 3D-EGS-31322 reveals the presence of a nearby neighboring source. We have clarified that there is no contamination from the neighboring source based on the slit orientation, dithering pattern, and 1D spatial profile modeling.

$\sim70\%$ of the galaxies show clear NaD absorption profiles upon visual inspection.
However, given the low resolution of NIRSpec PRISM mode observations, we are unable to provide tight constraints to the NaD EWs or velocity offsets to distinguish between ISM/stellar atmosphere absorption vs galactic outflow driven absorption.
Similarly, we also find that Mgb absorption of some of our galaxies ($\sim30\%$) to be strong enough to be visually identified at the prism resolution. 
We further discuss the enhanced NaD profiles and the Mgb detections of our sources in Section \ref{sec:NaD}.

\section{Reconstructing Star Formation Histories} \label{sec:sfh}

In this Section we detail the methodology used to analyze the star-formation histories of our galaxies. 
The modelling of SFHs require prior assumptions of the stellar population models, how various ISM and AGN related quantities are handled, and how the SFHs are parameterized \citep[e.g.][]{Leja2019a,Carnall2019a}. 
For this analysis,  we utilize two different SED fitting codes. Firstly we utilize \fastpp\ using the same SFH assumptions as outlined in \citet{Schreiber2018}, so there is a direct comparison of the SFHs of our sample to that reported in \citet{Schreiber2018}, \citet{Nanayakkara2024a}, and \citet{Glazebrook2024a}. 
Next, we use \prospector\ SED fitting code \citep{Johnson2021a} to investigate the difference in the recovered SFHs when using different assumptions for SFHs priors, stellar population models, AGN, and emission line contributions. 
We use a base set of assumptions following \citet{Nanayakkara2024a,Glazebrook2024a} to compare between \fastpp\ and \prospector\ results. 
Finally, we also explore how well the SFHs can be recovered using \prospector\  for a sample of $z\sim3$ massive quiescent galaxies simulated by the Illustris TNG simulation \citep{Pillepich2018a} and compare the mass buildup of our observed galaxies with massive quiescent galaxies from Illustris TNG, Magneticum \citep{Lustig2023a}, and SHARKv2.0 \citep{Lagos2024a} simulations.

For both spectral fitting codes, when JWST NIRCam imaging is available, we utilize photometry from all NIRCam bands that cover the galaxy, as presented in Table \ref{table:sample_photometry}. As aforementioned, three sources have no current public JWST/NIRCam coverage. For these galaxies we use multi band photometric data from the 3DHST survey as presented in \citet{Skelton2014} catalogues.  In both cases,  the spectra are calibrated to the observed photometry, so there are no additional scaling factors introduced in SED fitting to match the observed photometry to fitting.

All spectral fitting also utilize the observed NIRspec spectra, thus, spectra are jointly fit with the photometry. 
Given the large wavelength coverage of prism mode, the spectral resolution and dispersion is heavily non-linear. Thus, the model spectra are convolved with the line spread function and then is resampled to the non-linear dispersion of the prism based on the NIRCam source profile on slit as outlined in \citet{Nanayakkara2024a}. Dispersion for a uniformly illuminated slit as provided by STScI is used for the three sources with no NIRCam coverage.

Spectral fitting is performed twice for galaxies that have uniform $1-5\mu$m NIRSpec coverage (no part of the spectral trace falls in the NIRSpec detector gaps\footnote{see \url{https://jwst-docs.stsci.edu/jwst-near-infrared-spectrograph/nirspec-operations/nirspec-mos-operations/nirspec-mos-wavelength-ranges-and-gaps}}). 
We first consider a base spectral fitting model, where the observed spectra are trimmed at $<0.7\mu$m in rest-frame wavelength. This allows a direct comparison with results presented in \citet{Glazebrook2024a,Nanayakkara2024a} and is also similar to the spectral fitting reported by \citet{Carnall2024a}. 
Next, in Section \ref{sec:spectral_range}, we briefly investigate systemic differences on the recovered average SFHs when the rest-frame NIR spectral range is folded into spectral fitting. 
As shown by Figures  \ref{fig:full_qu_sample_spectra_set_1} and  \ref{fig:full_qu_sample_spectra_set_2}, there are two galaxies (3D-EGS-27584 and ZF-COS-20133) with limited wavelength coverage. 
This is due to the NIRSpec detector gap overlapping with the dispersed light from the source.
For these two sources spectral fitting is limited to this available wavelength window. Photometric data are not trimmed for any of the spectral fitting scenarios.

\begin{table*}
\scriptsize
\centering
\caption{JWST/NIRCam photometry of sources that have NIRCam coverage. All photometry are total fluxes and are normalized to AB=25.}
\label{table:sample_photometry}
\begin{tabular}{||c||c|c||c|c||c|c||c|c||c|c||c|c||c|c||}
\toprule
ID & \multicolumn{2}{c||}{F115W} & \multicolumn{2}{c||}{F150W} & \multicolumn{2}{c||}{F200W} & \multicolumn{2}{c||}{F277W} & \multicolumn{2}{c||}{F356W} & \multicolumn{2}{c||}{F410M} & \multicolumn{2}{c||}{F444W} \\
\cline{2-15}
   & $f$ & $e$ & $f$ & $e$ & $f$ & $e$ & $f$ & $e$ & $f$ & $e$ & $f$ & $e$ & $f$ & $e$ \\
\midrule
ZF-COS-10559 & 0.147 & 0.029 & 0.337 & 0.025 & 1.258 & 0.021 & 3.086 & 0.021 & 3.715 & 0.021 & 4.139 & 0.033 & 4.279 & 0.029 \\
ZF-UDS-6496 & 0.474 & 0.040 & 1.211 & 0.033 & 5.415 & 0.030 & 10.226 & 0.030 & 12.519 & 0.033 & 13.483 & 0.047 & 14.455 & 0.043 \\
ZF-UDS-3651 & 0.627 & 0.026 & 1.121 & 0.021 & 4.208 & 0.021 & 6.789 & 0.021 & 8.904 & 0.024 & 10.255 & 0.035 & 11.011 & 0.032 \\
ZF-COS-18842 & 0.704 & 0.032 & 1.199 & 0.027 & 4.332 & 0.025 & 6.468 & 0.023 & 7.563 & 0.024 & 8.463 & 0.036 & 9.143 & 0.032 \\
ZF-COS-20115 & 0.607 & 0.053 & 1.608 & 0.046 & 8.625 & 0.040 & 14.678 & 0.040 & 18.338 & 0.043 & 21.221 & 0.062 & 23.467 & 0.059 \\
ZF-UDS-4347 & 0.522 & 0.026 & 1.056 & 0.022 & 3.801 & 0.021 & 5.203 & 0.019 & 5.809 & 0.020 & 6.326 & 0.028 & 6.664 & 0.026 \\
3D-EGS-27584 & 0.208 & 0.013 & 0.622 & 0.014 & 3.828 & 0.015 & 9.114 & 0.020 & 14.206 & 0.026 & 19.240 & 0.038 & 22.081 & 0.037 \\
ZF-UDS-8197 & 0.382 & 0.028 & 0.737 & 0.024 & 2.936 & 0.022 & 4.952 & 0.022 & 5.349 & 0.022 & 6.152 & 0.035 & 6.688 & 0.031 \\
3D-UDS-35168 & 0.323 & 0.023 & 0.689 & 0.019 & 2.024 & 0.018 & 2.841 & 0.017 & 3.222 & 0.018 & 3.547 & 0.026 & 3.733 & 0.023 \\
ZF-COS-20133 & 0.295 & 0.016 & 0.340 & 0.015 & 1.533 & 0.014 & 4.030 & 0.014 & 4.020 & 0.013 & 4.599 & 0.019 & 4.820 & 0.017 \\
3D-EGS-31322 & 1.007 & 0.011 & 2.485 & 0.013 & 10.153 & 0.018 & 12.898 & 0.021 & 14.251 & 0.022 & 15.844 & 0.028 & 16.852 & 0.025 \\
3D-EGS-18996 & 2.752 & 0.014 & 6.637 & 0.018 & 17.990 & 0.026 & 20.083 & 0.027 & 21.881 & 0.028 & 23.716 & 0.033 & 24.941 & 0.033 \\
3D-EGS-34322 & 0.182 & 0.017 & 0.550 & 0.021 & 1.308 & 0.018 & 2.481 & 0.015 & 3.476 & 0.018 & 4.811 & 0.027 & 5.478 & 0.027 \\
ZF-UDS-7329 & 0.593 & 0.046 & 1.907 & 0.039 & 7.403 & 0.036 & 13.750 & 0.037 & 18.593 & 0.043 & 22.583 & 0.066 & 24.075 & 0.060 \\
ZF-UDS-7542 & 0.370 & 0.044 & 1.327 & 0.036 & 4.341 & 0.033 & 6.811 & 0.057 & 9.462 & 0.036 & 11.924 & 0.060 & 13.005 & 0.054 \\
ZF-COS-14907 & 0.586 & 0.018 & 2.340 & 0.016 & 6.182 & 0.017 & 8.404 & 0.018 & 10.022 & 0.020 & 11.815 & 0.026 & 12.202 & 0.024 \\
\bottomrule
\end{tabular}
\end{table*}

\subsection{Parametric SFHs with {\tt fast++}}\label{sec:fastpp}

\fastpp\ is a spectral fitting code written in {\tt C++}, with similar functionality to {\tt IDL FAST} code \citet{Kriek2009}. 
The SED fitting procedure utilized by us here largely mirrors that of \citet{Schreiber2018} with a few exceptions and upgrades.
\fastpp\ {\tt v1.5.0} implements a LSF functionality, where the model spectra can be convolved to a user defined wavelength dependent $\sigma$ of the Gaussian LSF. This can be provided using the {\tt SPEC\_LSF\_FILE} option.  
In terms of the stellar population properties, similar to \citet{Schreiber2018}  we use \citet{Bruzual2003} high resolution stellar population models with Padova 1994 stellar tracks \citep{Bertelli1994a} with STELIB \citep{LeBorgne2003} and BaSeL v3.1 \citep{Westera2002a} spectral libraries. 
We use a \citet{Chabrier2003} IMF and a \citet{Calzetti2000} dust law following what was used by \citet{Schreiber2018}.
In \citet{Schreiber2018} the stellar metallicity of the models was kept at \zsol. In this analysis we are performing spectral fitting utilizing the full rest-frame optical wavelengths. Thus, to take optimal use of available data, we allow the stellar metallicity of the models top vary between the full available grid, which is $Z=0.004, 0.008, 0.02, 0.05$ (20-250\% $Z_{\odot}$). 
When transitioning from a fixed-metallicity run to one where metallicity is allowed to vary, the best-fit SFHs can change due to degeneracies between age and metallicity \citep{Conroy2013}. Therefore, direct one-to-one comparisons of the SFHs between our results and those of \citet{Schreiber2018} are not possible.

The SFH parameters used in the fitting is also kept same as \citet{Schreiber2018}, which is specifically suited to model quiescent galaxies at $z>3$. The model use two epochs to describe the SFHs. The main component include an exponentially increasing and decreasing SFH, where the two exponents ($e$-folding times; $\tau_{\text{rise}}$ and $\tau_{\text{decl}}$) are kept as free parameters. 
Additionally, the lookback time that defines the boundary between exponentially increasing and decreasing ($t_{\text{burst}}$) is also allowed to vary freely. This allows the flexibility in the SFH to capture the slow/fast rising SFHs, constant SFHs, and slow/fast quenching SFHs. 
Thus, the primary mode of the SFH can be defined as follows:

\begin{equation}
SFR_{\text{Base}}(t) \propto \begin{cases} 
e^{(t_{\text{burst}}-t)/\tau_{\text{rise}}} & \text{for } t > t_{\text{burst}}, \\
e^{(t-t_{\text{burst}})/\tau_{\text{decl}}} & \text{for } t \leq t_{\text{burst}}
\end{cases}
\label{eq:sfr_base}
\end{equation}

where $SFR_{\text{Base}}(t)$ is the base SFR at lookback time t. However, this base SFR is unable to capture the most recent star-formation episode of a galaxy to where IR and sub-mm observations are sensitive to. Specially, stacked \emph{Herschel} and ALMA measurements are unable to be recovered by  this base SFR parameters. 
In order to alleviate this, a SFR multiplicative factor, $R_{\text{SFR}}$ was introduced to the model to be activated  close to the time of observation of the galaxy. The time window for which this multiplicative factor was applied ($t_{\text{free}}$) to was allowed to vary freely between the last $10-300$ Myr of the galaxies SFH. Thus, the final SFR of the galaxy at lookback time $t$ ($SFR(t)$) can be mathematically expressed as: 

\begin{equation}
SFR(t) = SFR_{\text{Base}}(t) \times \begin{cases} 
1 & \text{for } t > t_{\text{free}}, \\
R_{SFR} & \text{for } t \leq t_{\text{free}}.
\end{cases}
\label{eq:sfr_extra}
\end{equation}
We refer the readers to \cite{Schreiber2018,Schreiber2018b} for further detailed justification of this SFR within the context of massive and quiescent galaxies at $z>3$.

\begin{table}
\centering
\caption{The stellar population and SFH parameters used in \fastpp\ spectral fitting.}
\label{table:fastpp_sfh_param}
\begin{tabular}{l rc c c }
\hline
Free Parameter & Lower Bound & Upper Bound & Step Size \\
\hline
$t_{burst}$ (Gyr)  & 0.01 & tH(z) & $\log_{10}(0.05)$    \\
$\tau_{rise}$ (Gyr) & 0.01 & 3 & $\log_{10}(0.1)$   \\
$\tau_{decl}$ (Gyr) & 0.01 & 3 & $\log_{10}(0.1)$   \\
$R_{SFR}$          & $10^{-2}$ & $10^5$ & $\log_{10}(0.1)$ \\
$t_{free}$ (Myr)   & 10 & 300 & $\log_{10}(0.5)$   \\
\Av &	0		&	6		&	0.1		\\
\hline
\hline
\multicolumn{1}{l}{Free Parameter} & \multicolumn{3}{l}{Values} \\
\hline
\multicolumn{1}{l}{Z} & \multicolumn{3}{l}{0.004, 0.008, 0.02, 0.05} \\
\hline
\hline
\multicolumn{1}{l}{Fixed Parameter} & \multicolumn{3}{l}{Value} \\
\hline
\multicolumn{1}{l}{SSP Model} & \multicolumn{3}{l}{\citet{Bruzual2003}} \\
\multicolumn{1}{l}{IMF} & \multicolumn{3}{l}{\citet{Chabrier2003}} \\
\multicolumn{1}{l}{Attenuation Curve} & \multicolumn{3}{l}{\citet{Calzetti2000}} \\
\multicolumn{1}{l}{$z$} & \multicolumn{3}{l}{$z_{spec}$ from JWST/NIRSpec} \\
\hline
\end{tabular}
\end{table}

Table \ref{table:fastpp_sfh_param} details a summary of the stellar population and SFH parameters used in this analysis. 
All SFR related priors are varied in logarithmic steps as outlined in the table. 
Observed photometry is fit simultaneously with the observed spectra.
\citet{Bruzual2003} models included with \fastpp\ does not include emission line contributions. Thus, for galaxies with strong emission lines (such as \Hbeta, \OIII, \Halpha, \SII), we remove regions where the strong emission lines fall as shown by Figures \ref{fig:full_qu_sample_spectra_set_1} and \ref{fig:full_qu_sample_spectra_set_2} and broad band photometry contaminated by the strong emissions lines are also removed. 
For galaxies with evolved stellar populations, the contribution of the nebular continuum to the optical flux is expected to be negligible compared to the stellar continuum \citep{Byler2017}. This effect strongly depends on stellar age, with younger stars producing a more prominent nebular continuum. The impact is particularly noticeable in the UV range but diminishes significantly at wavelengths longer than the Balmer break, where the nebular continuum becomes much weaker.
The best-fit spectrum is considered as the model with the lowest reduced $\chi^2$ value and its values are considered as the best-fit model parameters. 
1000 Monte Carlo simulations are performed to obtain the 68\%, 95\%, and 99\% confidence intervals of all the parameters obtained through \fastpp\ fitting. Both the observed photometry and spectral fluxes are permutated within their respective 1-$\sigma$ errors in each of the Monte Carlo iterations.

The best fit \fastpp\ models are shown in Figure \ref{fig:full_qu_sample_spectra_set_1} and \ref{fig:full_qu_sample_spectra_set_2}. Here, when available,  the full range between $1-5\mu$m of the NIRSpec spectrum is used for the fitting. As evident, all galaxies are fit to a very high degree of accuracy with a median reduced $\chi^2$ of 4.3. 
3D-EGS-18996 has the highest reduced $\chi^2$ of 8.2 which can be partly attributed to the increased flux of the best-fit model $\sim1\mu$m. 
\fastpp\ is able to model the Balmer/D4000\AA\ breaks accurately. 
At these redshifts, these features fall $\sim 2 \mu$m, where the NIRSpec/PRISM resolution is at the minimum and the highest degree of variation. 
Thus, we found that having an accurate model of the LSF input to \fastpp\ in the fitting process was crucial to obtain reasonable fits around this wavelength with good overall  $\chi^2$ values.

\begin{figure}
\includegraphics[scale=0.65, trim= 0.1 50 0.1 50, clip]{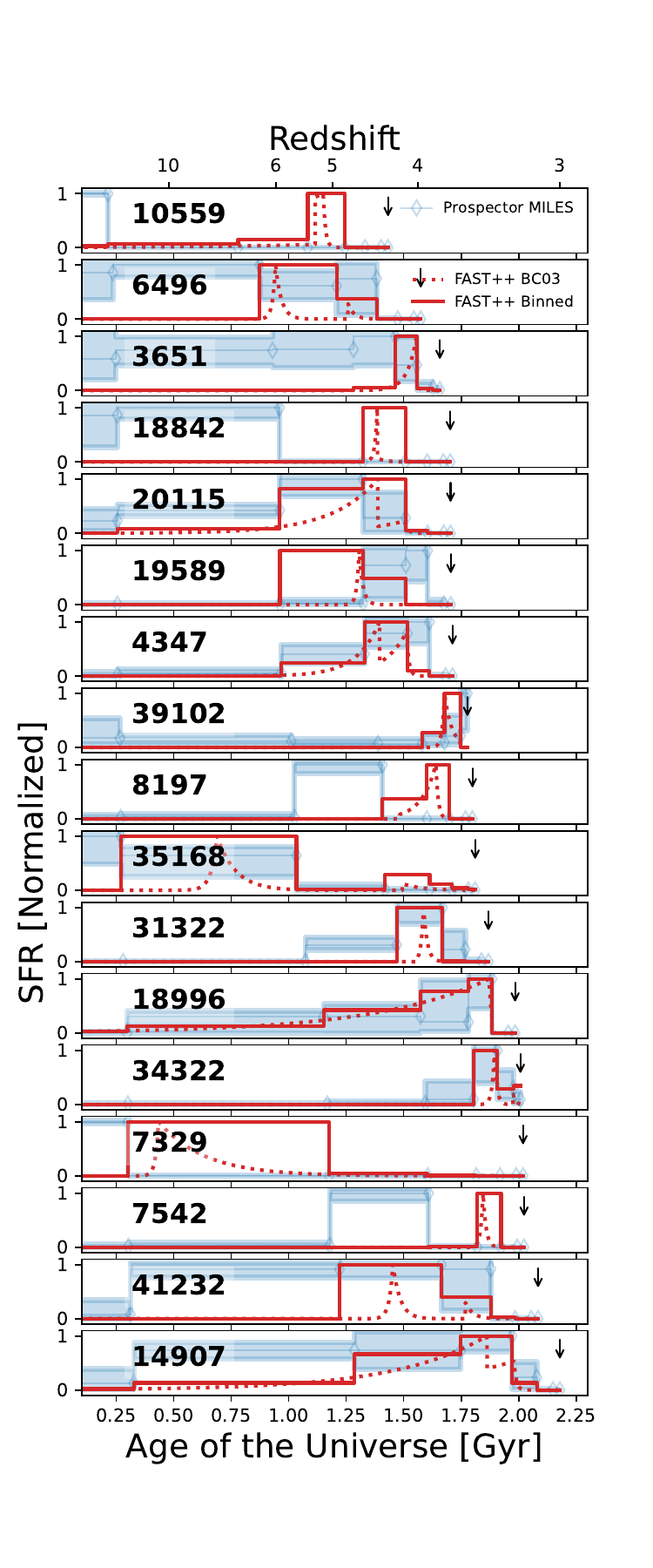}
\caption{The comparison between \prospector and \fastpp\ SFH reconstructions for our observed sample. Each panel represents a galaxy as labelled and is organized from the highest observed redshift to the lowest. 
The \prospector\  maximum a posteriori SFH for our base model is shown by blue with its associated 1-$\sigma$ errors shaded. The red dotted line is the SFH reconstructed using the best-fit parameters from the \fastpp\ fitting. The red solid line shows the \fastpp\ SFH reconstruction binned in the same time windows utilized by \prospector\ to aid direct comparison between \prospector\ and \fastpp. Arrows denote the observed redshift of the galaxy. The maximum SFR of each galaxy is normalized to unity to enhance the clarity of the image. 
\label{fig:full_qu_sample_sfh_comp}}
\end{figure}

\begin{figure}
\includegraphics[scale=0.70, trim= 10 55 20 65, clip]{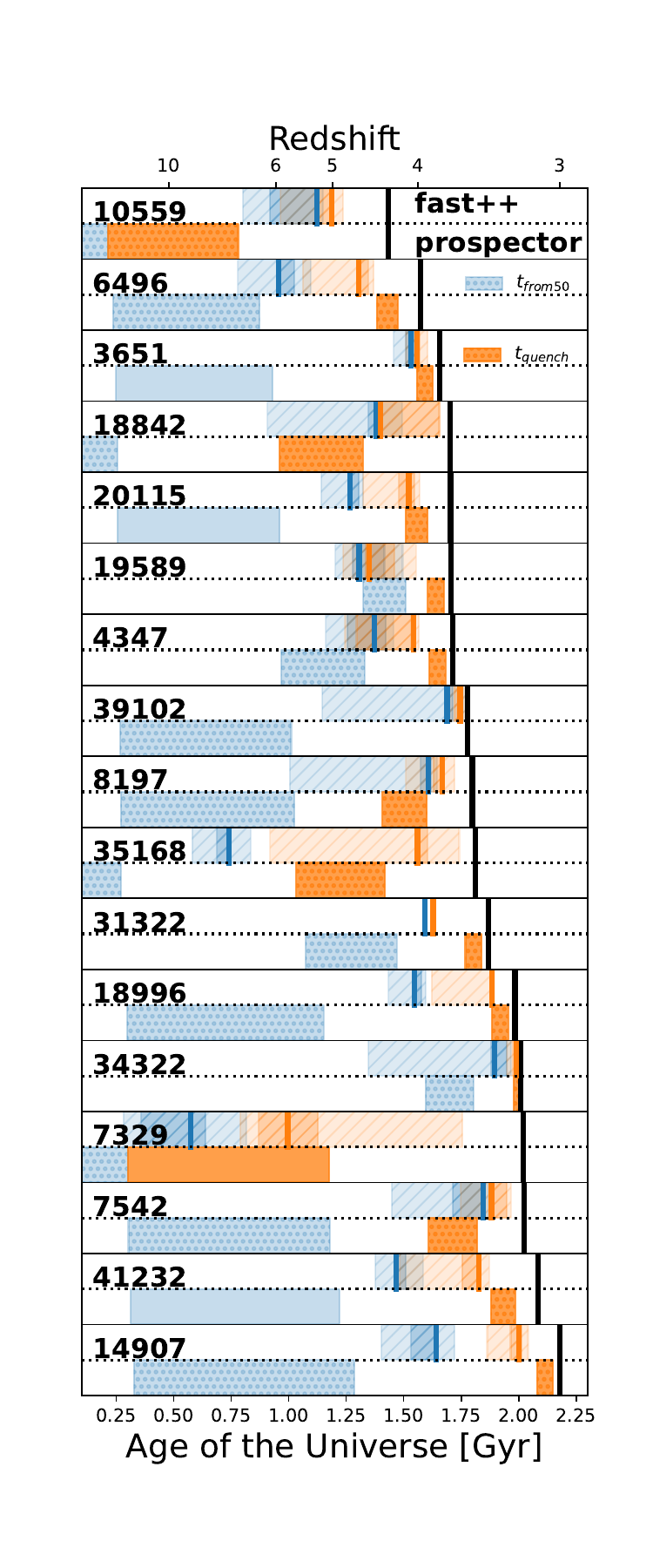}
\caption{Comparison of formation and quenching timescales between \fastpp\  and \prospector\  SED fitting codes.
Each panel represents a galaxy as labelled and is organized from the highest observed redshift to the lowest. Each galaxy panel is divided into two sub panels as denoted by the horizontal dotted lines. For each galaxy, the top panel show results from \fastpp\ and bottom panel shows results from \prospector.
\prospector\ fitting utilize a binned parameterization for the SFHs, thus the formed and quenched time is defined by the bin that satisfy the imposed criteria. The time window the galaxy reach the 50\% of its formed stellar mass is highlighted in blue and the quenching time window is highlighted in orange. If a galaxy does not satisfy the quenched criteria, the quenching time window is not shown for that galaxy. The vertical solid line denotes the observed redshift of the galaxy.
\label{fig:full_qu_sample_sfh_t50quench_comp}}
\end{figure}

The SFHs with \fastpp\ shows considerable variety between our galaxies and are shown by Figures \ref{fig:full_qu_sample_sfh_comp} and \ref{fig:full_qu_sample_sfh_t50quench_comp}. 
We define the formation time of a galaxy (\(z_{\mathrm{form50}}\)) as the redshift at which it formed 50\% of its total stellar mass by the time of observation. 
We define the quenching time (\(z_{\mathrm{quench}}\)) as the redshift marking the start of the longest contiguous period before observation during which the SFR fell below 10\% of the mean SFR during the galaxy's main formation phase. 
We consider a galaxy to be a fast quencher if it reaches \(z_{\mathrm{quench}}\) within 250 Myr of \(z_{\mathrm{form50}}\); otherwise, we consider it a slow quencher.

3D-UDS-7329 at $z=3.2$ is the first galaxy to have reached 50\% of its observed stellar mass after the Big Bang reaching $\sim6\times10^{10}$\msol\ within the first $\sim600$ Myr of the Universe ($\sim1.5$ Gyr in lookback time). It is also the first galaxy in our sample to have quenched its star-formation by $\sim1$ Gyr of the Universe ($\sim1$ Gyr in lookback time). 
3D-UDS-35168 has also a similar  50\% of observed stellar mass formation time ($\sim700$ Myr from the Big Bang), however, it has $\sim1$ dex smaller stellar mass and an extended SFH resulting it only being quenched $\sim250$ Myr before the time of observation. 
All of our other sources have only formed $\gtrsim1$ Gyr after the Big Bang ($z<6$). 

\begin{figure}
\includegraphics[scale=0.45, trim= 5 15 0.1 0, clip]{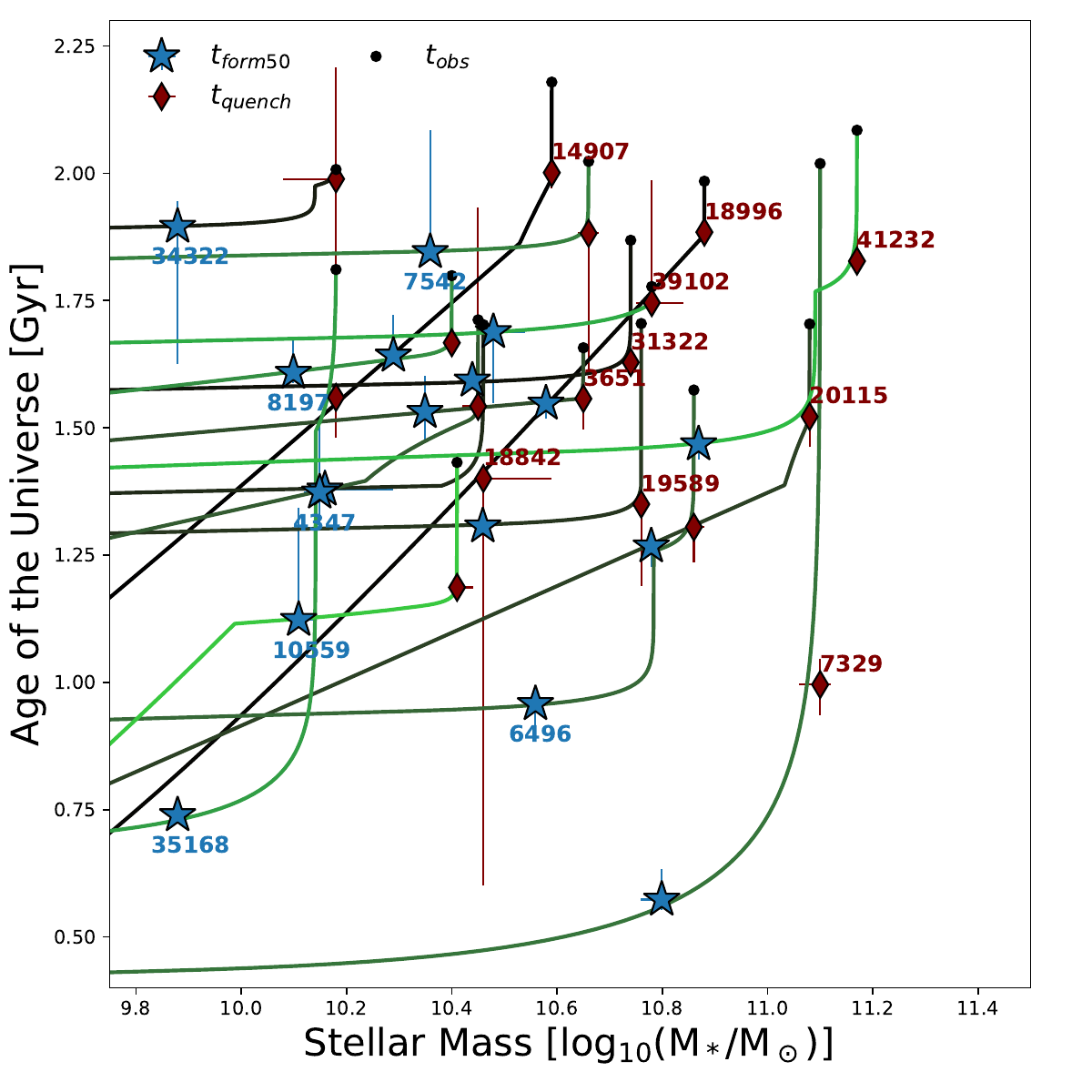}
\caption{50\% of the observed mass formation ($t_{form50}$) and quenching times ($t_{quench}$) for our sample as parameterized by \fastpp. Blue stars on the y-axis represent $t_{form50}$, with the corresponding 50\% observed stellar mass on the x-axis. Similarly, maroon diamonds indicate $t_{quench}$ and the observed stellar mass. Black circles denote the age of the Universe at the time of galaxy observation (with the x value fixed at the observed stellar mass). These points are connected by solid lines for each galaxy. For clarity, the confidence intervals of the \fastpp\ results are not shown in the figure and the solid lines are drawn using different colors. The figure illustrates that galaxies exhibit a variety of mass buildup pathways over time.
\label{fig:fastpp_tform_tquench_vs_mass}}
\end{figure}

ZF-COS-10559 is the highest redshift source in our sample with $z=4.3$. It also has the 2nd fastest quenching time at $\sim1.2$ Gyr after the Big Bang. However, given its high redshift, in look back time it has only quenched within the $\sim250$ Myr before it was observed. 
In fact, with the exception of 3D-UDS-7329 all our sources have only quenched within $\lesssim350$ Myr of the time of observation. The formation, quenching, and observed times of our sample is further visualized by Figure \ref{fig:fastpp_tform_tquench_vs_mass}.

The best-fit values obtained by \fastpp\ fitting are provided in Table \ref{table:fast_results}. 
Since the \citet{Calzetti2000} dust law assumes a simplified uniform screen model and lacks the 2175\AA\ dust bump, we have re-run \fastpp\ using the \citet{Kriek2013} dust model and find that the results are statistically consistent with each other.

\begin{table*}
\begin{threeparttable}
\scriptsize
\centering
\caption{Redshifts and best fit \fastpp\ parameters of our galaxies. $1\sigma$ upper and lower bounds derived using MCMC iterations for {\tt slinefit} spectroscopic redshift measurements and \fastpp\ inferred values are included with the best fit parameters.}
\label{table:fast_results}
\begin{tabular}{lccccccccc}
\toprule
Galaxy ID & $z_{spec}$ &  $M_*$   & Z           &  Av    & $\mathrm{SFR_{10}\tnote{1} }$              & $z_{quench}\tnote{2} $     & $z_{form50}\tnote{3} $       & $t_{\mathrm{SF}}\tnote{4} $  &  $\langle SFR \rangle_{main}\tnote{5} $ \\
          &             & $\rm{log_{10}}(M_\odot)$ &   & mag     & $\rm{log_{10}}(M_\odot/yr)$  &   &   & $\rm{log_{10}}(yr)$      & $\rm{log_{10}}(M_\odot/yr)$ \\
\midrule
\midrule
ZF-COS-10559 & $4.302^{4.304}_{4.298}$ & $10.410^{10.440}_{10.410}$ & $0.050^{0.043}_{0.015}$ & $0.300^{0.330}_{0.100}$ & $-9.240^{-5.510}_{-10.180}$ & $5.012^{5.264}_{4.993}$ & $5.237^{6.129}_{5.211}$ & $8.160^{8.580}_{7.620}$ & $2.280^{2.830}_{1.860}$ \\
\midrule
ZF-UDS-6496 & $3.976^{3.978}_{3.974}$ &  $10.860^{10.880}_{10.860}$ & $0.050^{0.045}_{0.023}$ & $0.000^{0.060}_{0.000}$ & $-2.510^{-2.400}_{-3.910}$ & $4.641^{4.659}_{4.528}$ & $5.937^{5.937}_{5.630}$ & $7.800^{8.700}_{7.800}$ & $3.090^{3.090}_{2.210}$ \\
\midrule
ZF-UDS-3651 & $3.807^{3.809}_{3.806}$ &  $10.650^{10.660}_{10.640}$ & $0.008^{0.007}_{0.004}$ & $1.300^{1.330}_{1.200}$ & $-1.830^{-1.170}_{-2.810}$ & $4.012^{4.012}_{3.984}$ & $4.069^{4.118}_{4.033}$ & $7.680^{7.690}_{7.200}$ & $2.950^{3.440}_{2.950}$ \\
\midrule
ZF-COS-18842 & $3.721^{3.723}_{3.720}$ & $10.460^{10.590}_{10.460}$ & $0.004^{0.039}_{0.004}$ & $0.500^{0.670}_{0.380}$ & $-1.900^{-0.840}_{-9.880}$ & $4.381^{4.399}_{3.812}$ & $4.437^{4.521}_{4.399}$ & $7.300^{8.620}_{7.250}$ & $3.180^{3.230}_{1.980}$ \\
\midrule
ZF-COS-20115 & $3.718^{3.719}_{3.718}$ &  $11.080^{11.080}_{11.070}$ & $0.050^{0.045}_{0.024}$ & $0.500^{0.510}_{0.410}$ & $-4.420^{-2.810}_{-5.910}$ & $4.090^{4.186}_{4.038}$ & $4.753^{4.753}_{4.639}$ & $8.450^{8.490}_{8.290}$ & $2.650^{2.810}_{2.610}$ \\
\midrule
ZF-COS-19589 & $3.717^{3.721}_{3.714}$ &  $10.760^{10.760}_{10.740}$ & $0.004^{0.016}_{0.004}$ & $1.200^{1.190}_{1.060}$ & $-9.230^{-5.680}_{-15.310}$ & $4.515^{4.561}_{4.235}$ & $4.637^{4.720}_{4.341}$ & $7.470^{7.680}_{7.400}$ & $3.310^{3.370}_{3.090}$ \\
\midrule
ZF-UDS-4347 & $3.703^{3.705}_{3.702}$ &  $10.450^{10.450}_{10.420}$ & $0.020^{0.017}_{0.004}$ & $0.400^{0.390}_{0.170}$ & $-6.110^{-4.130}_{-16.230}$ & $4.045^{4.670}_{4.072}$ & $4.452^{4.791}_{4.322}$ & $8.360^{8.360}_{7.550}$ & $2.110^{2.880}_{2.110}$ \\
\midrule
3D-EGS-27584 & $3.624^{3.626}_{3.622}$ & $11.240^{11.250}_{11.230}$ & $0.020^{0.019}_{0.010}$ & $1.300^{1.350}_{1.210}$ & $-10.710^{-3.790}_{-10.800}$ & $4.199^{4.215}_{4.115}$ & $6.546^{6.675}_{6.204}$ & $8.950^{8.960}_{8.920}$ & $2.330^{2.360}_{2.310}$ \\
\midrule
3D-UDS-39102 & $3.587^{3.587}_{3.586}$ &  $10.780^{10.840}_{10.750}$ & $0.020^{0.039}_{0.011}$ & $1.800^{1.870}_{1.660}$ & $0.040^{0.450}_{-1.280}$ & $3.642^{3.684}_{3.642}$ & $3.748^{3.752}_{3.702}$ & $7.740^{7.590}_{7.300}$ & $3.010^{3.480}_{3.150}$ \\
\midrule
ZF-UDS-8197 & $3.550^{3.550}_{3.550}$ &  $10.400^{10.410}_{10.390}$ & $0.008^{0.007}_{0.004}$ & $1.100^{1.090}_{0.970}$ & $-3.920^{-3.870}_{-6.900}$ & $3.788^{3.863}_{3.783}$ & $3.905^{3.974}_{3.887}$ & $7.860^{7.930}_{7.410}$ & $2.530^{3.010}_{2.470}$ \\
\midrule
3D-UDS-35168 & $3.529^{3.536}_{3.525}$ &  $10.180^{10.180}_{10.160}$ & $0.008^{0.007}_{0.004}$ & $0.000^{0.000}_{0.000}$ & $-0.270^{-0.190}_{-0.520}$ & $4.007^{4.019}_{3.918}$ & $7.242^{7.640}_{7.242}$ & $8.150^{8.360}_{8.120}$ & $2.070^{2.100}_{1.820}$ \\
\midrule
ZF-COS-20133 & $3.480^{3.481}_{3.479}$ & $10.340^{10.350}_{10.340}$ & $0.004^{0.004}_{0.004}$ & $0.700^{0.690}_{0.610}$ & $0.160^{0.530}_{-25.320}$ & $4.843^{4.983}_{4.720}$ & $5.143^{5.201}_{4.983}$ & $7.750^{7.930}_{7.600}$ & $2.630^{2.790}_{2.470}$ \\
\midrule
3D-EGS-31322 & $3.435^{3.436}_{3.435}$ &  $10.740^{10.750}_{10.740}$ & $0.050^{0.045}_{0.025}$ & $0.300^{0.280}_{0.220}$ & $-4.060^{-4.060}_{-7.560}$ & $3.864^{3.864}_{3.842}$ & $3.936^{3.950}_{3.936}$ & $7.460^{7.700}_{7.460}$ & $3.290^{3.290}_{3.050}$ \\
\midrule
3D-EGS-18996 & $3.259^{3.261}_{3.258}$ &  $10.880^{10.880}_{10.870}$ & $0.050^{0.045}_{0.023}$ & $0.000^{0.000}_{0.000}$ & $-4.260^{-2.800}_{-4.260}$ & $3.410^{3.410}_{3.403}$ & $4.033^{4.055}_{3.970}$ & $8.740^{8.760}_{8.710}$ & $2.160^{2.180}_{2.140}$ \\
\midrule
3D-EGS-34322 & $3.226^{3.230}_{3.223}$ &  $10.180^{10.190}_{10.080}$ & $0.020^{0.018}_{0.007}$ & $1.900^{2.020}_{1.750}$ & $1.140^{1.230}_{0.190}$ & $3.253^{3.271}_{3.240}$ & $3.393^{3.415}_{3.314}$ & $7.440^{7.830}_{7.430}$ & $2.710^{2.720}_{2.260}$ \\
\midrule
ZF-UDS-7329 & $3.207^{3.209}_{3.206}$ &  $11.100^{11.120}_{11.060}$ & $0.020^{0.018}_{0.010}$ & $0.300^{0.270}_{0.140}$ & $-2.440^{-0.980}_{-20.520}$ & $5.758^{6.390}_{5.219}$ & $8.760^{12.318}_{8.086}$ & $8.380^{8.710}_{8.340}$ & $2.770^{2.780}_{2.440}$ \\
\midrule
ZF-UDS-7542 & $3.204^{3.206}_{3.204}$ &  $10.660^{10.680}_{10.640}$ & $0.050^{0.045}_{0.022}$ & $1.400^{1.370}_{0.990}$ & $0.690^{0.850}_{-2.370}$ & $3.414^{3.424}_{3.314}$ & $3.472^{3.699}_{3.472}$ & $7.410^{8.780}_{7.410}$ & $3.250^{3.250}_{1.920}$ \\
\midrule
3D-UDS-41232 & $3.121^{3.122}_{3.119}$ &  $11.170^{11.170}_{11.160}$ & $0.050^{0.045}_{0.021}$ & $0.400^{0.390}_{0.310}$ & $-1.180^{0.100}_{-2.150}$ & $3.502^{3.629}_{3.502}$ & $4.214^{4.214}_{4.119}$ & $7.910^{8.160}_{7.850}$ & $3.290^{3.350}_{3.080}$ \\
\midrule
ZF-COS-14907 & $2.999^{3.000}_{2.998}$ &  $10.590^{10.600}_{10.580}$ & $0.050^{0.045}_{0.024}$ & $0.300^{0.290}_{0.180}$ & $-6.570^{-1.510}_{-6.490}$ & $3.235^{3.287}_{3.218}$ & $3.837^{4.064}_{3.837}$ & $8.720^{8.860}_{8.720}$ & $1.910^{1.910}_{1.770}$ \\
\midrule
\bottomrule
\bottomrule
\end{tabular}
 \begin{tablenotes}
	\item[1]{The SFR of the galaxy computed over a lookback time of 100 Myrs.}
	\item[2]{The redshift at which the galaxy is considered quenched. }
	\item[3]{The redshift at which the galaxy formed 50\% of its total stellar mass at the time of observation.}
	\item[4]{The length of time where 68\% of the total integrated SFR of the galaxy took place.}
	\item[5]{The average SFR in the time window defined by $\mathrm{t_{SF}}$. }
\end{tablenotes}
 \end{threeparttable}
\end{table*}


\subsection{Binned/Non-parametric SFHs with \prospector}\label{sec:prospector}

Next we use the \prospector\ SED fitting code \citep{Johnson2021a} to infer the SFHs of our galaxies utilizing a flexible approach to parameterize the SFHs. We use a {\tt continuity\_sfh} to parametrize the SFHs of our galaxies. While this is commonly known as a non-parametric SFH, in reality the inference of the SFH is made on fixed time bins in look back time for each galaxy.  
The amount of star-formation that can happen in each time bin is not constrained, thus this allows greater freedom in defining the formation history of galaxies. 
The SFR in each time bin represents the average star-formation happening in that time bin.
The SFH for our galaxies is constrained over 7 SFR bins similar to \citet{Leja2019a}. In lookback time, for each galaxy the first two bins are fixed between 0-30 Myr and 30-100 Myr and the final time bin is fixed to be between 85\%-100\% of the time of the Universe. The remaining 4 bins are distributed evenly between the 100 Myr to 85\% of the time of the Universe in equally spaced logarithmic time bins. We note that the time bins are slightly different to what was utilized for 3D-UDS-7329 in \citet{Glazebrook2024a}, which was driven by the rapid formation of that source within the first few hundred million years of the Universe.

We allow the stellar metallicity, dust optical depth, and the stellar mass to vary freely. 
The stellar metallicity is allowed to vary between $\mathrm{-2.5<log_{10}(Z/Z_\odot)<0.5}$, the dust optical depth following \citet{Calzetti1999} dust law is allowed to vary between $\mathrm{0<\tau_{dust}<4.0}$, and the stellar mass is allowed to vary between $\mathrm{7.0<log_{10}(M_*/M_\odot)<12.0}$. 
The SFR is allowed to vary freely between 7 time bins ($j$) and is parameterized using 6 $\mathrm{log_{10}(SFR)}$ ratios defined as $\mathrm{log_{10}(SFR_j / SFR_{j+1})}$.
We use a \citet{Chabrier2003} IMF, MILES stellar library \citep{Falcon-Barroso2011a}, and MIST stellar isochrones \citep{Paxton2015a,Choi2016,Dotter2016a}. 
To aid with direct comparisons with \fastpp\ we turn off nebular line and continuum emission and AGN contribution in \prospector. The effect of nebular contribution is considered in Section \ref{sec:prospector_em_lines}.

Similar to \fastpp, we run \prospector\ using all available photometry (either from JWST/NIRCam or 3DHST/ZFOURGE) and trim  the observed spectra at $<0.7 \mu$m rest-frame. We further mask out emission lines in the observed spectra using the same masks as used with \fastpp. 
The spectral fits agree well with the observations, with all fits showing a mean reduced \(\chi^2 \sim 3.9 \pm 1.7\). We find that galaxies with \(\gtrsim 1.5~Z_\odot\) tend to have slightly higher \(\chi^2\) values.

\begin{figure}
\includegraphics[scale=0.65, trim= 5 15 0.1 0, clip]{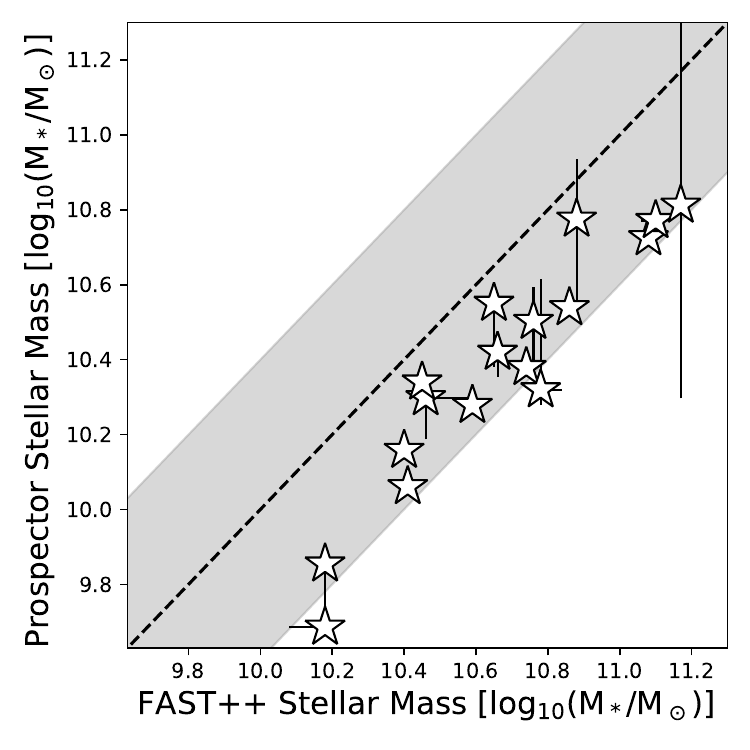}
\caption{Comparison of stellar masses obtained with \prospector\ (using the C3K stellar library) and \fastpp\ (using \cite{Bruzual2003} stellar population models). The black dashed line represents the $x=y$ relation, and the gray shaded area indicates a $\pm0.4$ dex offset.
\label{fig:stellar_mass_comp}}
\end{figure}

Figure \ref{fig:stellar_mass_comp} shows the stellar mass estimates obtained with \fastpp\ and \prospector. All masses agree within a $\sim0.4$ dex scatter, which is consistent with the expected scatter between \fastpp\ and \prospector\ stellar mass estimates \citep{Leja2019b}.

The reconstructed SFHs of the galaxies are shown in Figure \ref{fig:full_qu_sample_sfh_comp}.  
Driven by the parametric nature of the \fastpp\ SFHs, the buildup of stellar mass is gradual over time. In comparison \prospector\ has more variability in the SFH driven by the increased freedom in its SFH parameterization. 
While most galaxies exhibit good overall agreement in their SFH shapes between \prospector\ and \fastpp, the parametric nature of \fastpp's SFH modeling results in more narrowly defined star formation burst episodes.
Visually, ZF-COS-10559 and ZF-UDS-3651 show the largest deviations in their SFHs between \fastpp\ and \prospector.
ZF-COS-10559 according to \prospector\ has an early intense star-burst phase that makes $\sim99.9\%$ of the total formed mass. 
As we discuss in Section \ref{sec:prospector_agn}, ZF-UDS-3651 is a potential AGN with strong emission lines and ZF-COS-10559 hits the highest possible metallicities in both \fastpp\ and \prospector\ fitting. Both of these  could add extra complexity to the spectral analysis.

To determine formation time and quenching times as parameterized by the two SED fitting codes, we integrate the cumulative mass of the galaxies in the SFH bins and find the time bin where 50\% of the total formed mass at the time of the observation is made. 
We define this as the formation time window of the galaxy ($t_{form50}$). We then find the peak SFR of the galaxy and investigate whether the SFR dips below the necessary 10\% threshold of the maximum SFR between the time of the peak SFR to the observed time. If such time exists, we explore whether the galaxies meet this conditions continuously between the peak SFR and the time of observation. If so, we consider the galaxy to be quenched.

In Figure \ref{fig:full_qu_sample_sfh_t50quench_comp} we compare the formation and quenching time of our sample for \fastpp\ and \prospector\ SED fitting codes.  
Given the parametric nature of the SFH, \fastpp\ reports a best-fit formation and quenching times along with their Monte-Carlo uncertainties.
\prospector\ utilizes fixed time bins to infer the variation between the SFR ratios in adjoining time bins, thus, the formation and quenching time are defined in terms of the time bins and are highlighted in blue and orange, respectively. 
While the formation and quenching time windows agree well between the two codes for most of our sources, there is a general tendency for \prospector\ to have older $t_{form50}$ times (50\% of the mass formed at a higher-$z$) compared to \fastpp. 
Non-parametric SFHs from \prospector\ have been shown to point toward older SFHs compared to parametric SFHs in the $z<3$ Universe \citep[e.g.][]{Leja2020a}. 
Our assumption of a flat prior for the {\tt continuity\_sfh} in \prospector\ differs from the exponentially increasing SFH parameterization assumed by \fastpp\ at earlier times. Due to the flat prior, \prospector\ can place more mass formation at earlier times. We further investigate the impact of the assumed SFH prior shape in Section \ref{sec:sfh_prior}.

For galaxies that satisfy the quenching criteria in \prospector, the agreement between the \fastpp\ and \prospector\ quenching time is better than the formation time. 
One source, 3D-UDS-39102, is not classified as quiescent based on \prospector\ results. 3D-UDS-39102 has the S/N in our sample and shows only marginal evidence for a Balmer break based on NIRCam photometry. \fastpp\ results also suggest very recent quenching for this source ($<100$ Myr from the time of observation). 
\citet{Schreiber2018} also found  3D-UDS-39102 to have a lower bound for quenching time to be consistent with 0 Myr with \fastpp, however, our joint spectra + photometry fit from \fastpp\ suggest a 1-$\sigma$ lower bound of $\sim30$ Myr for quenching time. 
3D-EGS-34322 satisfies the quenching criteria only in the most recent SFH bin in \prospector, which is defined between 0 and 30 Myr.  This aligns well with the \fastpp\ quenching time of approximately 20 Myr.

Our results demonstrate that the parametric form of the SFH optimized for $z>3$ massive galaxies presented by \citet{Schreiber2018} largely provides consistent results to the SFHs parameterized using \prospector\ with a more flexible approach. 
However, there are other notable assumptions routinely made in SED fitting such as the choice of input stellar population models, how emission lines are considered, and how effects of an AGN are considered in SED fitting.
For example, the MIST stellar library used with \prospector\ includes effects of stellar rotation while the BC03 stellar population models we used for \fastpp\ do not consider these effects. 
Thus, it is imperative to understand the role different SED fitting assumptions such as the  input stellar libraries, emission line contributions, and AGN contributions play in determining the formation histories of our massive quiescent galaxies. 
Given \prospector\ by design affords flexibility to vary these parameters, we opted to use \prospector\ for the next steps of our analysis and only included galaxies that have full rest-frame optical coverage in our observed spectra.
A brief summary of the \prospector\ assumptions that we utilized is presented in Table \ref{table:prospector_assumptions}.

\begin{table*}
\begin{threeparttable}
\centering
\caption{Summary of SED fitting assumptions explored using \prospector.}
\label{table:prospector_assumptions}
\begin{tabular}{l c }
\hline
\hline
{\bf Free Parameter} & {\bf Range} \\
\hline
\hline
SFH bins (lookback time in Gyr)              &  0-30, 30-100, bin3-6, 0.85$t_{univ}$- $t_{univ}$    \\
Dust optical depth \citep{Calzetti1999}              &  $\mathrm{0<\tau_{dust}<4.0}$ \\
Stellar mass                &  $\mathrm{7.0<log_{10}(M_*/M_\odot)<12.0}$ \\
Ionization parameter\tnote{1} & $-4.0<U<-1.0$ \\
AGN bolometric luminosity fraction\tnote{1} & $10^{-5}<f_{agn}<3.0$ \\
AGN optical depth for individual clouds\tnote{1} \citep{Nenkova2008a} & $5<\tau_{agn}<150$ \\
Marginalize emission lines \tnote{1,2} \citep{Johnson2021a}& $100<\sigma<20000$ km/s \\
\hline
\hline
\multicolumn{2}{l}{MILES stellar library \citep{Falcon-Barroso2011a} \& MIST stellar isochrones \citep{Paxton2015a,Choi2016,Dotter2016a}}\\
\hline
Stellar metallicity & $\mathrm{-2.5<log_{10}(Z/Z_\odot)<0.5}$, \zsol=0.0142  \\
\hline
\hline
\multicolumn{2}{l}{C3K stellar library \citep{Conroy2009} \& MIST stellar isochrones}\\
\hline
Stellar metallicity & $\mathrm{-2.5<log_{10}(Z/Z_\odot)<0.5}$, \zsol=0.0142 \\
\hline
\hline
\multicolumn{2}{l}{BPASSv2.0 stellar population models \citep{Eldridge2017}}\\
\hline
Stellar metallicity & $\mathrm{-2.3<log_{10}(Z/Z_\odot)<0.3}$, \zsol=0.0200  \\
\hline
\hline
{\bf Fixed Parameter} & {\bf Value} \\
\hline
\hline
IMF & \citet{Chabrier2003} \\
SFH prior & {\tt continuity\_sfh} with a flat prior\tnote{3}\\
Spectral smoothing ({\tt smooth\_type}) & LSF \\
emcee sampling \citep{Foreman-Mackey2013}   &  {\tt nwalkers}=1024, {\tt nburn}=[16, 32, 64], {\tt niter}=2056\\
\hline
\hline
\end{tabular}
 \begin{tablenotes}
	\item[1]{Runs are also computed completely turning off this parameter.}
	\item[2]{Only the following lines are marginalized over: \Hbeta, \OIIId, \NIId, \Halpha, \SII}
	\item[3]{Results are also compared with $\text{SFR(t)}\propto e^{t/\tau}$ prior with $\tau=t_{univ}/4$, where $t$ is time in age of the Universe and $t_{univ}$ is the age of the Universe corresponding to the spectroscopic redshift of the galaxy \citep{Leja2019a}.}
\end{tablenotes}
\end{threeparttable}
\end{table*}


\subsection{The Role of input synthetic stellar populations}\label{sec:prospector_ssps}

We investigate three different stellar population models that are inbuilt with {\tt FSPS} \citep{Conroy2009}. 
Firstly, we use the default MILES stellar library \citep{Falcon-Barroso2011a} with MIST stellar isochrones \citep{Paxton2015a,Choi2016,Dotter2016a} to generate the model spectra used by \prospector\ to make inferences on the observed spectra and photometry of our sources. 
MILES library has a higher resolution of $R\sim3000$ in the optical and a lower resolution at $\lambda>7500$\AA. 
Secondly, we use the C3K stellar library used in {\tt FSPS} \citep{Conroy2009}. Here the spectral templates are sampled at a $R\sim3000$ between $2750<\lambda<9100$. At $z\sim3.0$, this cutoff translates to $3.64\mu$m in the observed frame. 
Finally, we also consider BPASSv2 stellar population models \citep{Eldridge2017}. BPASS spectral resolution is comparable to C3K, but has effects of binary stellar evolution incorporated in stellar evolution. BPASS spectra are sampled at 1\AA\ uniform sampling but underlying stellar atmospheric models will have a lower resolution, possibly limited by the resolution of the BaSeLv.31 stellar library.

\begin{figure}
\includegraphics[scale=0.65, trim= 0 0 0 0 , clip]{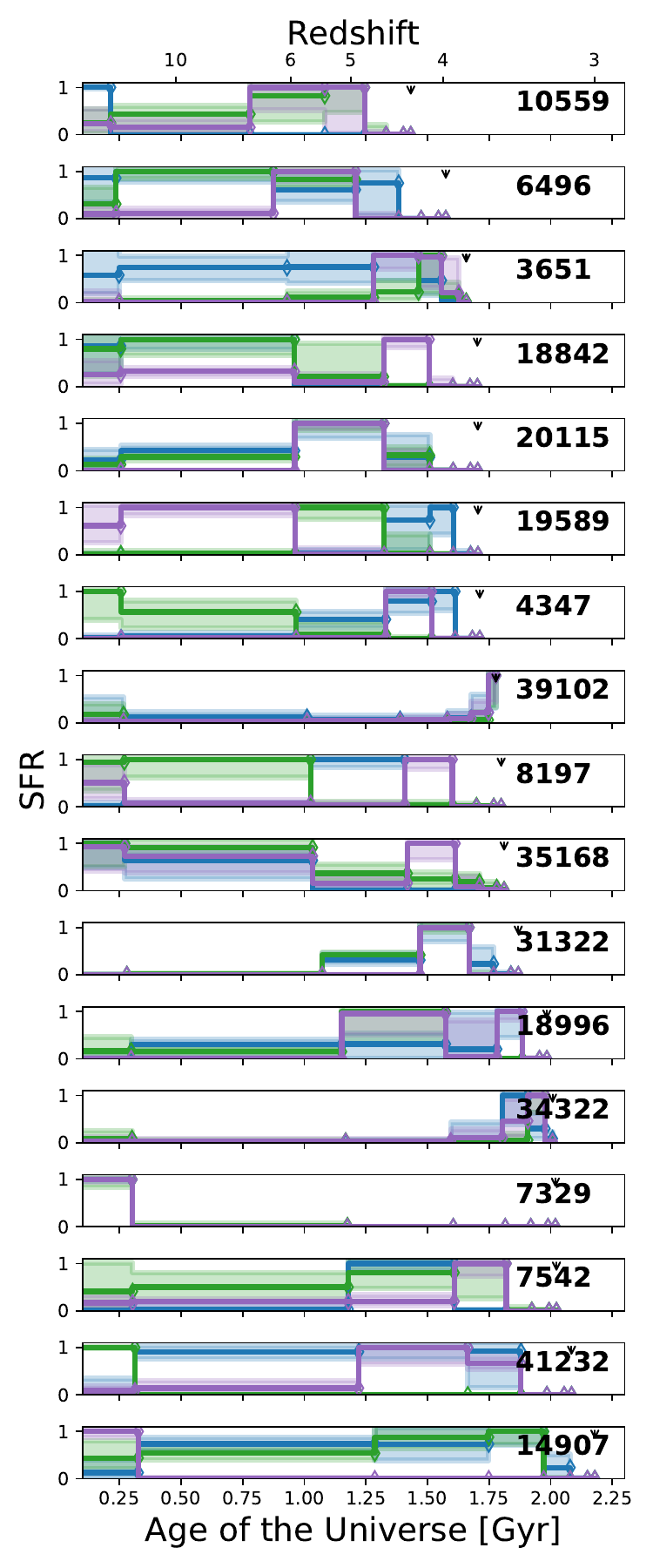}
\caption{Reconstruction of the SFHs of our sample using the MILES \citep{Falcon-Barroso2011a} stellar library (in blue), C3K \citep{Conroy2009} stellar library (in green), and BPASS \citep{Eldridge2017} stellar population (in purple) models. 
Each panel represents a galaxy as labeled and is organized from the highest observed redshift to the lowest. 
The SFHs are shown by the solid lines with associated 1-$\sigma$ errors shaded. Diamond symbols show the start and end of each time bin. Arrows denote the observed redshift of the galaxy. 
\label{fig:full_qu_sample_pros_comp_ssps_trim}}
\end{figure}

Fitting is performed with similar parameters as explained in Section \ref{sec:prospector} using the three spectral libraries/stellar population models. 
SFHs reconstructed using the MILES stellar library \citep{Falcon-Barroso2011a} are shown in blue, the C3K stellar library \citep{Conroy2009} in green, and the BPASS stellar population models \citep{Eldridge2017} in purple.
The residuals of the best-fits are shown in Appendix Figure \ref{fig:best-fit-residuals}. 
Observed spectra are trimmed at rest-frame $<0.7\mu$m. \prospector\ recovered maximum a posteriori SFHs are shown in Figure \ref{fig:full_qu_sample_pros_comp_ssps_trim}. The reduced $\chi^2$ of spectral fits between the three models agree well within $1\sigma$ of each other.  
All of our  sources  show largely consistent formation histories between the 3 models. 
ZF-UDS-4347 is a notable exception where the C3K models prefer a very early formation while the other models prefer a later growth.

\begin{figure}
\includegraphics[scale=0.7, trim= 0.1 55 0.1 65, clip]{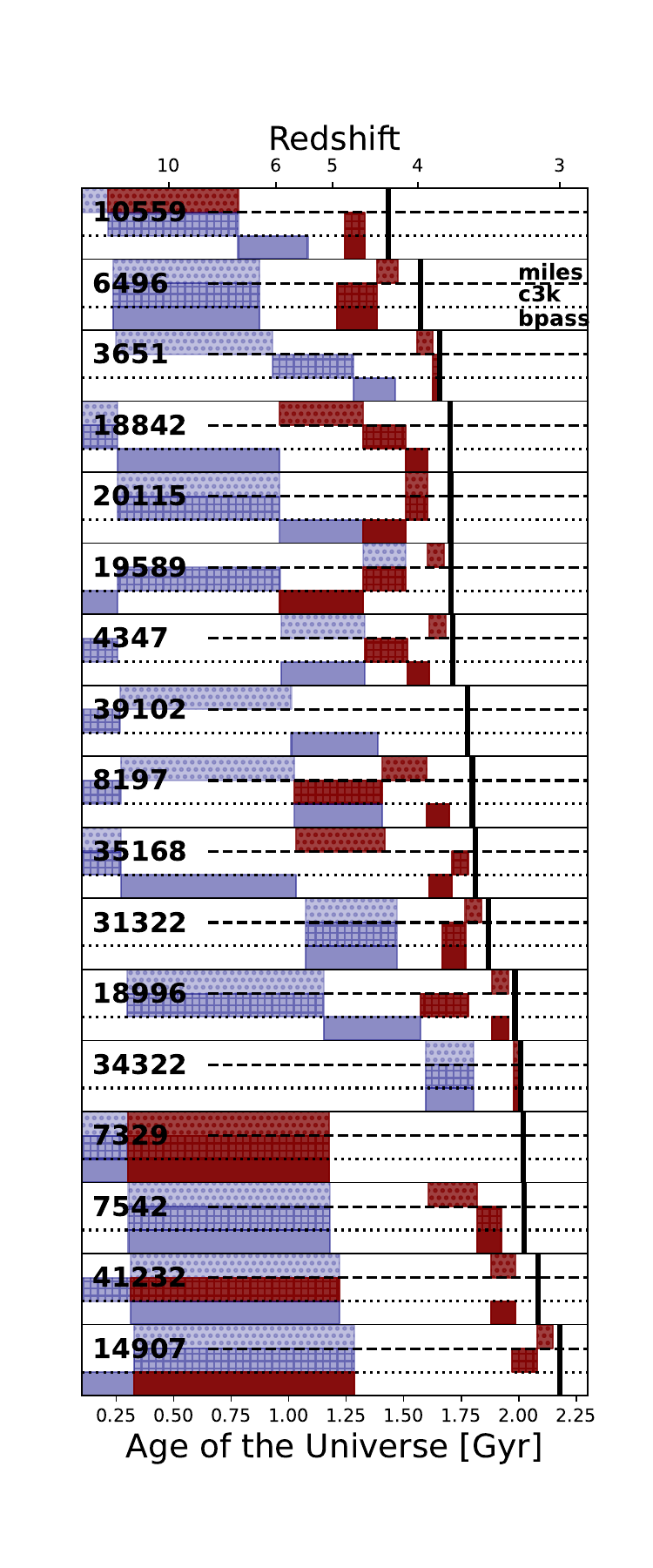}
\caption{Similar to Figure \ref{fig:full_qu_sample_sfh_t50quench_comp}, but a comparison of formation and quenching timescales between different SSP models used within \prospector. For each galaxy, the top panel show results from MILES stellar library, the middle panel shows results when using C3K stellar library, and the lower panel shows results when BPASS SSPs are used. 
The time window for which the galaxy reaches 50\% of its formed stellar mass is highlighted in purple and the quenching time window is highlighted in maroon with slightly varying shades between stellar libraries/SSPs to aid in comparison. If a galaxy does not satisfy the quenched criteria, the quenching time window is not shown for that galaxy. 
\label{fig:full_qu_sample_sfh_t50quench_comp_ssps}}
\end{figure}

We further explore the variation in results between these three models using the definitions for galaxy formation and quenching time scales as discussed in Section \ref{sec:prospector}. 
We show the reconstructed formation and quenching time scales for our galaxies in Figure \ref{fig:full_qu_sample_sfh_t50quench_comp_ssps}. 
The formation and quenching times between the three models largely agree with each other. 
The notable exception is ZF-UDS-4347, where the C3K models prefer an earlier formation time compared to the other models. 3D-UDS-39102 is considered not quenched by all three models. Similarly, the quenching time bin for 3D-EGS-34322 is also in the latest time bin for all three models.
There are no instances where the choice of model determines whether a galaxy is classified as quenched or not.

Galaxy evolution is a statistical study. Galaxy evolution models by design are tuned to reproduce a diverse range of galaxy scaling relations that describe the average properties of the observable Universe \citep{Crain2023a}. Therefore, it is plausible that the models find it challenging to reproduce the properties of individual sources that are outliers from the general galaxy population.
In Figure \ref{fig:full_qu_sample_pros_comp_ssps_trim} we have shown that the input stellar population models used in the fitting can have some effect on SFH recovery in an individual galaxy basis. 
In the context of our analysis, we argue that the average formation histories of massive galaxies are more important than individual reconstructions of their formation histories.

\begin{figure}
\includegraphics[scale=0.70, trim= 0.1 10 0.1 10, clip]{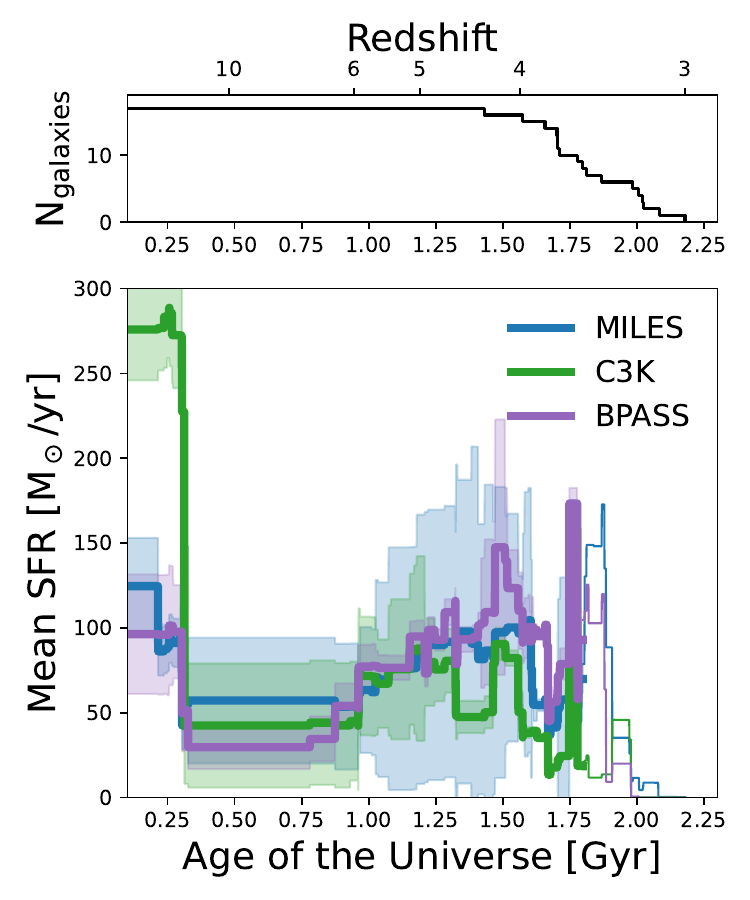}
\caption{
The {\bf top} panel shows the number of galaxies that cover the given age/redshift which is used to compute the average SFH of the full sample. 
The {\bf bottom} panel shows the \prospector\ recovered average SFHs of our sample using the MILES \citep{Falcon-Barroso2011a} stellar library,  C3K \citep{Conroy2009} stellar library, and  BPASS \citep{Eldridge2017} stellar population models. The 1-$\sigma$ scatter of the mean parameterized by the normalized absolute deviation is 
highlighted for each stellar model fit. 
Once the sample is reduced to less than 50\%, the lines are displayed with reduced thickness, and the scatter is not shown.
\label{fig:full_qu_sample_pros_average_sfhs_trim}}
\end{figure}

To investigate this further, we explore if the 3 different stellar libraries/stellar population models used in our analysis provide consistent results to the average formation history of our sources. 
We resample the SFR of each galaxy in 1 Myr increments up to 2200 Myr and compute the average SFR at each time step for our sample, weighting by the number of galaxies in each time step. 
The 1-$\sigma$ scatter at each time step, calculated using the normalized absolute deviation, is considered as the uncertainty in the SFH.
In Figure \ref{fig:full_qu_sample_pros_average_sfhs_trim} we show the evolution of the average SFH for our galaxies as a function of cosmic time. 
The average SFR evolution across the three models is in good agreement. All models show a consistent increase in SFR for ages greater than $1$~Gyr, peaking around $1.5$~Gyr. This peak is followed by a decline in the average SFHs due to a reduced number of galaxies beyond $1.5$~Gyr and the fact that our sample consists of massive quiescent galaxies with limited SFRs in the last few hundred Myr.
All models show an enhancement of SFR in the first $\sim250$ Myr, with the C3K model exhibiting the largest increase.

From Figure \ref{fig:full_qu_sample_pros_average_sfhs_trim} it is clear that regardless of the spectral library/stellar population model used with \prospector, the average SFHs are statistically consistent with each other. 
The major deviation is only observed in the first $\sim250$ Myr of the Universe. 
The NIRSpec prism largely covers the rest-frame optical regime. With continued star-formation features from younger stellar populations are likely to dominate the observed spectrum. This, combined with the lower spectral resolution means that spectral fitting will have less sensitivity to F and G star features that would indicate the nature of the oldest stellar populations in a galaxy. 
Therefore, a larger uncertainty is expected in this first $\sim250$ Myr window which may translate to deviations between the different models.

\subsection{The role of the spectral range used in full spectral fitting}\label{sec:spectral_range}

To investigate the effect of utilizing the rest-frame NIR in spectral fitting, we use the \prospector\ C3K stellar library to fit the spectra of the galaxies, covering the full 1–5~$\mu$m spectral range. 
We utilize C3K for the comparison here because its spectral resolution is higher at longer wavelengths compared to the other models. Specifically, at wavelengths $>9100\,\text{\AA}$, the critically sampled spectral resolution of the C3K models remains at $R \sim 500$ \footnote{\href{https://github.com/cconroy20/fsps/blob/82a873508d500ca353bbb922459bf928498f7a72/SPECTRA/C3K/readme.md}{See README here}}, which is comparable to the resolution of the observed data \citep{Nanayakkara2024a}.

\begin{figure*}
\includegraphics[scale=0.7, trim= 0.1 10 0.1 10, clip]{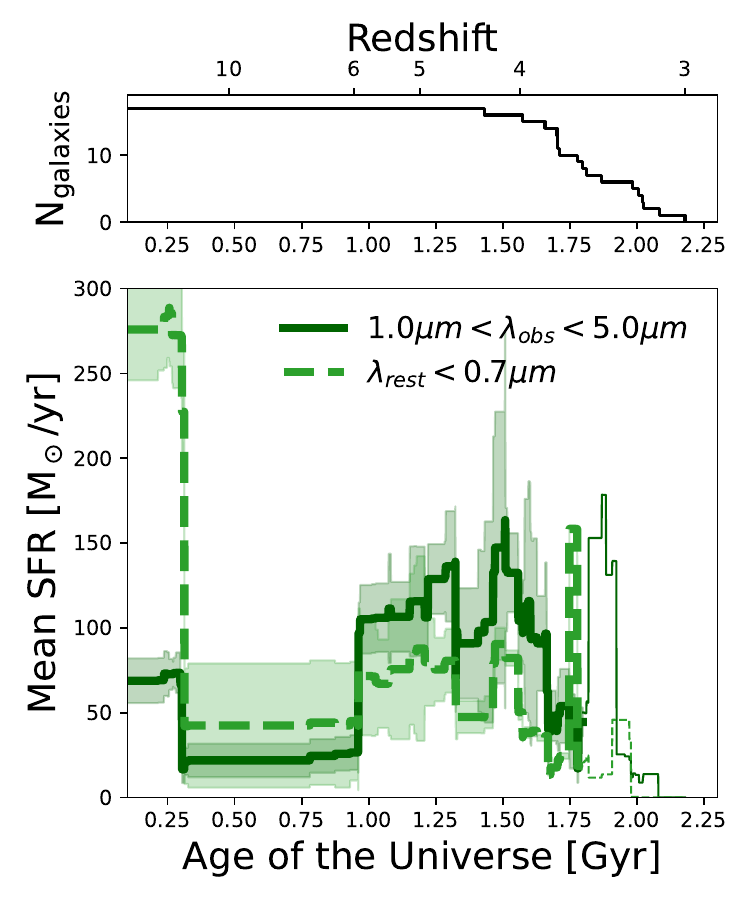}
\includegraphics[scale=0.7, trim= 0.1 10 0.1 10, clip]{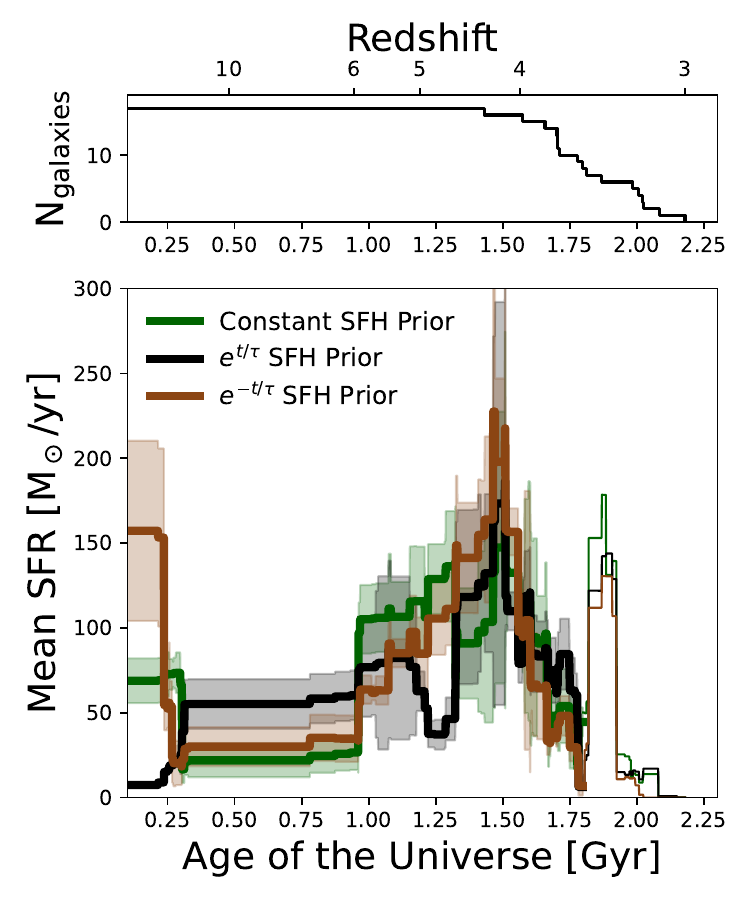}
\caption{Similar to Figure \ref{fig:full_qu_sample_pros_average_sfhs_trim}: {\bf Left:} Shows the average \prospector\ C3K model SFH reconstructions for our observed sample when the full available spectrum between 1-5$\mu m$ is utilized for spectral fitting.  {\bf Right:} The comparison between average \prospector\ C3K model SFH reconstructions with a flat, an exponentially increasing ($e^{t/\tau}$), and an exponentially decreasing ($e^{-t/\tau}$)  {\tt continuity\_sfh} prior. 
\label{fig:full_qu_sample_pros_average_sfhs_fullspec}}
\end{figure*}

The best fit \prospector\ models are shown in Figure \ref{fig:full_qu_sample_spectra_set_1} and \ref{fig:full_qu_sample_spectra_set_2}. 
The reconstructed average SFHs of the galaxies are shown in Figure \ref{fig:full_qu_sample_pros_average_sfhs_fullspec}. The overall shape of the reconstructed SFHs shows good agreement between the $\lambda_{\text{rest}} < 0.7\,\mu\text{m}$ fits and the full $\lambda_{\text{observed}} = 1$–$5\,\mu\text{m}$ fits. The largest deviation is observed in the earliest time bin, which parameterizes the first $\sim250$ Myr of the Universe. When the full spectral range is utilized in the fitting, the enhancement of the SFR at the earliest times decreases, bringing the average SFR of C3K into close agreement with the SFRs observed for the MILES and BPASS models as shown by Figure \ref{fig:full_qu_sample_pros_average_sfhs_trim}.

\subsection{The role of the SFH prior}\label{sec:sfh_prior}

The non-parametric SFH spectral fitting of our analysis performed with \prospector, uses a {\tt continuity\_sfh} model with a flat prior. We utilize seven fixed time bins, as detailed in Table \ref{table:prospector_assumptions}. 
We use smaller time steps for the most recent bins to achieve tighter constraints on recent SFH episodes, while employing logarithmically spaced bins for intermediate epochs to efficiently cover the Universe's age with sufficient resolution.
The variability of the SFRs between the bins is parameterized using the logarithmic ratio of the SFR in adjacent bins, defined as:
\begin{equation}
	\log \text{SFR\_Ratio} = \log_{10}(SFR(i)/SFR(i+1))
\end{equation}
where $i$ represents the closest time bin in lookback time. A constant prior with a mean of 0, a standard deviation of $\sigma = 0.3$, and 2 degrees of freedom, following a Student's T-distribution, is used to sample the posteriors in the Markov Chain Mote Carlo. This resembles a constant SFH prior.

In this Section, we test the role of the assumed SFH prior used in SED fitting. 
Firstly, motivated by the average accretion rate of halo mass over cosmic time \citep[][]{Dekel2013a,Turner2025a}, we apply an exponentially increasing SFH prior for the $\log \text{SFR\_Ratio}$ to explore how it affects the average SFHs of our observed sample. Thus, we modify the constant mean of 0 prior in each of the 6  $\log \text{SFR\_Ratio}$  bins following $SFR(t_i)=e^{t_i/\tau}$, with $\tau=t_{univ}/4$, where $t_i$ is the mean time in age of the Universe and $t_{univ}$ is the age of the Universe corresponding to the spectroscopic redshift of the galaxy \citep{Leja2019a}. All other parameters are unaltered. 
We use the \prospector\ C3K stellar library to fit the spectra of the galaxies utilizing the updated prior covering the full 1–5~$\mu$m spectral range.

\begin{figure*}
\includegraphics[scale=0.60, trim= 0 0 0 0 , clip]{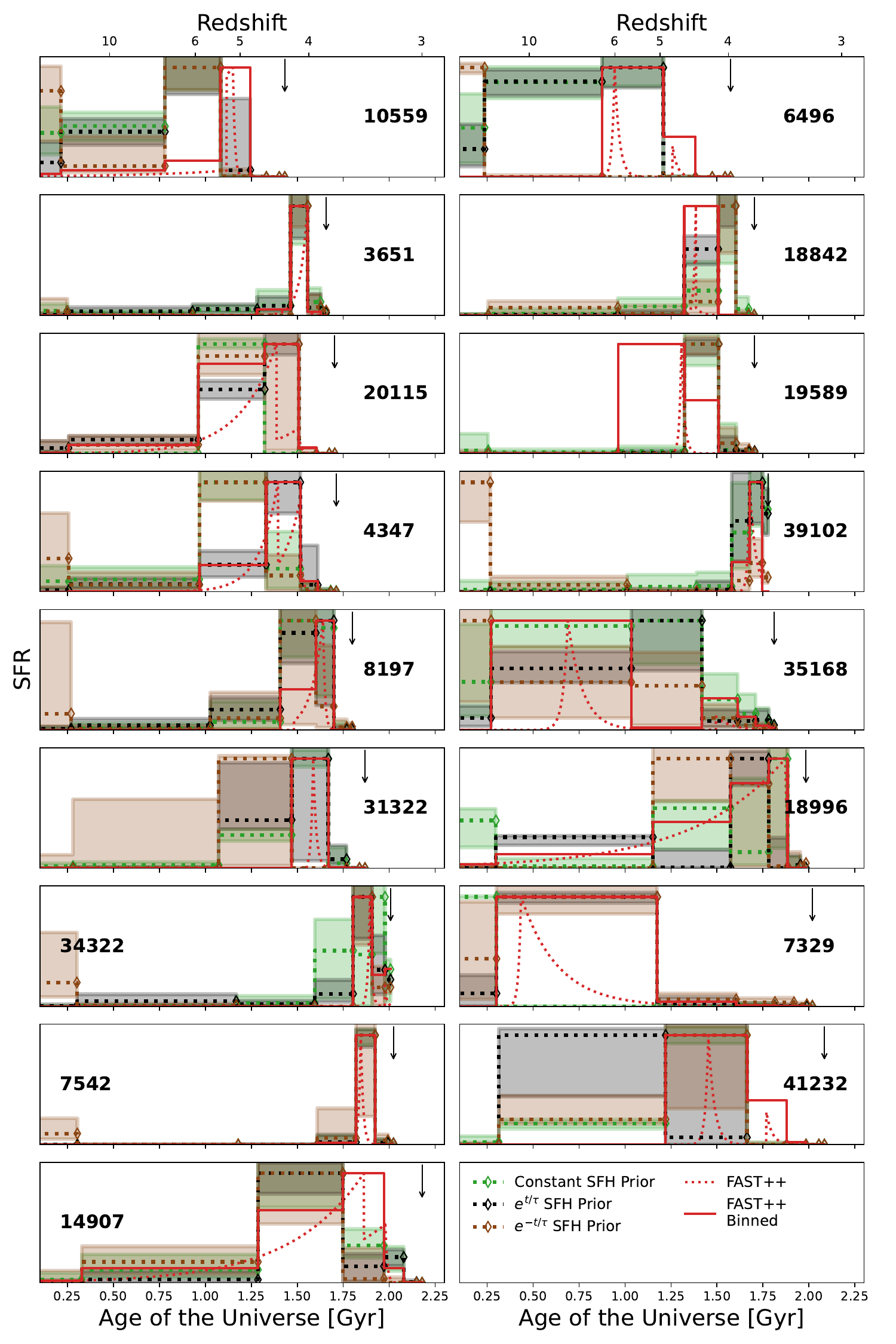}
\caption{Reconstruction of the SFHs of our sample with \prospector\ C3K models with different assumptions about the SFH prior.  
Similar to Figure \ref{fig:full_qu_sample_pros_comp_ssps_trim} each panel represents a galaxy as labeled.  
The SFHs are shown by the solid lines with associated 1-$\sigma$ errors shaded. Diamond symbols show the start and end of each time bin. Arrows denote the observed redshift of the galaxy. Following Section \ref{sec:spectral_range}, the full observed spectral range is utilized for the fitting here. The \fastpp\ results from Figure \ref{fig:full_qu_sample_sfh_comp} are also shown for comparison.
\label{fig:full_qu_sample_sfh_prior_comp}}
\end{figure*}

The reconstructed average and individual SFHs of the galaxies with flat and exponentially increasing priors are shown by Figures \ref{fig:full_qu_sample_pros_average_sfhs_fullspec} and \ref{fig:full_qu_sample_sfh_prior_comp}. 
The overall shapes of the reconstructed SFHs show good agreement between the two priors. Both SFHs exhibit a slight dip in the mean SFR around $\sim1.25$–$1.50$ Gyr, with the dip occurring at slightly earlier times in the fits using the exponentially increasing prior. 
The most significant shift between the two SFHs is in the first $250$ Myr.
The constant SFH prior shows a slight increase in the mean SFR at $<250$ Myr while the exponentially increasing SFR shows a slight decrease in the SFR at a similar amplitude. 
This shift can be attributed to galaxies such as 3D-EGS-18996 and ZF-UDS-7329, where the mass formed in the oldest time bin (the highest-$z$ bin) is reduced by the introduction of the exponentially increasing prior.

Similarly, we also investigate the effect an exponentially decreasing prior, $SFR(t_i)=e^{-t_i/\tau}$, has on the average SFHs of the observed galaxies. 
As shown by Figure \ref{fig:full_qu_sample_pros_average_sfhs_fullspec}, the average evolution of the SFR $>250$ Myr is largely consistent with the flat and exponentially increasing prior. The peak SFRs in all three models are similar, where a gradual incline of the SFRs can be observed from $>1$Gyr onward. 
The major discrepancy is at the first $\sim250$ Myr of the Universe, where the exponentially decreasing SFR shows the highest increase in SFRs at the earliest times. 
From Figure \ref{fig:full_qu_sample_sfh_prior_comp}, it is evident that an increasing fraction of galaxies tend to prefer more mass formation in the earliest time bin when the exponentially decreasing prior is introduced in \prospector.

Overall, the three assumed SFH priors in \prospector\ agree well with the \fastpp\ parametric form at later times. 
Visually, the exponentially increasing SFH prior shows the closest results to \fastpp, which is expected given that \fastpp\ also utilizes an exponentially increasing parametric form for the SFH.
Based on our tests, it is evident that the prior assumed for the posteriori sampling has an effect at older times where the diagnostic power of the age sensitive features may be limited. 
This can have an effect on how the earliest stages of the SFH of galaxies are probed with full spectral fitting techniques. 
At later ages where the observed spectral features show diagnostic sensitivity, consistent results to the SFHs can be obtained independent of the assumed prior.

\subsection{The role of nebular contribution} \label{sec:prospector_em_lines}

In both \fastpp\ and \prospector\ spectral fitting so far, we have masked out emission lines in the observed galaxies and used models that do not account for contributions from nebular continuum and nebular emission lines. In this Section, we investigate whether allowing nebular contributions as a free parameter in the spectral fitting would significantly affect the reconstruction of the SFHs.

The {\tt Flexible Stellar Population Synthesis} code \citep[{\tt FSPS},][]{Conroy2009}, utilized by \prospector, computes emission lines using the {\tt cloudy} photoionization code \citep{Ferland2017}, which are stored in a pre-computed grid \citep{Byler2017}. The age and gas-phase metallicity of the input spectra are matched to generate nebular continuum and emission lines for the {\tt FSPS} stellar population model. This combined spectrum is then used for inference with the observed data. The emission lines are powered solely by the ionizing photons from the stellar population models, with no AGN contribution included in the ionizing spectrum.

To account for possible contributions from nebular emission lines, especially given the high-EW limits observed in the NIRSpec prism mode, we marginalize over the {\tt cloudy} input grid with variable ionization parameters. We remove the emission line masks applied to the observed spectra and allow for emission line contributions in \prospector\ by letting the gas ionization parameter vary freely from $-4.0 < U < -1.0$. The gas-phase metallicity is fixed to match the stellar metallicity. Each line is fitted with a single-component Gaussian and convolved with the instrument's LSF.

\begin{figure}
\includegraphics[scale=0.80, trim= 40 55 20 65, clip]{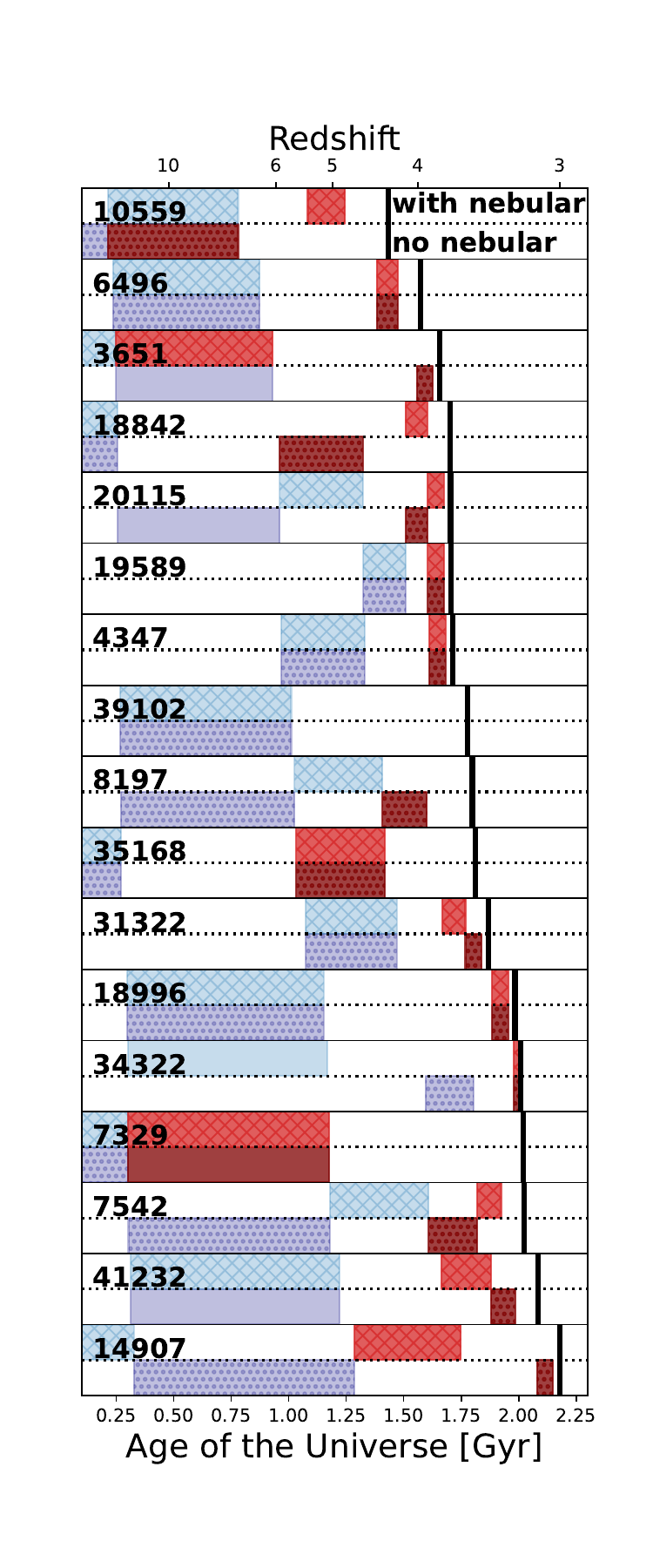}
\caption{Similar to Figure \ref{fig:full_qu_sample_sfh_t50quench_comp} but both panels show results from \prospector. The upper panels consider effects of nebular contribution in the spectral fitting, while the lower panels do not. 
If a galaxy does not satisfy the quenched criteria, the quenching time window is not shown for that galaxy. 
\label{fig:full_qu_sample_sfh_t50quench_comp_elines}}
\end{figure}

Figure \ref{fig:full_qu_sample_sfh_t50quench_comp_elines} shows a comparison of the formation and quenching time for the galaxies when \prospector\ is run with and without nebular contribution as a free parameter. We note that when nebular contributions are not considered, the emission lines in the observed spectra are masked. 
For the majority of galaxies, both runs produce similar formation and quenching times. The largest deviation in formation time is observed for 3D-EGS-34322, where the formation time window is approximately 500 Myr earlier than when nebular contributions are considered. A similar offset in quenching time is seen for ZF-UDS-3651. In both runs, 3D-UDS-39102 is not considered quenched. When emission line contributions are included, 3D-UDS-8197 is also classified as not quenched by \prospector.
The \prospector\ SFH reconstruction of 3D-UDS-8197 indicates that the galaxy was classified as quiescent during a period before the observation but has since experienced an increase in SFH, classifying it as star-forming at the time of observation. The rest-frame $UVJ$ colors, however, classify this galaxy as quiescent, which is discussed further in Section \ref{sec:quiescence}. Additionally, this galaxy exhibits strong and broad emission lines likely driven by AGN activity. Given that the {\tt cloudy} photoionization models used in \prospector\ are solely driven by star formation, they tend to elevate the SFR to match the observed spectrum, thereby failing to account for possible AGN contributions to the emission lines.
The limitations of this approach are further discussed and addressed in Section \ref{sec:prospector_agn}.

\subsection{The Role of how an AGN is handled}\label{sec:prospector_agn}

In this Section, we explore the role of adding AGN contribution as a free parameter to SED fitting for SFH recovery. 
There are two free parameters that are introduced to the \prospector\ fitting to account for AGN contributions. 
As summarized in Table \ref{table:prospector_assumptions}, the first is the AGN bolometric luminosity fraction which is defined as a fraction of the stellar bolometric luminosity. 
This is allowed to vary from $10^{-5}<f_{agn}<3.0$. The second parameter related to AGNs is the optical depth for individual clouds as parameterized by \citet{Nenkova2008a}. This is computed in the $V$-band and is allowed to vary from $5<\tau_{agn}<150$. 
The contribution of AGNs to the emission lines is not accounted for in the \prospector\ fitting. Consequently, the emission lines are not physically motivated by an underlying stellar or AGN ionizing spectrum, and thus the SFH is associated solely with the stellar continuum.
Effects of AGNs will be addressed further in a forthcoming paper by \emph{M. Martínez-Marín et al (in prep)}.

\begin{figure}
\includegraphics[scale=0.70, trim= 30 50 10 10, clip]{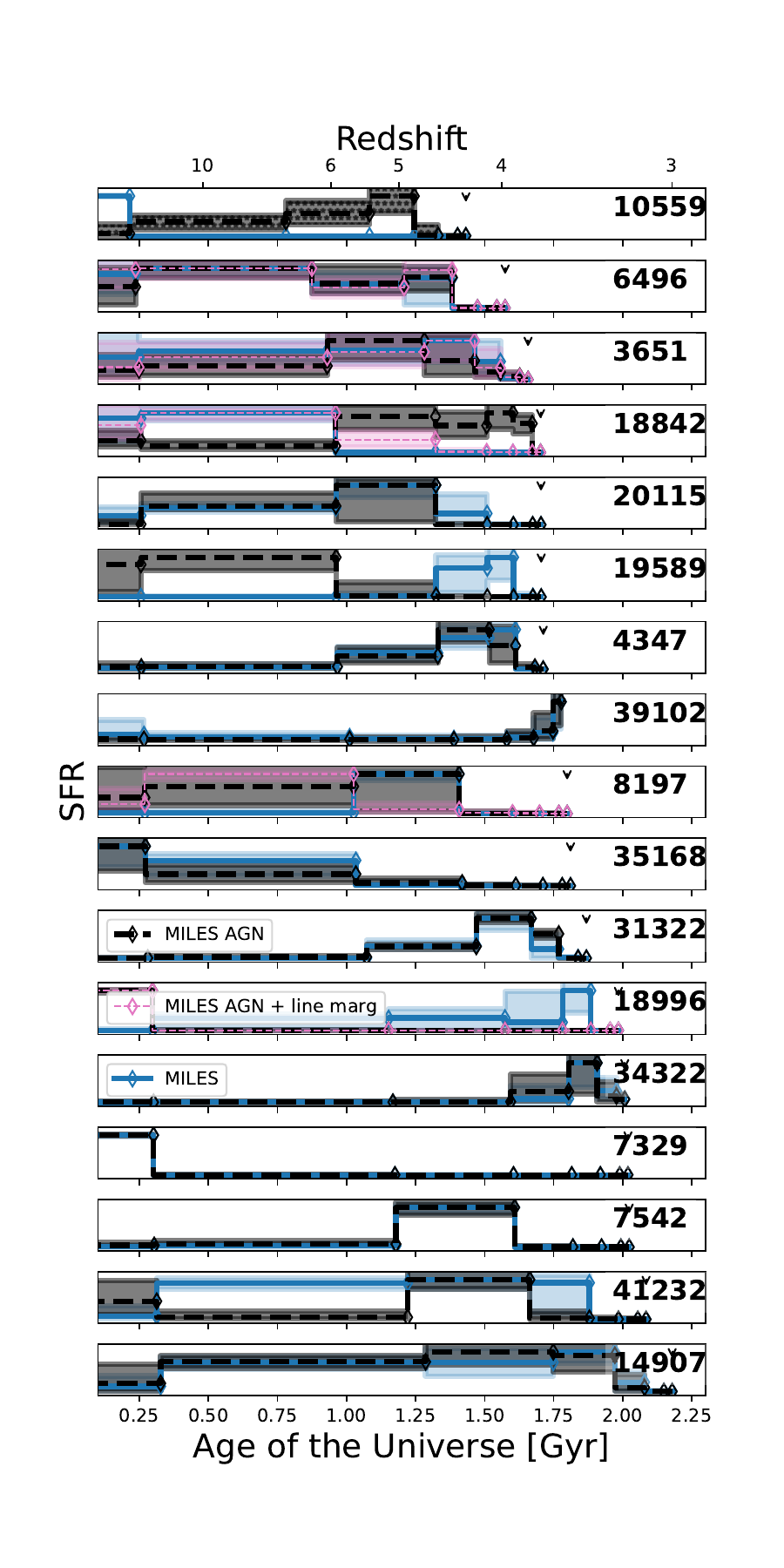}
\caption{
Reconstruction of the SFHs of our sample with (black) and without (blue) AGN effects from \prospector.
Six of our galaxies with prominent optical emission lines are also fit utilizing the emission line marginalization option in \prospector. The reconstructed SFHs of these galaxies are shown in pink.
The associated 1-$\sigma$ errors of the reconstructed SFHs are shaded using the same colors.
Each panel represents a galaxy as labeled and is organized from the highest observed redshift to the lowest. 
\label{fig:full_qu_sample_pros_agns}}
\end{figure}

Based on Figure \ref{fig:full_qu_sample_pros_agns}, for most galaxies, we find no statistically significant deviations in the reconstructed SFHs when AGN contributions are included. However, considering the overall shape of the SFH, both ZF-COS-10559 and ZF-COS-19589 show notable deviations between the two runs. In ZF-COS-10559, once AGN contributions are considered, the SFH shows a gradual incline, reducing the contribution of an early burst to the final stellar mass. Conversely, in ZF-COS-19589, the AGN contributions suggest that more mass was formed at earlier times.

\begin{figure*}
\includegraphics[scale=0.70, trim= 0.1 10 10 10, clip]{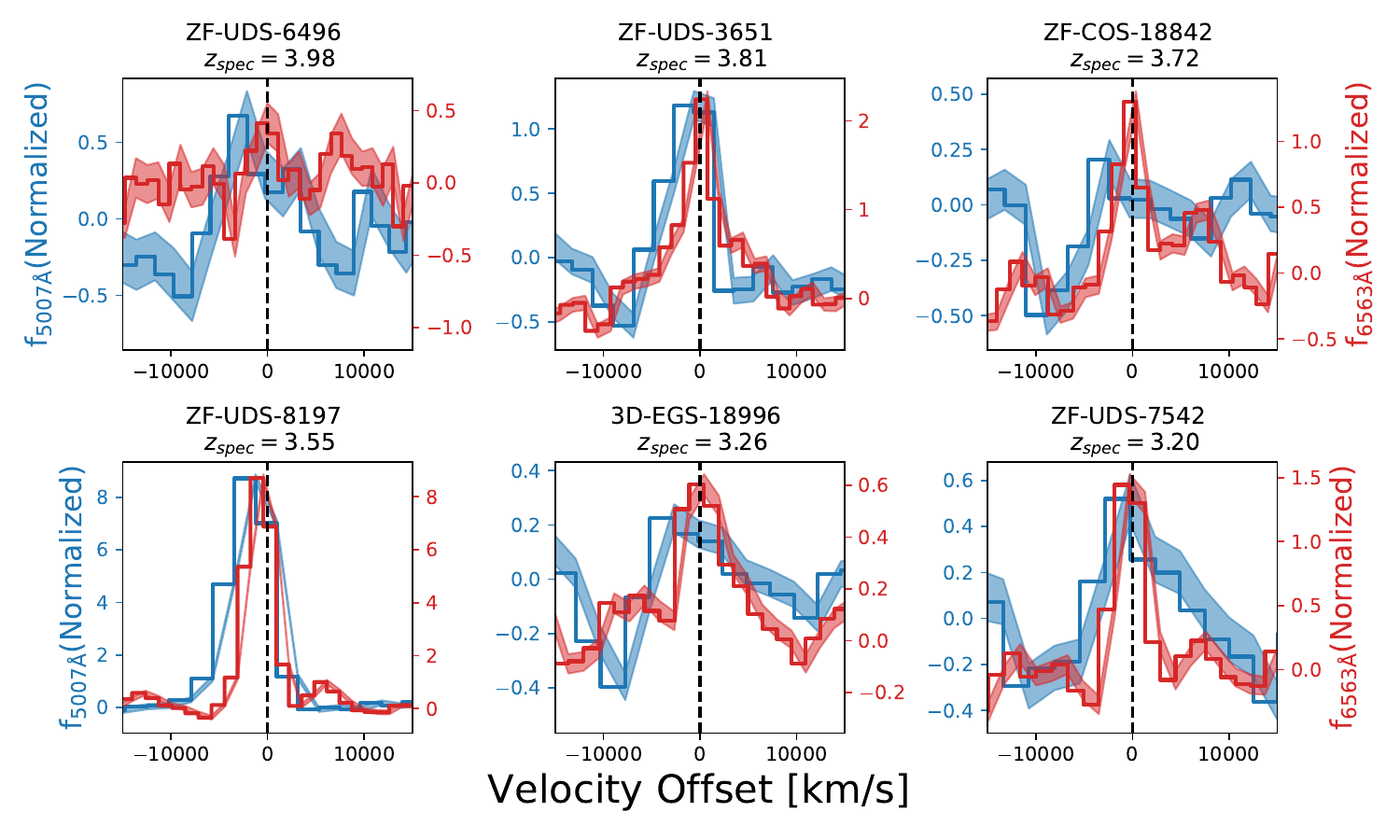}
\caption{
\OIII\ and \Halpha\ emission line regions of the spectrum for sources in our sample with significant emission line detections (S/N$>5$ and line equivalent width $>10$\AA\ for either of the two lines). The rest-frame velocity is defined at the \slinefit\ best fit redshift which is shown on top of each panel with the corresponding galaxy ID. In blue we show the \OIII\ emission line region (left axis) and in red we show the \Halpha\ emission line region (right axis). Spectra are trimmed $\pm15,000\ km/s$ from the rest-frame. Spectra are $K(F200W)$ band normalized and the local continuum is subtracted for visual clarity. The 1-$\sigma$ error region of the spectra are highlighted in the corresponding color. 
\label{fig:halpha_oiii_zoom}}
\end{figure*}

There are 6 galaxies in our sample with either significant detections of \Halpha\ or \OIII\ emission lines. We define significant detection as a line with S/N$>5$ and an observed line equivalent width (EW) of $>10$\AA, and we show these galaxies in Figure \ref{fig:halpha_oiii_zoom}.  
Visual selection of the spectra also confirm the lines of these sources to be clearly visible over the local continuum level.
In ZF-UDS-3651 and ZF-UDS-8197, both \Halpha\ and \OIII\ satisfy the significant detection criteria, in others only one of the 2 lines satisfies this selection. 
Visual inspection of ZF-UDS-7542 might suggest \Halpha\ to also be significantly detected, however, it is observed at an EW$\sim6$\AA, and thus is lower than our EW cut. 
With a nominal $R\sim200$ in PRISM mode observations, our velocity resolution is $\gtrsim 1500$km/s in the $\sim K$ band. Therefore, we are unable to provide tight constraints to the observed line widths of these sources, except for 3D-EGS-18996. Both \OIII\ and \Halpha\ are broad for this galaxy with \slinefit\ providing a best-fit line width of $\sim6000$km/s.

The ionizing spectrum of the \prospector\ {\tt cloudy} grids are limited to stars, hence, emission lines lack contribution from AGN sources. 
While there are only minimal changes to the recovered SFHs when AGN effects are added, it is possible that this limitation of the input {\tt cloudy} photoionization grids has an effect on the \prospector\ results. 
Given the prominence of the emission lines of the 6 galaxies highlighted in Figure \ref{fig:halpha_oiii_zoom}, it is possible for them to have a contribution from an AGN. 
While we cannot completely rule in favour of strong AGN contribution for these sources, it is necessary to determine the limitations of using {\tt cloudy} grids that only have contributions from stars to our recovered SFHs.

We explore this by removing the emission line masks in the observed spectra and allowing freedom in \prospector\ to fit emission lines independently of the input ionization conditions. 
In this setting, emission lines are considered purely Gaussian in nature and are marginalized over line width prior to obtain the maximum a posteriori solution.  
We utilize the {\tt nebular\_marginalization} template in \prospector\ and select the following strong optical emissions to marginalize over: \Hbeta, \OIIId, \NIId, \Halpha, \SII. Given the prism resolution \Halpha\ and \NIId\ are not resolved.

As shown in Figure \ref{fig:full_qu_sample_pros_agns}, the recovered SFH reconstructions are largely statistically consistent between the two \prospector\ runs that consider AGN effects. 
We also find that the $f_{agn}$ and $\tau_{agn}$ parameters to be statistically consistent between the two runs.
When lines are marginalized $f_{agn}$ varies between 0.25 to 0.93 with ZF-UDS-8197 showing highest $f_{agn}$ value. 
Similarly, $\tau_{agn}$ vary between 5.9 and 7.9 with 3D-EGS-18996 showing the largest value. 
However, we note that constraining both the ionizing spectrum and the overall SED shape together with star forming and AGN contributions is necessary and would provide tighter constraints to the nature of the AGN in galaxies such as ours. 
We revisit this in our forthcoming paper \emph{Martínez-Marín et al (in prep)} and envision that novel advancements in machine-learning-assisted spectral fitting techniques would be able to address this in future \citep[e.g.][]{Li2024a}.

\subsection{Recovery of mock star-formation histories from IllustrisTNG}\label{sec:prospector_recovery}

So far we have investigated the role of the form and prior of the assumed SFH parameterization, the input stellar library/stellar population model, the wavelength window of the observed spectrum used for spectral fitting,  nebular emission, and AGNs in reconstructing the SFHs of our massive quiescent galaxy sample.
The results indicate that the final best fit or maximum a posteriori SFH solution in general has good agreement between these variations.
The average SFH of the sample is also shown to be statistically consistent.
However, the main limitation here is that we have no ground truth to establish the accuracy of the recovered SFHs.

To assess this limitation, we utilize mock observations to investigate whether the input SFHs can be recovered accurately by \prospector. 
If \prospector\ works as intended, the input SFHs of our models should be recovered through the fitting. This is because \prospector\ utilizes a comprehensive Bayesian framework that effectively leverages the available wavelength coverage and S/Ns to constrain the SFH parameters. By accurately modeling the observed spectral energy distributions and accounting for uncertainties, \prospector\ is expected to reproduce the input SFHs when the underlying assumptions and observational conditions are met.
In order to use realistic SFHs to perform mock observations, we utilize the IllustrisTNG simulations \citep{Pillepich2018a,Springel2018a} to select 283 massive galaxies ($M>10^{11} M_\odot$) at $z=3$ from the TNG300 suite that are expected to be quiescent based on the definition of specific star formation rate (sSFR)$< 1.5 \times 10^{-10}$/yr from \citet{Schreiber2018}.
We note that  TNG300 simulations show good agreement for the number density of massive quiescent galaxies at $z\sim3-4$ \citep{Valentino2023a}.

Once the galaxies are selected from TNG300, the SFHs are computed from the mass-weighted distribution of stellar formation times of all star particles bound to the subhalo, as was done in \citet{Chittenden2023a}.
This gives the SFHs a greater time resolution than the snapshot data: 1.4 Myr per age bin, compared with a median snapshot time difference of 74 Myr, up to $z=3$. 
Furthermore, unlike the stellar mass obtained from the merger tree, this field is derived from the stellar ages of all particles of the subhalo. 
Therefore, we can compute an intrinsic SED from a fine-grained composite stellar population. 
Additionally, it effectively captures the star formation from all progenitors, not solely the main progenitor branch, therefore tracing the stellar mass contribution of all merger events throughout the galaxy's history.
Finally, this results in realistic SFHs for 283 sources.

Mock spectra for the 283 TNG300 selected sources are computed using the {\tt FSPS} code. 
The MILES stellar library with MIST stellar isochrones and \citet{Kroupa2001_conf} IMF is used for this purpose. Metallicity is fixed at 50\% \zsol, and a $\tau_{dust}=0.5$ is applied to the spectra following a \citet{Calzetti1999} dust law. 
The evolution of the SFR with cosmic time for each galaxy is input into the stellar population code using the {\tt set\_tabular\_sfh} option. 
Spectra are generated at the rest-frame and are redshifted to their observed redshift. 
A physical velocity of 300 km/s is applied to the spectra following results from \citet{Esdaile2021a}. 
We convert the NIRSpec PRISM instrument LSF to velocity and assert that the intrinsic resolution of the SSPs is lower than that. 
We then apply the LSF velocity smoothing to the spectra after subtracting the SSP spectral library resolution. 
Noise is added to the spectra following a random normal distribution such that there is a median S/N of 50, which is similar to our observations. 
The photometry for the mock spectrum is computed in the following JWST NIRCam bands using the {\tt sedpy\footnote{https://github.com/bd-j/sedpy}} package: F115W, F150W, F200W, F277W, F356W, F410M, F444W. 
For simplicity, we don't add emission lines to the mock spectra nor use it as a free parameter in \prospector\ fitting.

\begin{figure*}
\includegraphics[scale=0.60, trim= 12 0 0 10, clip]{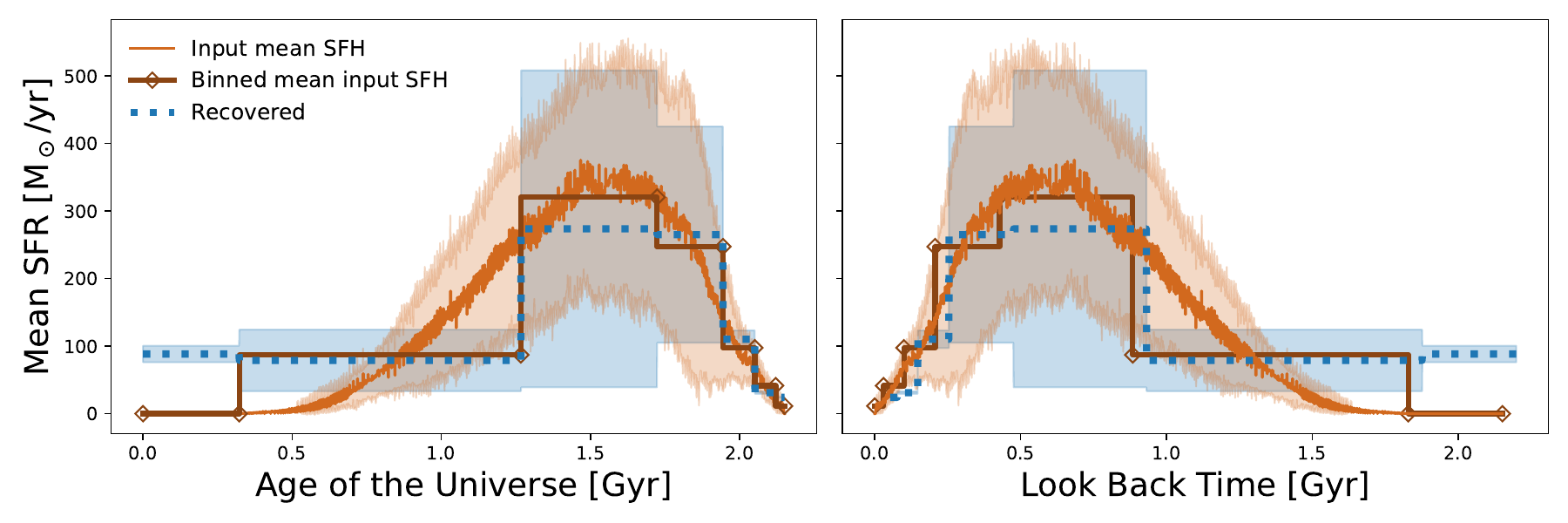}
\caption{
The average input and \prospector\ recovered SFHs of the $z=3$ massive quiescent galaxies selected from the TNG300 simulation. The mean input SFH of the simulated galaxies is shown in brown with its 1-$\sigma$ scatter shaded in light brown. The mean input SFH binned in the same age bins as used for \prospector\ is shown by the dark brown line. The mean SFH recovered by \prospector\ is shown by the blue dotted line along with the 1-$\sigma$ scatter highlighted in light blue. {\bf Left} panel shows age of the Universe and the {\bf Right} panel shows the look back time at $z=3$.
\label{fig:pros_recovery_sims}}
\end{figure*}

The mock spectrophotometry is fit using \prospector\ with similar parameters as outlined in Section \ref{sec:prospector}. 
The recovered average SFHs are shown in Figure \ref{fig:pros_recovery_sims}. 
The \prospector\ recoveries show good agreement with the average SFH of the input galaxies, with the growth of the peak of the SFRs $t=\sim1.5$ Gyr and the subsequent quenching phase matched well. 
The mock tests demonstrate that the earliest time bin defined around the first 300 Myrs in the Universe is not well constrained by the SFH recovery. As discussed in Section \ref{sec:sfh_prior} we attribute this offset at the earliest bin to the use of a flat constant SFH prior with \prospector. 
However, our tests with realistic SFHs from TNG300 massive quiescent sources demonstrates that on average, current SED fitting models are able to recover the SFHs in the $z>3$ Universe with good sensitivity to stellar ages, even with low resolution NIRSpec PRISM data.

\subsection{SFH comparison with cosmological simulations}\label{sec:comparison_with_cos_sims}

In order to compare the average formation histories of massive quiescent galaxies in our observed sample with the predictions from cosmological simulations, we utilize four cosmological suites as outlined below. To assist in comparison between the simulations, we use a constant definition of $\text{M}_*>3\times10^{10} \text{M}_\odot$ (mean mass of the observed sample) and $sSFR < 1.0 \times 10^{-10}$/yr as the selection for massive quiescent galaxies.

\begin{figure*}
\includegraphics[scale=0.7, trim= 12 0 0 10, clip]{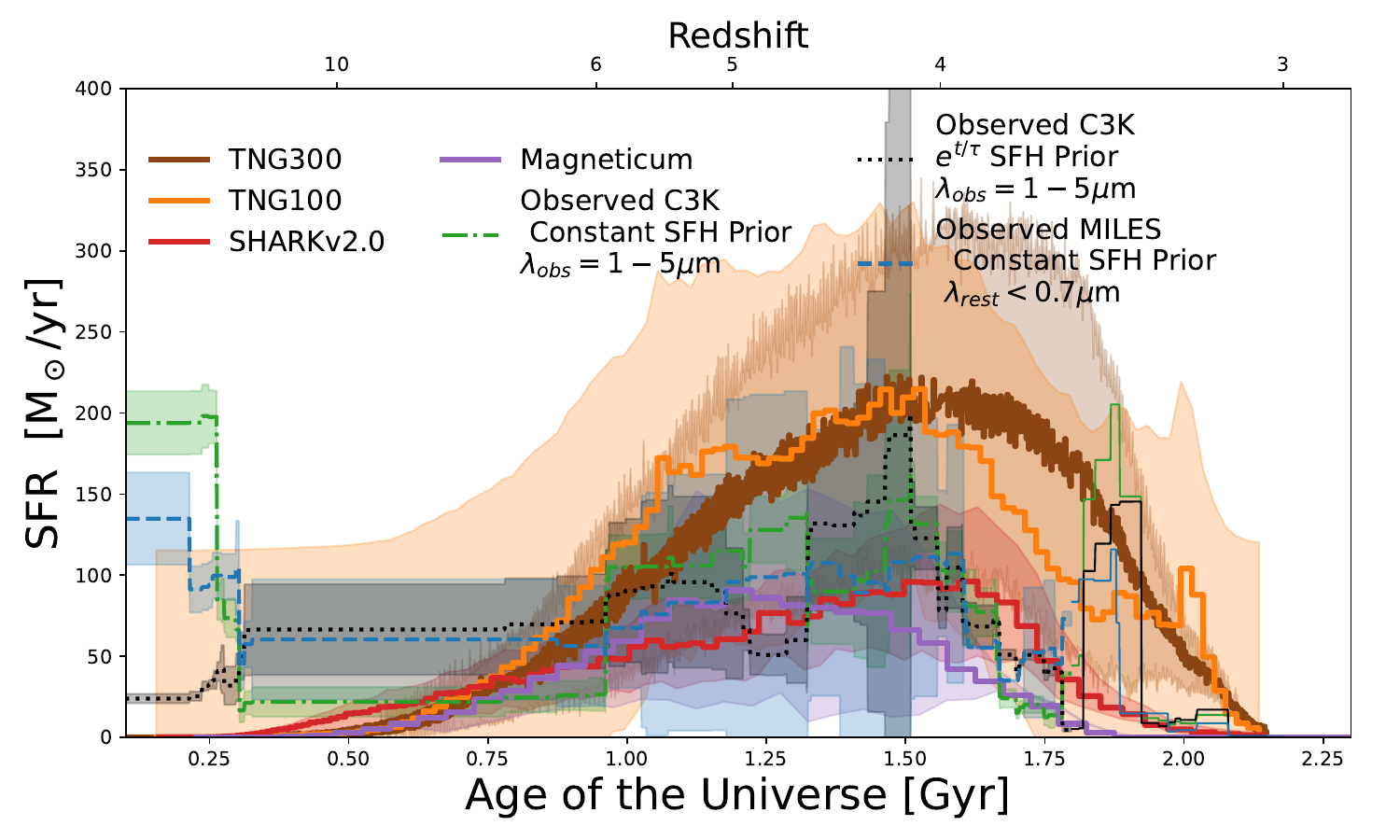}
\caption{
Comparison of the SFHs from Illustris TNG100, TNG300, SHARKv2, and Magneticum cosmological simulations with the reconstructed average SFHs of the quiescent galaxies presented by our study. The variation in peak SFRs between different simulations and observed data is due to the differing total stellar masses of the samples considered as shown by Figure \ref{fig:sims_ssfr_vs_mass}. The $1\sigma$ scatter, parameterized by the normalized absolute deviation, for the observed data and simulations is shaded in their respective colors.
\label{fig:sfh_comp_with_sims}}
\end{figure*}

First we select galaxies from the TNG300 cosmological suite \citep{Pillepich2018b}. TNG300 has a size of $302.6^3$Mpc$^3$ and we find 556 sources that are classified as massive quiescent based on our definition at $z\sim3$. 
The SFHs of these galaxies are computed as detailed in Section \ref{sec:prospector_recovery}. 
In Figure \ref{fig:sfh_comp_with_sims}, we show the comparison between the average SFHs of our massive quiescent sample (see Section \ref{sec:quiescence}) with the average SFHs of the TNG300 massive quiescent sample. 
Regardless of the stellar population models or the utilized SFH prior, the buildup of the average SFH of the observed sample largely follows what is predicted by TNG300.
The significant deviation is during the first $\sim250$ Myr of the Universe.
As we noted in Section \ref{sec:prospector_ssps}, this initial time window also shows the largest dependency for the input stellar library/stellar population models. 
As discussed in Section \ref{sec:sfh_prior}, utilizing an exponentially increasing SFH prior over a constant SFH prior, the average SFHs at the earliest time bins can be reduced. 
However, even with the exponentially increasing SFH prior, the offsets between the SFHs of the models and the data are not statistically consistent with each other.

Utilizing the same IllustrisTNG cosmological suite, we use the SFHs of massive quiescent galaxies from the TNG100 simulations \citep{Pillepich2018b} to compare with the observations. 
Compared to TNG300, TNG100 has a smaller size of $110.7^3$Mpc$^3$  but has $\sim1$ dex higher particle resolution size. Given the limited volume of the TNG 100 simulations we only find 43 galaxies that satisfy our massive quiescent definition at $z\sim3$. 
We show the average SFHs of these galaxies in Figure \ref{fig:sfh_comp_with_sims}. 
While both TNG100 and TNG300 exhibit largely similar mass growth, the SFRs in TNG100 show a slightly earlier rise and fall. 
Additionally, TNG100 shows a rapid increase in SFR at  $\sim2$ Gyr before its rapid decline to $z=3.0$.

The lower particle size of TNG100 compared to TNG300 allows for finer resolution in feedback processes and halo mass growth mechanisms \citep{Nelson2018a,Springel2018a,Chittenden2023a}. The higher particle resolution in TNG100 enables more detailed mapping of the gravitational potential of individual stellar and dark matter particles. As a result, TNG100 produces deeper (higher amplitude) potential wells compared to TNG300.
Since star formation is coupled with the simultaneous cooling of gas and an increase in pressure, TNG100 achieves more efficient star formation, leading to the slightly earlier bias in mass growth relative to TNG300. Additionally, the finer resolution of feedback mechanisms in TNG100 allows for more effective regulation of star formation.

Next, we use SHARKv2.0 \citep{Lagos2024a} semi-analytical models to compare the formation histories of massive quiescent galaxies at $z\sim3$. 
SHARK has a volume of 464$^3$ cMpc$^3$ and we find 502 galaxies that satisfy our massive quiescent galaxy criteria at $z\sim3$. For each galaxy we investigate the SFR as a function of cosmic time and compute the mean SFH of the sample. 
We show this in Figure \ref{fig:sfh_comp_with_sims}. 
While SHARK galaxies show a marginal increase in the average SFHs at earlier times,  similar to the TNG suite, at $>250$ Myr the shape of mass buildup of SHARK galaxies largely resembles our observed population.

In Figure \ref{fig:sfh_comp_with_sims} we further show the average formation histories of $z\sim3$ massive quiescent galaxies from the  Magneticum Pathfinder hydrodynamical cosmological simulations \citep{Remus2023a}. These sources are selected from the snapshot 6 (corresponding to $z=3.4$) of Magneticum Box 3 with a boxsize of 128cMpc/h. 184 central galaxies satisfy our massive quiescent galaxy criteria. 
The average mass buildup of the Magneticum sample is similar to the other simulations.
 However, the Magneticum sample reaches the peak SFRs at earlier times compared to the other simulations. 
Thus, it is evident that Magneticum has quenching mechanisms that are in general switched on at earlier times compared to those of the other simulations. 
This results in providing a higher number density of massive quiescent sources at $z\sim4$ \citep{Kimmig2025a}. 
We attribute the early decline in the SFR in the Magneticum simulations to the more flexible implementation of feedback mechanisms.

In Magneticum, quenching is attributed to both AGN and star-formation feedback, allowing galaxies in underdense regions to expel gas more efficiently and quench star formation \citep{Kimmig2025a,Remus2023a}. Since quenching in Magneticum only requires a rapid isotropic gas inflow to trigger a star-burst, a rapid mass buildup may immediately precede quenching. 
As Magneticum differentiates cold and hot accretion onto the black hole \citep{Steinborn2015a}, this allows for continuous accretion as long as cold dense gas falls in isotropically, even while a hot outflow is starting to be driven by the AGN. This leads to earlier black hole growth and stronger feedback \citep[see][]{Steinborn2015a}.
Thus, while AGN feedback plays a significant role in quenching,  there is less of a requirement for the black hole to reach a characteristic stellar mass to drive kinetic mode quenching.

The quenching of star formation in TNG300 galaxies at $z \sim 4$ is attributed to the kinetic mode of the AGN being triggered when the central black hole reaches a critical stellar mass \citep{Hartley2023a}. Once the kinetic mode energy exceeds 1\% of the thermal mode energy in the AGN, the kinetic mode AGN expels the remaining star-forming gas in the galaxies, leading to quenching \citep{Weinberger2018a}. 
Therefore, in TNG300, black holes must exceed a certain mass threshold before triggering quenching mechanisms, which results in a delayed onset of quenching compared to Magneticum.

Most of the quenching in SHARK happens through the AGN jet-mode feedback, which can start acting as long as jets are produced and there is a hot halo against which to produce work. The latter condition makes it hard for SHARK to start quenching a significant amount of massive galaxies at earlier times, producing the later quenching of galaxies compared to Magneticum.

The normalization difference between the peak SFRs of different simulations is due to the stellar mass distribution of the different galaxies. 
The mean mass of the TNG suite is $\sim3-4 \times$ higher compared to SHARK and Magneticum. The observed mean mass of our quiescent galaxies agrees well with SHARK and Magneticum. 
Thus, the absolute normalization in SFR for our observations is similar to that of SHARK and Magneticum, while TNG100 and TNG300 have a marginally higher normalization as is evident from  Figure \ref{fig:sfh_comp_with_sims}. 
This is further shown by Appendix Figure \ref{fig:sims_ssfr_vs_mass}.

Further analysis of the formation and quenching mechanisms across a range of hydrodynamical and semi-analytical models is presented in \citep{Lagos2025a}.

\section{Discussion} \label{sec:discussion}

\subsection{The formation histories of the massive quiescent galaxies }\label{sec:formation_histories}

Our analysis utilized novel JWST NIRSpec $1-5\mu$m spectroscopy of ground based $K$ selected $z>3$ massive quiescent galaxy candidates to explore their formation mechanisms. We showed consistent SFHs between different SED fitting codes, SFH parameterizations, input stellar population models, and other ISM and AGN related properties. Our sample showed a variety of formation and quenching timescales, ranging from the first billion years of the Universe to few megayears before the time of observation. 
In this Section, we first briefly discuss the advancements made in the $z>3$ Universe from previous NIR facilities and then explore the advancements we are now able to make with JWST.

In the pre-JWST era, deep observations from ground NIR facilities provided a first look into the properties of $z>3$ massive quiescent galaxies. 
In \citet{Glazebrook2017} we presented the first deep Keck/MOSFIRE $K$ band spectroscopy of a $z=3.7$ massive quiescent galaxy which was photometrically identified by NIR medium band imaging data from the ZFOURGE survey \citep{Straatman2016}. 
Spectral fitting inferred the galaxy to be $\sim500-1000$ Myr old and to have have assembled $\sim10^{11}$\msol within a $\lesssim250$ Myr window. This implied a very high star-formation efficiency for galaxies at $z>5$, which was not observed in direct UV observations of sources at these early times \citep[e.g.][]{Smit2012a,Salmon2016}. 
A similar view of the early formation timescales was also established by the study of massive strong emission line dominated galaxies at $z>3$ \citep[e.g.][]{Marsan2017}.

In \citet{Schreiber2018} we preformed a mass complete analysis of the $z>3$ massive quiescent candidates utilizing photometric selections from ZFOURGE \citep{Straatman2016} and 3DHST \citep{Skelton2014} surveys. Keck/MOSFIRE $K$ and $H$ band spectroscopy of 24 candidates was presented by \citet{Schreiber2018}. 12 sources were spectroscopically confirmed, out of which 2 were found to be $z\sim2$ interlopers. The remaining sources, which primarily constituted the fainter end of the $K$-band selection eluded confirmation (also see Section 5.6.2 and Appendix E of \citet{Forrest2020a}), even with deep $\sim10$h of Keck/MOSFIRE observations. 
A reconstruction of the SFHs of the 22 candidates showed a variety of formation and quenching time scales, but stringent limits were not possible due to the limited S/N and the limited spectral coverage of the ground based data.

Utilizing deep KECK/MOSFIRE and VLT/XSHOOTER observations, many surveys of  $z>3$ massive quiescent galaxies were conducted by several teams. 
This result transformed our view of rapid formation of massive sources in the early Universe. 
The MAGAZ3NE survey presented spectroscopic confirmation for 16 photometrically selected $>10^{11}$\msol\ galaxies at $z>3$ selected from the UltraVISTA survey \citep{Forrest2020a}, out of which $\sim5$ were expected to be quiescent. 
The ages of the massive quiescent candidates were constrained using the D4000\AA\ and H-$\delta$ spectral features which suggested that most of the stellar mass of these galaxies formed $<1$ Gy from the time of observations. Tighter constraints were obtained with multi-band SED fitting, which constrained the formation window for MAGAZ3NE quiescent sources to be between $4<z<5$ with intense star-formation up to $\sim3000$\msol/yr followed by rapid quenching. 
\citet{Valentino2020} presented spectroscopy of three $\sim10^{11}$\msol\ quiescent galaxies at $z\sim3.7-4.0$. They found the galaxies to have likely formed most of their stellar masses by $z\sim5$ and subsequently quenched at $z\lesssim5$ after experiencing peak SFRs of $\sim3500$\msol/yr.  

\citet{Carnall2020a} presented a sample of 10 robust $z>3$ photometrically selected massive quiescent candidates selected from the CANDELS UDS and GOODS-S fields \citep{Grogin2011,Koekemoer2011}. Spectroscopy for the sources was limited to the rest-UV, where Ly-$\alpha$ emission was confirmed for 2 candidates, while another showed evidence for a clear Lyman break (the latter source was spectroscopically confirmed to be a $z=4.7$ massive quiescent galaxy \citep{Carnall2023b}). 
The reconstructed SFHs for the robust sources suggested that the galaxies assembled most of their stellar masses within the first billion years of the Universe. 

\citet{Antwi-Danso2025a} presented spectroscopy of three $>10^{10}$\msol\ quiescent galaxy candidates at $z=3.3-4.7$, selected using split-$K$ band imaging from the FENIKS survey \citep{Esdaile2021b}. Two of the sources are expected to have formed at $z\sim4$ and undergone subsequent rapid quenching. One of the other galaxies (albeit with very poor spectral quality) suggests a formation time window at $z\sim9$. This is similar to what was observed for ZF-UDS-7329 \citep[also see][]{Glazebrook2024a,Nanayakkara2024a,Carnall2024a}. However, the star formation window spans $>1$Gyr with the source only quenching within $\sim300$ Myr of the observation time. This can be compared to $>1$ Gyr old stellar population observed for ZF-UDS-7329, through the direct detection of the D4000\AA\ spectral feature.

Open questions remain on the pathway to quiescence. Our observations have shown that, even with ground $K$ band selections assisted with $>3\mu$m Spitzer imaging we are able to photometrically select the oldest massive quiescent sources in the Universe \citep{Glazebrook2024a,Nanayakkara2024a}. 
One of the common themes between most of the reconstructed SFHs of $z>3$ massive quiescent galaxies across various surveys is the intense SFR at early time. 
This is tied to the rapid buildup of stellar masses and suggests that most of these sources could have had extensive star-formation in $z>5$. Thus, a natural evolution of highly star-forming and obscured sub-mm galaxies at $z>5$ to $z\sim3-5$ massive quiescent galaxies can be expected \citep{Valentino2020,Long2023a}. 
The open question is whether all $z\sim3-5$ massive quiescent galaxies undergo a sub-mm galaxy phase at $z>6$ or whether all sub-mm galaxies end up becoming massive quiescent galaxies by $z\sim3$. 
Further studies linking the two populations with robust number density constraints are required to conclusively determine the evolutionary pathways of these two galaxy populations.

\subsection{What does it mean to be quiescent?}\label{sec:quiescence}

Color selection techniques, specially the rest-frame $U-V$ vs $V-J$ color distribution of galaxies have been commonly utilized to select quiescent galaxies in the extragalactic Universe \citep[e.g.][]{Williams2009,Spitler2014,Straatman2014,Schreiber2018,Carnall2020a}. 
Quiescent galaxies with strong Balmer/D4000\AA\ breaks show redder $U-V$ colors and bluer $V-J$ colors. Dusty galaxies show redder $V-J$ colors, hence in this space quiescent galaxies occupy a region that helps break the degeneracy of the red colors that are also observed in dusty star-forming galaxies. 
The quiescent quadrant can also hold dusty post-starburst galaxies which can smooth the Balmer break of post-starburst galaxies to look like the D4000\AA\ break mimicking an older quiescent galaxy. 
JWST now opens up a new redshift and magnitude (e.g. spectroscopy of $K$ faint sources) frontier which eluded  ground based observations. 
Thus we need to address whether our traditional definitions for quiescence still holds in this era and whether the traditional color selection techniques used in identifying quiescent systems in the $z>3$ Universe are able to effectively select quiescent sources with high purity.

\begin{figure*}
\includegraphics[scale=0.65, trim= 0 0 0 0, clip]{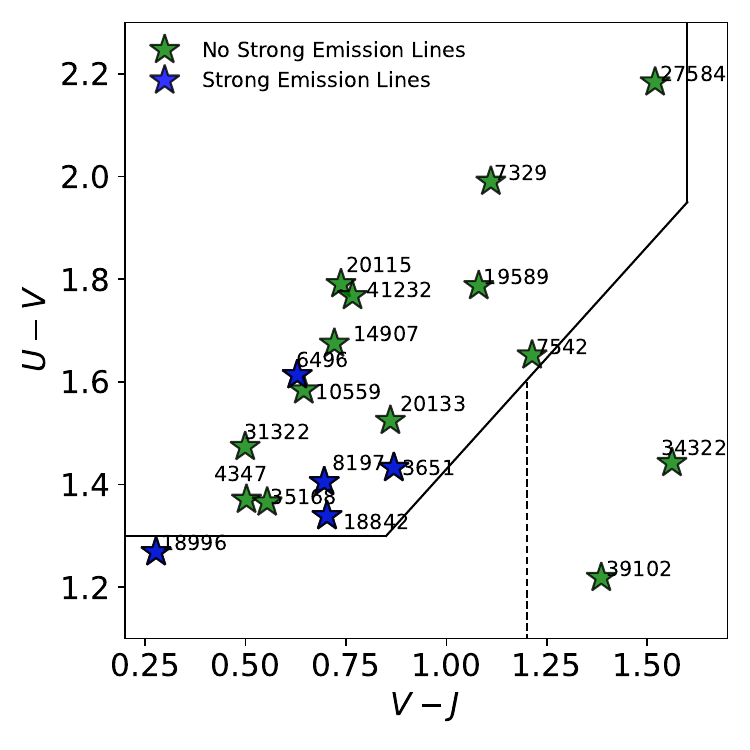}
\includegraphics[scale=0.65, trim= 0 0 0 0, clip]{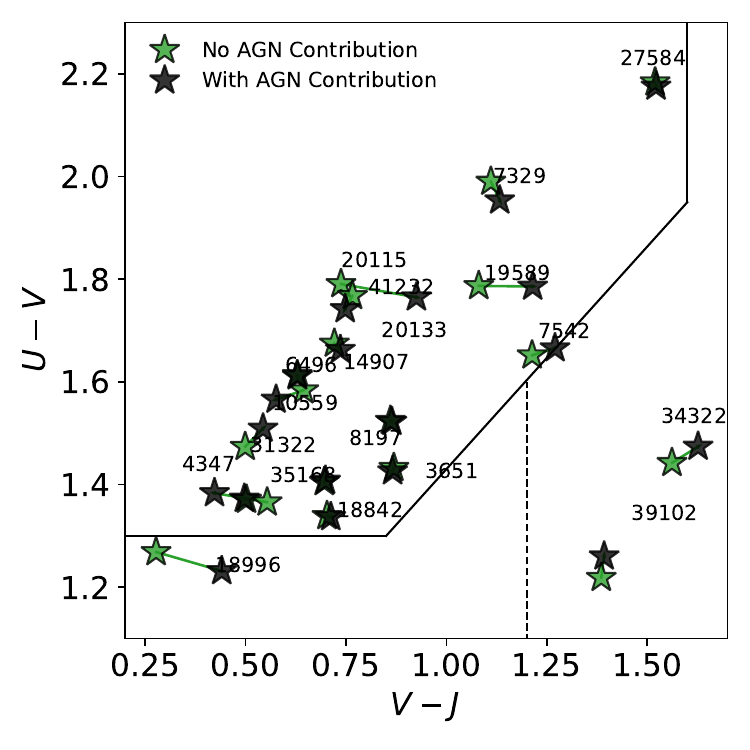}
\caption{
{\bf Right:} The rest-frame $U-V$ vs $V-J$ color distribution of our galaxies. The quiescent, red, and blue star-forming regions in the color-color space are cordoned following \citet{Spitler2014} criteria. Rest-frame colors are derived using  \prospector\ C3K best-fit models utilizing nebular contributions. Galaxies with strong emission lines are highlighted separately from the rest of the population. 
{\bf Left:} Similar to the right panel but shows the change in best-fit model colors obtained with and without AGN effects as parameterized by \prospector. The shift in location in the $U-V$ vs $V-J$ color space is marked by the green lines.
\label{fig:UVJ}}
\end{figure*}

In Figure \ref{fig:UVJ} we show the rest-frame $U-V$ vs $V-J$ color distribution of our galaxies. We use the C3K models fitted to the full spectrum with \prospector\ including emission line contributions for this purpose. 
The demarcations for quiescent, red (dusty) star-formation, and non-dusty (blue) star-formation follows \citet{Spitler2014} criteria. Within $\pm0.1$mag, except for two galaxies, all of our  sources fall in the quiescent region. This corresponds to an accuracy of approximately 90\% for our sample and confirms that photometric selections are highly efficient in identifying quiescent galaxy candidates at $z>3$.

The dustiest source in our sample is 3D-EGS-34322, with $\tau_{\text{dust}} \sim 1.7$. 3D-UDS-39102 also exhibits a higher dust content, with $\tau_{\text{dust}} \sim 1.1$. Both of these sources lie in the dusty star-forming region. Our oldest quiescent galaxy, ZF-UDS-7329, shows a low level of dust, while ZF-COS-20115 has the highest dust content among the quiescent sources, with $\tau_{\text{dust}} \sim 1.5$.

\begin{figure*}
\includegraphics[scale=0.75, trim= 10 0 0 0, clip]{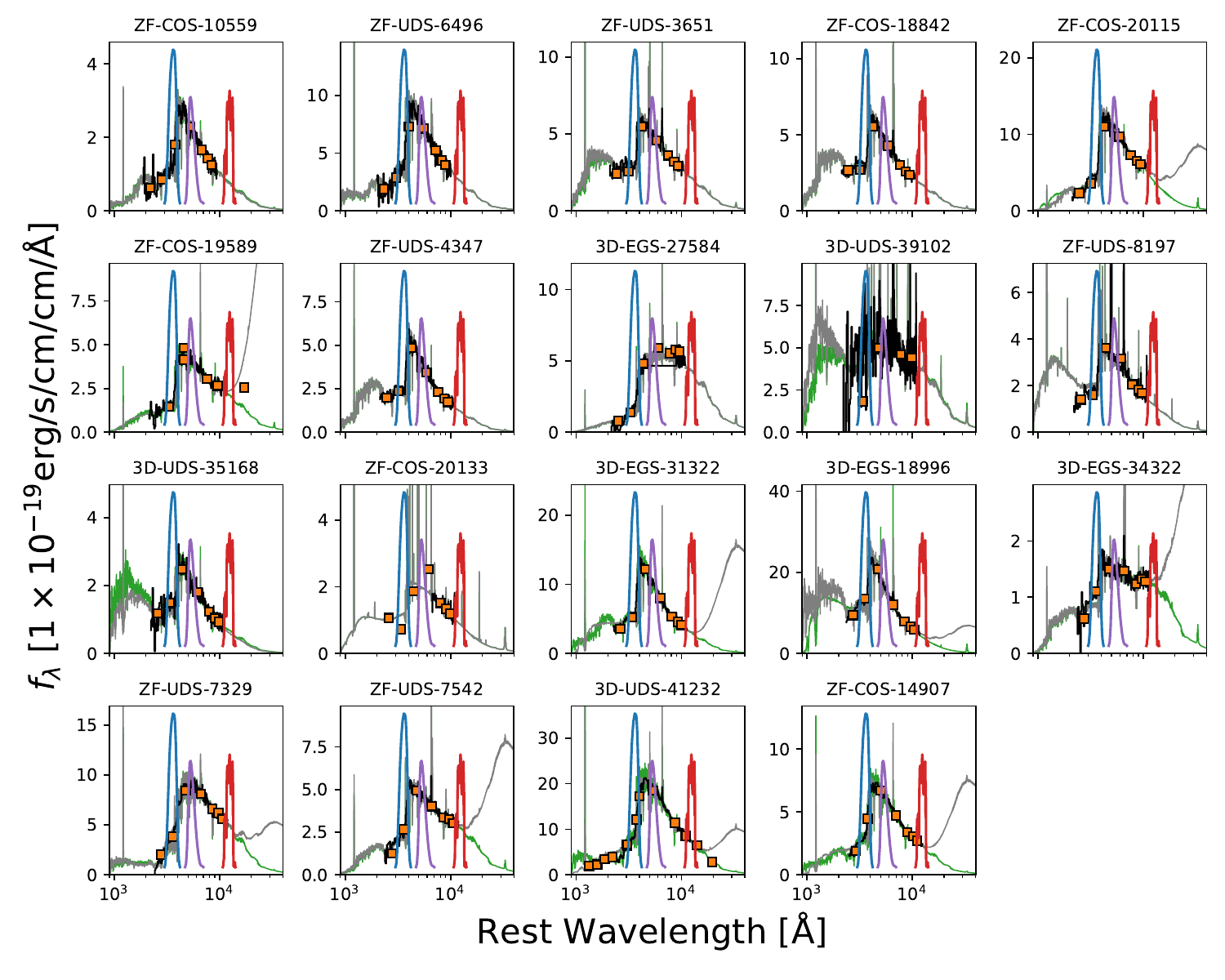}
\caption{
Rest-frame spectral energy distributions (SED) of our galaxies. The rest-frame best-fit SEDs from \prospector\ with and without AGN effects are shown by grey and green, respectively.  The observed photometry with S/N$>5$ is shown by the orange data points. 
In black we show the observed NIRSpec spectrum. 
In blue, purple, and red we show the $U, V$, and $J$ filter transmission functions, respectively, used to compute the rest-frame $U-V$ vs $V-J$ colors shown by Figure \ref{fig:UVJ}. 
\label{fig:seds}}
\end{figure*}

Galaxies with strong emission lines as identified as possible AGNs in Section \ref{sec:prospector_agn}  primarily fall in the quiescent region. 
To investigate the effects of strong lines in the $UVJ$ classification, we show the \prospector\ best-fit rest-frame SEDs of our  galaxies in Figure \ref{fig:seds}. 
ZF-UDS-8197 has the strongest emission lines in our sample, and it is clearly visible that the strong \Halpha\ and \OIII\ emission lines fall outside the $V$ band. If these lines fell within the $V$ band, the galaxy's position in the $U-V$ vs. $V-J$ color space would likely shift diagonally left toward the blue, star-forming region. Since these colors are defined in the rest frame, the fact that strong optical emission lines fall outside the $V$ band helps in more efficiently identifying quiescent galaxies photometrically, as there is less optical emission line contamination and a greater focus on the Balmer break.

Adding effects of AGNs to \prospector\ fitting only has a small role for the rest-frame $UVJ$ colors. 
There are four galaxies that show an absolute shift of $>0.1$dex in the color-color space. There is a mean $\sim0.06\pm0.06$ offset in rest-frame  $U-V$ vs $V-J$ colours for the galaxies. The maximum shift observed is 0.19 dex. 
None of the galaxies move away from quiescent and/or star-forming regions when AGN effects are considered in the fitting. 

To investigate the effect that AGN contributions may have on rest-frame $UVJ$ colors, we present the \prospector\ best-fit rest-frame SEDs used for the rest-UVJ color analysis in Figure \ref{fig:seds}.
For 3D-UDS-35168, 3D-UDS-41232, and 3D-UDS-39102, the rest-UV shows a visual deviation between the two best-fit models. However, these variations are not captured by the $U$ band. 
The larger deviations in spectral shape are in the rest NIR regime of the SEDs and $\sim50\%$ of our sample show significant variations in the NIR spectral shape. 
For the majority of these galaxies the shift is beyond the region covered by observed photometry.  
The larger observed color shifts in galaxies such as ZF-COS-20115 and 3D-EGS-18996 are driven by the increase in the rest-frame $J$ band due to the contributions in the AGN in the SEDs. 
However, only the very early stages of this shift are captured by the $J$ band.  
Galaxies such as 3D-EGS-34322 also show significant variation in the NIR part of the best-fit SED when AGN effects are included in \prospector. However, this variation is not captured by the rest-frame $J$ band, thus the absolute shift in colors are $\lesssim0.1$ mag.

For most of our sources, the reddest $F444W$ NIRCam band does not cover the rest-frame J band; thus, direct observational constraints are limited in constraining this turnover.
We have three sources with no JWST/NIRcam imaging. For these sources as mentioned in Section \ref{sec:sample}, we use photometric data from the ZFOURGE and 3DHST surveys. Spitzer imaging in these surveys goes beyond the rest-frame $J$ band. 
Therefore, for galaxies such as ZF-COS-19589, Spitzer IRAC Channel 4 imaging provides additional constrains to the shape of the SED, albeit with relatively lower S/N which reduces its constraining power. 
Based on photometric coverage, it is evident that imaging in the mid infra-red bands with instruments such as JWST/MIRI is required in $z\sim3-5$ galaxies to provide tighter constraints to the contribution of AGNs to the observed SEDs of the galaxies.

\begin{figure*}
\includegraphics[scale=0.65, trim= 0 0 0 0, clip]{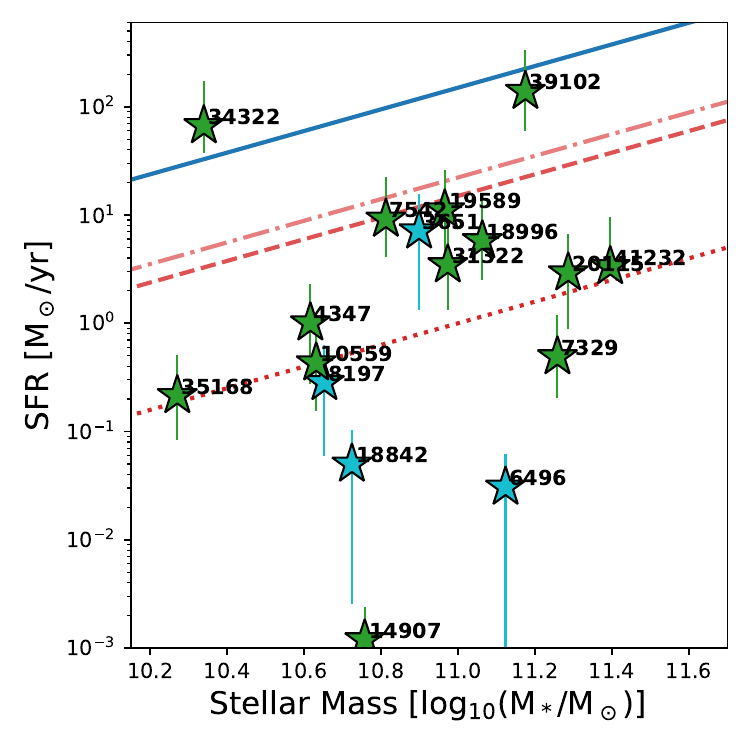}
\includegraphics[scale=0.65, trim= 0 0 0 0, clip]{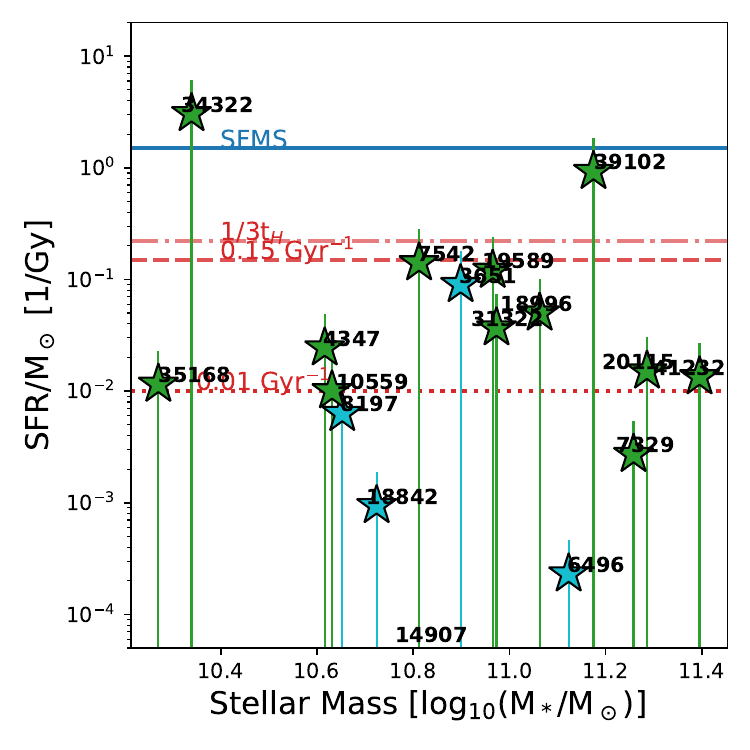}
\caption{
{\bf Left:} the star formation vs stellar mass relation and {\bf Right:} the specific star-formation rate vs stellar mass relation of our sample. The SFRs are computed by averaging the total star-formation in last 100 Myr before the time of the observation and are as parameterized by \prospector. Similar to Figure \ref{fig:full_qu_sample_pros_agns}, galaxies selected as AGNs based on prominent rest-frame optical emission lines are shown by cyan stars and the other sources are shown by green stars. In both panels, the star-forming main sequence ($\mathrm{sSFR_{MS}=1.5 Gyr^{-1}}$) from \citet{Schreiber2018} for massive $3<z<4$ ZFOURGE galaxies is shown by the solid blue line. In red we show several definitions for quenching presented by \citet{Schreiber2018}: the dash dotted line shows the $sSFR = 1/(3 t_H)$ (where $t_H$ is the age of the Universe at $z\sim4$), the dashed line shows the $sSFR=0.15\ Gyr^{-1}$, and the dotted line shows the $sSFR=0.01\ Gyr^{-1}$.  
\label{fig:sSFRs}}
\end{figure*}

To confirm whether our galaxies are truly quiescent, we use the \prospector\ best-fit parameters to examine the sSFRs of our sample and the distribution of the quiescent candidates within the star-forming stellar mass relation \citep{Noeske2007}.
In Figure \ref{fig:sSFRs}, we show the distribution of our sample in this space. 
The main locus for star-forming galaxies is defined based on the average SFR of the massive ($3\times10^9<M_*/M_\odot<10^{10}$) $3<z<4$ galaxies in the ZFOURGE survey as defined by \citet{Schreiber2018}. 
The definition of quiescence is explored using three different methods. 
Firstly, the cut at the highest sSFR is obtained with $sSFR = 1/(3 t_H) \sim 0.22\ \text{Gyr}^{-1}$, where $t_H$ is the age of the Universe at $z\sim4$. 
Secondly, the $sSFR=0.15\ \text{Gyr}^{-1}$ cut is utilized, which is $\times10$ lower than the main star-forming locus as defined above.
Finally, we also use an  $sSFR=0.01\ \text{Gyr}^{-1}$ definition which is generally defined as the absolute cut in sSFR to obtain \emph{red and dead galaxies} \citep[e.g][]{De-Lucia2024a}.

All but the two galaxies that fell within the (red) star-forming region in the $UVJ$ color selection fall within at least one of the quenching definitions. As noted by \citet{Schreiber2018} this highlights that the $UVJ$ color selection is effectively able to identify fully quenched galaxies and galaxies with very low residual star-formation. We expect the massive galaxies in the latter selection are on route to be fully quenched. 
It is also important to highlight the distinction between our massive $z\sim3-5$ quiescent galaxies and the \emph{napping} galaxies identified by JWST at higher redshifts \citep[e.g.][]{Looser2024a,Strait2023a}. 
Given that these sources have young stellar populations ($\lesssim100$ Myr), they have considerable UV flux and therefore cannot be photometrically selected by $UVJ$ color selection techniques. 
Thus, spectroscopy is crucial to discover\ such sources. 
Furthermore, these galaxies are considerably smaller ($\sim10^5-10^8$\msol), and rejuvenation at later times is expected to further build the stellar masses. 
Recent analysis of $z>6$ star-forming galaxies shows evidence for such stochastic SFH at early times \citep{Ciesla2024a,Endsley2024a,Pallottini2023a}.  
Further complications arise due to spatial color variations observed within some massive galaxies \citep{Setton2024a}, which hints that central regions of massive quiescent galaxies may have formed at earlier times with residual star-formation observed in the outskirts. 
High resolution observations of massive and quiescent $z>3$ galaxies with spatially resolved spectroscopy is required distinguish between different quenching pathways for galaxies in the early Universe by exploring age gradients and outflow signatures \citep{Park2024a}.

\subsection{Prominence of  NaD absorption and AGN in massive quiescent galaxies }\label{sec:NaD}

\begin{figure*}
\includegraphics[scale=0.45, trim= 0.1 10 0.1 10, clip]{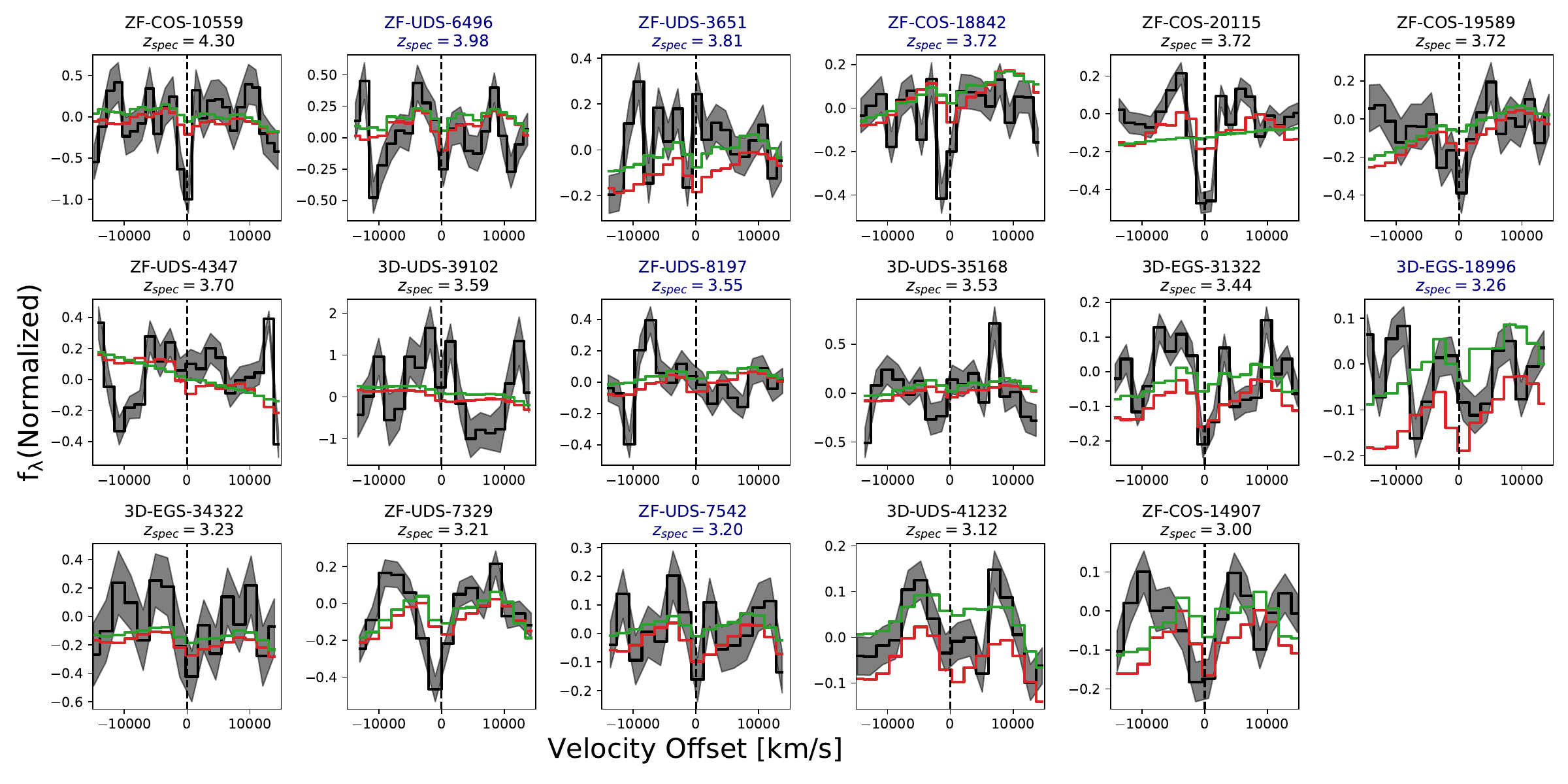}
\caption{NaD absorption profiles of our sample. Spectra are shown respective to their expected rest-frame velocity of NaD based on the {\tt slinefit} spectroscopic redshift. The associated errors are highlighted in grey and the best fit \fastpp\ and \prospector\ C3K models are shown in red and green respectively. The name of galaxies with strong \Halpha\ and/or \OIII\ emission (shown by Figure \ref{fig:halpha_oiii_zoom}) are shown by blue.
\label{fig:full_qu_sample_spectra_NaD}}
\end{figure*}

Even within the lower resolution of the Prism observations, a sub-sample of our galaxies shows clear NaD absorption features. 
In Figure \ref{fig:full_qu_sample_spectra_NaD} we show the spectra zoomed in around the region of NaD. There is no coverage for ZF-COS-20133 and 3D-EGS-27584 due to the detector gaps in NIRSpec. 

Strong NaD absorption has been observed in quiescent galaxies at $z\sim2$ \citep{Belli2024a, Park2024a} and in old massive quiescent galaxies at $3.2<z<4.6$ \citep{Carnall2024a}. The high EW of NaD compared to the best fit stellar models and the blue shifted nature of the NaD feature \citep{Belli2024a} has been argued as evidence for AGN driven outflows which may have resulted in abruptly quenching the star-formation.  
Two of our sources, ZF-UDS-7329 and ZF-UDS-6496 are observed with the NIRSpec medium resolution gratings  by \citet{Carnall2024a} and show a clear enhancement of NaD compared to their best-fit model. 
\citet{Carnall2024a}  also found that  ZF-UDS-7329 has an enhanced abundance of [Mg/Fe] compared to solar values with best-fit results from {\tt alf} \citep{Conroy2012,Conroy2018a} suggesting an [Mg/Fe] of $0.42^{+0.19}_{-0.17}$. 
While Na is not an $\alpha$ element, the dominant production mechanisms of both Na and Mg are core collapse supernovae and asymptotic giant branch stars \citep{Kobayashi2020a}. Thus, enhancement of Na in the ISM in galaxies can generally mirror an enhancement of $\alpha$ process elements in galaxies. 
We also note that linking $\alpha$ enhancements to the formation timescales of galaxies can be complicated due to the possibility of strong outflows expelling elements such as Mg, leading to a deficiency in the measured $\alpha$ elements \citep[e.g.][]{Beverage2025a}.

Recent cosmological hydrodynamical simulations show that AGN feedback plays a dominant role regulating star-formation leading to quenching at $z>3$. Therefore, observational signatures of current and/or past effects of AGNs should be visible in observed spectroscopy.  
As discussed in Section \ref{sec:prospector_agn}, there are 6 galaxies in our sample with prominent broad emission lines which indicate possible AGN activity. 
As shown by Figure \ref{fig:full_qu_sample_spectra_NaD}, a subset of our sample also show prominent NaD absorption, including 4 of the 6 galaxies identified as possible AGN.
These two observational signatures tied together could hint current or past AGN driven activity in most of the galaxies in our sample. 
Qualitative analysis of broad AGN and NaD outflow signatures require higher resolution spectra than what we currently have in the Prism mode. 
Additionally, deeper, higher-resolution data will allow better constraints using line ratio diagnostic analyses, such as those presented by \citet{Baldwin1981}, to tighten our understanding of what powers these emission lines. Furthermore, deeper X-ray data will also add value by further constraining the nature of the AGN in these massive quiescent galaxies at $z>3$.

Ground based spectroscopy of $z\sim3-4$ massive star-forming galaxies also show evidence for AGNs, suggesting an enhanced fraction of AGN in higher stellar mass galaxies in the early Universe \citep[e.g.][]{Martinez-Marin2024a}.  
Therefore, a picture of AGN feedback driven rapid quenching is emerging to explain the formation mechanisms of $z>3$ massive quiescent galaxies. 
Future mass complete spectroscopic analysis of massive $z>3$ galaxies would provide tight constraints to AGN fractions and their outflow signatures in the early Universe. 
However, a possible complication does arise with the recent discovery of extremely compact and red objects in the early Universe \citep[e.g.][]{Labbe2025a,Greene2024a}, which are attributed to a previously unseen population of galaxies with AGNs and possibly progenitor bulges of lower redshift massive quiescent galaxies. 
The X-ray and spectral properties of these sources are different to nominal AGN \citep[e.g.][]{Ananna2024a,Wang2024a}, thus deeper constraints from x-ray and rest-frame optical and near-infrared spectroscopy is crucial to further constrain the properties of these galaxies and investigate their role in regulating galaxy growth in the early Universe.

\subsection{The number densities of $z>3$ massive quiescent galaxies }\label{sec:number_densities}

In the pre-JWST era, the number densities were largely constrained using photometric selection techniques. Candidates were selected with rest-$UVJ$ color selection techniques. Deep ground based spectroscopy was used for spectroscopic confirmations and photometric number densities were statistically adjusted based on the contamination rate obtained by the spectroscopic observations \citep[e.g][]{Schreiber2018}. 

The parent sample of our analysis presented by \citet{Schreiber2018} found a number density of $2.0\pm 0.3 \times 10^{-5} Mpc^{-3}$ for $3<z<4$ massive quiescent galaxies with a $K$ band magnitude cut at $K<24.5$. 
\citet{Carnall2020a} found a similarly consistent number density of $1.7 \pm 0.4 \times 10^{-5} Mpc^{-3}$ when similar selection criteria were applied to their photometric sample. 
However, when robust constrains were added to the redshift and sSFR posteriors, the number density of the \citet{Carnall2020a} sample was reduced to $6.8 \pm 2.4 \times 10^{-6} Mpc^{-3}$.

Building upon the number densities presented by \citet{Schreiber2018}, we find that 2 further sources are classified as star-forming based on our analysis presented in Section \ref{sec:quiescence}. Both 3D-EGS-34322 and 3D-UDS-39102 did not have ground based spectroscopic confirmations. In the initial 24 massive quiescent galaxy candidates presented by \citet{Schreiber2018}, five galaxies are considered to be contaminants.  ZF-COS-20032 and 3D-UDS-27939 were found to be $z<2.5$ dusty interlopers. 
Based on the MOSFIRE spectroscopic observations  \citet{Schreiber2018} found that ZF-COS-20133 and ZF-UDS-8197 had rest-frame $U-V$ vs $V-J$ colors that moved them out of the quiescent region due to high-EW \OIII\ emission lines. 3D-EGS-18996 was found also to have moved to the star-forming region due to significant underestimation of the photometric redshift.

In Section \ref{fig:UVJ} we recomputed rest-frame $UVJ$ colors for the JWST/NIRSpec sample. This includes all three galaxies that were ruled out to be $z\sim3-4$ star-forming galaxies by \citet{Schreiber2018}. The rest-frame $UVJ$ colors and the sSFRs of all three of these sources classify as quiescent. 
We note that we consider 3D-EGS-18996 rest $UVJ$ quiescent due to it falling within $<0.1$ mag from the quiescent region in the rest-frame $U-V$ vs $V-J$ color space. 3D-EGS-34322 and 3D-UDS-39102 are star-forming sources and can be removed as contaminants. 
While ZF-COS-10559 is quiescent, it is at $z=4.3$, which is beyond the $z\sim3-4$ region probed by \citet{Schreiber2018}. Thus, once we also remove this source, we end up with the same 80\% purity for quiescent sources and a number density of $(1.4\pm0.3)\times10^{-5}Mpc^{-3}$ as  \citet{Schreiber2018}. 
Thus our analysis highlights that pre-JWST number densities of massive $z\sim3-4$ quiescent galaxies are largely still consistent with post JWST results. 
Current cosmological and semi-analytical models are able to reproduce these number densities at $z\sim3$ \citep{Lagos2024a,Valentino2023a} but challenges lie at $z>4$ \citep{Weller2025a}.

\section{Summary and Conclusions} \label{sec:conclusions}

In this work we presented JWST/NIRSpec spectroscopy of 19 $3.0<z<4.5$ massive quiescent galaxy candidates. This completes the spectroscopic followup of the 24 $z>3-4$ $K$-selected massive quiescent galaxy candidates in the HST legacy fields presented by \citet{Schreiber2018}. 
We find that the 12 galaxies previously unconfirmed by ground-based spectroscopy are at $z>3$, highlighting an accuracy of over 80\% in the photometric redshifts for selecting Balmer break galaxies in the $3.0 < z < 4.5$ epoch.

We use \fastpp\ and \prospector\ SED fitting codes to analyze the formation history of our galaxies and compare with predictions from cosmological simulations. Our main conclusions are as follows:

\begin{itemize}

\item 17 of the 19 galaxies are classified to be quiescent based on the rest-frame $U-V$ and $V-J$ color selection techniques [Figure \ref{fig:UVJ}] and various quenching definitions utilizing sSFR [Figure \ref{fig:sSFRs}] rates at $z\sim3$.

\item The quenched sample shows a variety of formation and quenching time scales [Figure \ref{fig:full_qu_sample_sfh_t50quench_comp}]. The oldest galaxy in our sample formed $\sim6\times10^{10}$\msol by $z\sim9$ and quenched for $\gtrsim 1$ billion years by the time of the observation at $z=3.2$.

\item The formation histories reconstructed using parametric SFH prior assumptions from \fastpp\ and binned/non-parametric SFH assumptions from \prospector\ are largely similar [Figures \ref{fig:full_qu_sample_sfh_comp} and \ref{fig:full_qu_sample_sfh_t50quench_comp}]. 

\item The choice of stellar libraries/stellar population models [Figures \ref{fig:full_qu_sample_pros_comp_ssps_trim}, \ref{fig:full_qu_sample_sfh_t50quench_comp_ssps}, and
\ref{fig:full_qu_sample_pros_average_sfhs_trim}], the wavelength range used in spectral fitting [Figure \ref{fig:full_qu_sample_pros_average_sfhs_fullspec}], and how the effects of nebular emission [Figure \ref{fig:full_qu_sample_sfh_t50quench_comp_elines}] and AGN [Figure \ref{fig:full_qu_sample_pros_agns}] are handled have minimal impact on the reconstructed shape of the SFHs for most our sources. While some galaxies show significant deviations, the average SFHs of the observed galaxies also remain largely consistent. However, we find that the assumed SFH prior [Figures \ref{fig:full_qu_sample_pros_average_sfhs_fullspec} and \ref{fig:full_qu_sample_sfh_prior_comp}] plays a role in determining the SFR at the earliest times, particularly in galaxies where the diagnostic power of the observed spectra may be limited.

\item $\sim6$ of our galaxies show high-EW broad \Halpha\ and/or \OIII\ emission lines signifying possible AGN activity in these quiescent galaxies [Figure \ref{fig:halpha_oiii_zoom}].

\item $\sim8$ of our galaxies shows NaD absorption features signifying an Na enhancement in our sources and possible outflows [Figure \ref{fig:full_qu_sample_spectra_NaD}]. While these could be attributed to AGN and/or star-formation driven outflows that may have contributed to the rapid cessation in star-formation for our galaxies, we cannot conclusively investigate this due to the low resolution Prism mode observations.

\item With mock JWST/NIRSpec Prism mode observations of TNG300 $z\sim3$ massive quiescent galaxies, we show that the average SFHs are recovered accurately by \prospector\ [Figure \ref{fig:pros_recovery_sims}]. 

\item The average SFHs of $z\sim3$ massive quiescent galaxies from the Illustris TNG100, TNG300, SHARKv2, and Magneticum samples are consistent with the average SFR of our quiescent galaxies [Figure \ref{fig:sfh_comp_with_sims}]. This indicates that current hydrodynamical simulations can, on average, reproduce the observed formation histories of $z > 3$ massive quiescent galaxies. However, there are variations in the different feedback modes across simulations, leading to overall differences in the quenching timescales of the simulated galaxies.

\end{itemize}

Our results highlight that JWST has now opened up a new window to exploring the properties of stellar populations and formation histories of massive quiescent galaxies in the $z>3$ universe. 
The spectroscopic advancements compared to  ground-based observations tighten the formation and quenching timescales of these populations and suggest that AGNs may play a significant role in quenching the first massive galaxies in the early Universe. 
While methods used for SFH analysis may influence the formation histories of individual galaxies, the population-wide statistics show good agreement. The main uncertainty arises at older times (at the highest redshifts), where the diagnostic power of stellar features in the galaxy spectra may be limited.
Future higher resolution, spatially resolved spectroscopy from JWST NIRSpec and MIRI integral field spectrographs, as well as ground-based adaptive optics-enabled integral field spectroscopy from 10m-30m class telescopes, will provide tighter constraints on the relationship between stars and gas in these first massive quiescent galaxies, advancing our understanding of efficient star-formation quenching mechanisms in the early universe.
Higher-resolution data will also be essential for determining the nature of the underlying AGNs powering these galaxies. Additionally, future deep ALMA observations will be key to investigating whether any remaining neutral gas exists around these sources. This gas may have been expelled by AGN and could later cool and fall back to form stars, potentially rejuvenating the galaxy. Combined with deep X-ray imaging, future observations will provide a comprehensive understanding of the nature and role of AGNs in regulating galaxy evolution during the first $\sim2$ billion years of the Universe.

\facility{JWST: (NIRSpec)}

\begin{acknowledgments}
We thank Jarle Brinchmann, Adam Carnall, Kartheik Iyer, and Joel Leja for insightful discussions. 
This work is based on observations made with the NASA/ESA/CSA James Webb Space Telescope. The data were obtained from the Mikulski Archive for Space Telescopes at the Space Telescope Science Institute, which is operated by the Association of Universities for Research in Astronomy, Inc., under NASA contract NAS 5-03127 for JWST. These observations are associated with program JWST-GO-2565. The specific observations analyzed can be accessed via \dataset[doi: 10.17909/rkd4-gr92]{https://doi.org/10.17909/rkd4-gr92}.
T.N., K. G., H.G.C, C.J. and L.K. acknowledge support from Australian Research Council Laureate Fellowship FL180100060. 
This project made use of {\tt astropy} \citep{Astropy2018}, {\tt matplotlib} \citep{Hunter2007}, and {\tt pandas} \citep{Pandas2020}. 
In preparing this manuscript, we utilized ChatGPT, a language model developed by OpenAI, for assistance with {\tt python} and UNIX shell scripts to used in the analysis. Furthermore, ChatGPT was used to refine the manuscript's clarity through minor corrections to English grammar.
(Some of) The data products presented herein were retrieved from the Dawn JWST Archive (DJA). DJA is an initiative of the Cosmic Dawn Center (DAWN), which is funded by the Danish National Research Foundation under grant DNRF140.
\end{acknowledgments}

\appendix
\restartappendixnumbering
\section{Appendix Figures} \label{sec:appendix_sims}

\begin{figure*}
\includegraphics[scale=0.85, trim= 40 10 0.1 10, clip]{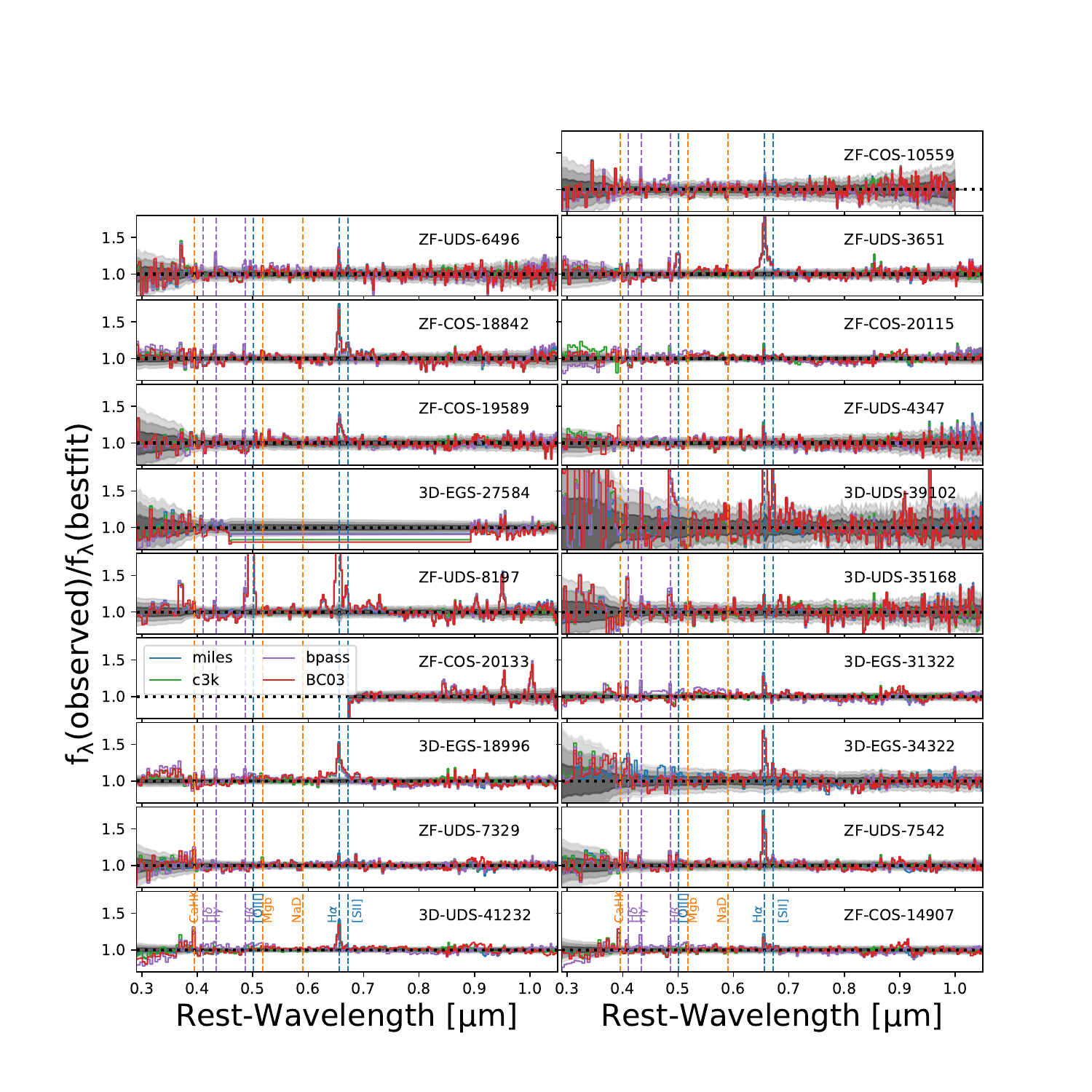}
\caption{\fastpp\ and \prospector\ best-fit residuals for our galaxies are shown here. The observed spectra are divided by the best-fit spectra. The \fastpp\ fits use the \citet{Bruzual2003} stellar population models, while the \prospector\ fits use the MILES, C3K, and BPASS models, as outlined in Section \ref{sec:sfh}. Prominent rest-frame optical features are marked by vertical dashed lines. The 1, 2, and 3-$\sigma$ error levels for the residuals are shaded in grey with decreasing transparency, respectively. y=1 line is marked with a dotted line for reference. 
\label{fig:best-fit-residuals}}
\end{figure*}

\begin{figure*}
\includegraphics[scale=0.85, trim= 0.1 10 0.1 10, clip]{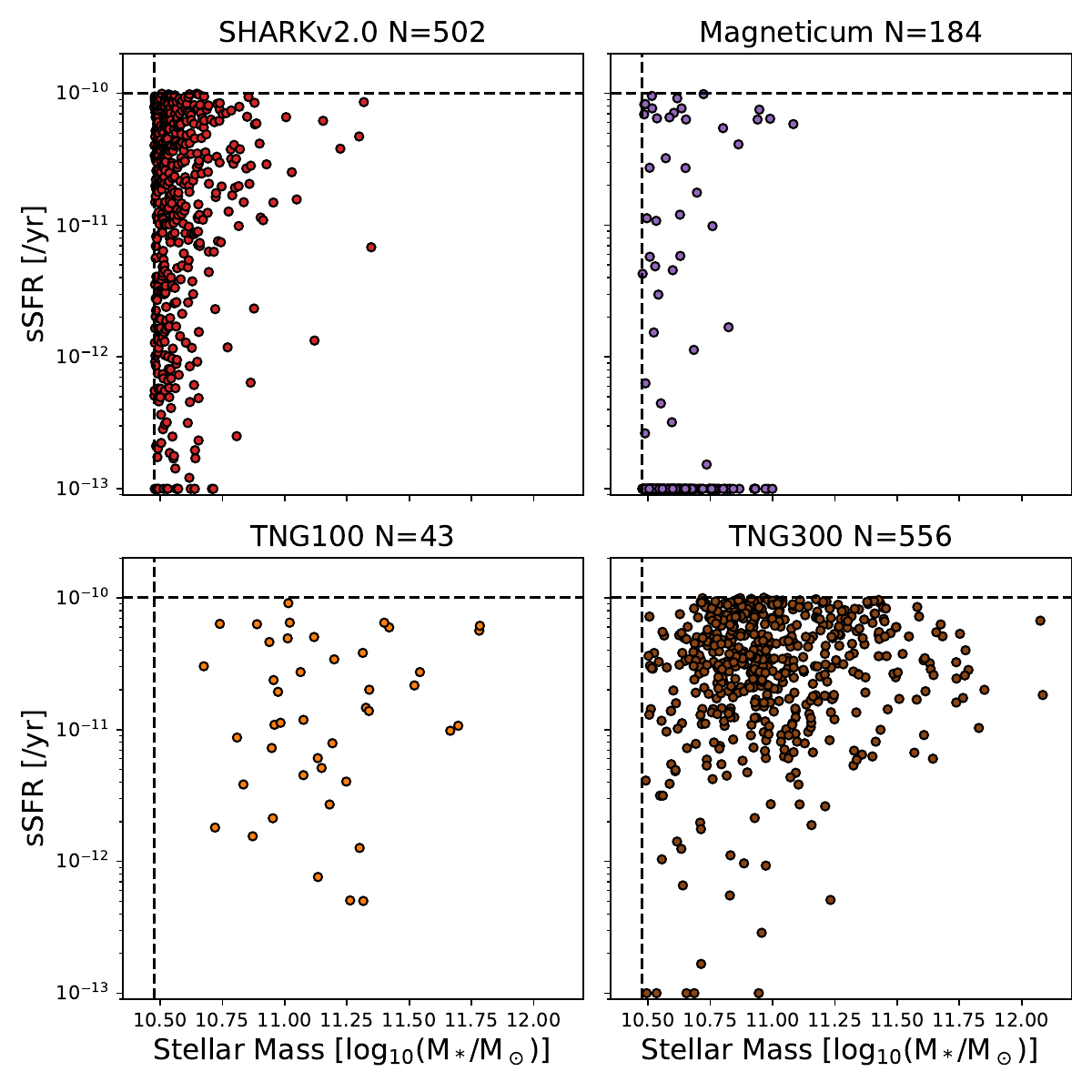}
\caption{The sSFRs and stellar masses of $z\sim3$ massive quiescent galaxies from cosmological simulations used in our analysis. Panels are as labelled and galaxies used for reconstructing the formation histories as detailed in Section \ref{sec:comparison_with_cos_sims} and Figure \ref{fig:sfh_comp_with_sims} are shown here. The horizontal and vertical dashed lines show the sSFR ($10^{-10}$/yr) and stellar mass ($3\times10^{10}$\msol) cut we have used in simulations to select the massive quiescent galaxies. 
\label{fig:sims_ssfr_vs_mass}}
\end{figure*}

The best-fit residuals for our galaxies are shown by Appendix Figure \ref{fig:best-fit-residuals}. 
Appendix Figure \ref{fig:sims_ssfr_vs_mass} shows the sSFRs and stellar masses of $z\sim3$ massive quiescent galaxies from cosmological simulations used in our analysis.

\bibliographystyle{aasjournal}
\bibliography{bibliography.bib}

\begin{thebibliography}{}
\expandafter\ifx\csname natexlab\endcsname\relax\def\natexlab#1{#1}\fi
\providecommand{\url}[1]{\href{#1}{#1}}
\providecommand{\dodoi}[1]{doi:~\href{http://doi.org/#1}{\nolinkurl{#1}}}
\providecommand{\doeprint}[1]{\href{http://ascl.net/#1}{\nolinkurl{http://ascl.net/#1}}}
\providecommand{\doarXiv}[1]{\href{https://arxiv.org/abs/#1}{\nolinkurl{https://arxiv.org/abs/#1}}}

\bibitem[{{Ananna} {et~al.}(2024){Ananna}, {Bogd{\'a}n}, {Kov{\'a}cs},
  {Natarajan}, \& {Hickox}}]{Ananna2024a}
{Ananna}, T.~T., {Bogd{\'a}n}, {\'A}., {Kov{\'a}cs}, O.~E., {Natarajan}, P., \&
  {Hickox}, R.~C. 2024, \apjl, 969, L18, \dodoi{10.3847/2041-8213/ad5669}

\bibitem[{{Antwi-Danso} {et~al.}(2025){Antwi-Danso}, {Papovich}, {Esdaile},
  {Nanayakkara}, {Glazebrook}, {Hutchison}, {Whitaker}, {Marsan}, {Diaz},
  {Marchesini}, {Muzzin}, {Tran}, {Setton}, {Kaushal}, {Speagle}, \&
  {Cole}}]{Antwi-Danso2025a}
{Antwi-Danso}, J., {Papovich}, C., {Esdaile}, J., {et~al.} 2025, \apj, 978, 90,
  \dodoi{10.3847/1538-4357/ad8b30}

\bibitem[{{Astropy Collaboration} {et~al.}(2018){Astropy Collaboration},
  {Price-Whelan}, {Sip{\H o}cz}, {G{\"u}nther}, {Lim}, {Crawford}, {Conseil},
  {Shupe}, {Craig}, {Dencheva}, {Ginsburg}, {VanderPlas}, {Bradley},
  {P{\'e}rez-Su{\'a}rez}, {de Val-Borro}, {Aldcroft}, {Cruz}, {Robitaille},
  {Tollerud}, {Ardelean}, {Babej}, {Bach}, {Bachetti}, {Bakanov}, {Bamford},
  {Barentsen}, {Barmby}, {Baumbach}, {Berry}, {Biscani}, {Boquien}, {Bostroem},
  {Bouma}, {Brammer}, {Bray}, {Breytenbach}, {Buddelmeijer}, {Burke},
  {Calderone}, {Cano Rodr{\'{\i}}guez}, {Cara}, {Cardoso}, {Cheedella},
  {Copin}, {Corrales}, {Crichton}, {D'Avella}, {Deil}, {Depagne}, {Dietrich},
  {Donath}, {Droettboom}, {Earl}, {Erben}, {Fabbro}, {Ferreira}, {Finethy},
  {Fox}, {Garrison}, {Gibbons}, {Goldstein}, {Gommers}, {Greco}, {Greenfield},
  {Groener}, {Grollier}, {Hagen}, {Hirst}, {Homeier}, {Horton}, {Hosseinzadeh},
  {Hu}, {Hunkeler}, {Ivezi{\'c}}, {Jain}, {Jenness}, {Kanarek}, {Kendrew},
  {Kern}, {Kerzendorf}, {Khvalko}, {King}, {Kirkby}, {Kulkarni}, {Kumar},
  {Lee}, {Lenz}, {Littlefair}, {Ma}, {Macleod}, {Mastropietro}, {McCully},
  {Montagnac}, {Morris}, {Mueller}, {Mumford}, {Muna}, {Murphy}, {Nelson},
  {Nguyen}, {Ninan}, {N{\"o}the}, {Ogaz}, {Oh}, {Parejko}, {Parley}, {Pascual},
  {Patil}, {Patil}, {Plunkett}, {Prochaska}, {Rastogi}, {Reddy Janga},
  {Sabater}, {Sakurikar}, {Seifert}, {Sherbert}, {Sherwood-Taylor}, {Shih},
  {Sick}, {Silbiger}, {Singanamalla}, {Singer}, {Sladen}, {Sooley},
  {Sornarajah}, {Streicher}, {Teuben}, {Thomas}, {Tremblay}, {Turner},
  {Terr{\'o}n}, {van Kerkwijk}, {de la Vega}, {Watkins}, {Weaver}, {Whitmore},
  {Woillez}, {Zabalza}, \& {Astropy Contributors}}]{Astropy2018}
{Astropy Collaboration}, {Price-Whelan}, A.~M., {Sip{\H o}cz}, B.~M., {et~al.}
  2018, \aj, 156, 123, \dodoi{10.3847/1538-3881/aabc4f}

\bibitem[{{Baldwin} {et~al.}(1981){Baldwin}, {Phillips}, \&
  {Terlevich}}]{Baldwin1981}
{Baldwin}, J.~A., {Phillips}, M.~M., \& {Terlevich}, R. 1981, \pasp, 93, 5,
  \dodoi{10.1086/130766}

\bibitem[{{Belli} {et~al.}(2024){Belli}, {Park}, {Davies}, {Mendel}, {Johnson},
  {Conroy}, {Benton}, {Bugiani}, {Emami}, {Leja}, {Li}, {Maheson}, {Mathews},
  {Naidu}, {Nelson}, {Tacchella}, {Terrazas}, \& {Weinberger}}]{Belli2024a}
{Belli}, S., {Park}, M., {Davies}, R.~L., {et~al.} 2024, \nat, 630, 54,
  \dodoi{10.1038/s41586-024-07412-1}

\bibitem[{{Bertelli} {et~al.}(1994){Bertelli}, {Bressan}, {Chiosi}, {Fagotto},
  \& {Nasi}}]{Bertelli1994a}
{Bertelli}, G., {Bressan}, A., {Chiosi}, C., {Fagotto}, F., \& {Nasi}, E. 1994,
  \aaps, 106, 275

\bibitem[{{Beverage} {et~al.}(2025){Beverage}, {Slob}, {Kriek}, {Conroy},
  {Barro}, {Bezanson}, {Brammer}, {Cheng}, {de Graaff}, {F{\"o}rster
  Schreiber}, {Franx}, {Lorenz}, {Mancera Pi{\~n}a}, {Marchesini}, {Muzzin},
  {Newman}, {Price}, {Shapley}, {Stefanon}, {Suess}, {van Dokkum}, {Weinberg},
  \& {Weisz}}]{Beverage2025a}
{Beverage}, A.~G., {Slob}, M., {Kriek}, M., {et~al.} 2025, \apj, 979, 249,
  \dodoi{10.3847/1538-4357/ad96b6}

\bibitem[{{Bruzual} \& {Charlot}(2003)}]{Bruzual2003}
{Bruzual}, G., \& {Charlot}, S. 2003, \mnras, 344, 1000,
  \dodoi{10.1046/j.1365-8711.2003.06897.x}

\bibitem[{{Byler} {et~al.}(2017){Byler}, {Dalcanton}, {Conroy}, \&
  {Johnson}}]{Byler2017}
{Byler}, N., {Dalcanton}, J.~J., {Conroy}, C., \& {Johnson}, B.~D. 2017, \apj,
  840, 44, \dodoi{10.3847/1538-4357/aa6c66}

\bibitem[{{Calzetti}(2000)}]{Calzetti1999}
{Calzetti}, D. 2000, in Building Galaxies; from the Primordial Universe to the
  Present, ed. F.~{Hammer}, T.~X. {Thuan}, V.~{Cayatte}, B.~{Guiderdoni}, \&
  J.~T. {Thanh Van}, 233

\bibitem[{{Calzetti} {et~al.}(2000){Calzetti}, {Armus}, {Bohlin}, {Kinney},
  {Koornneef}, \& {Storchi-Bergmann}}]{Calzetti2000}
{Calzetti}, D., {Armus}, L., {Bohlin}, R.~C., {et~al.} 2000, \apj, 533, 682,
  \dodoi{10.1086/308692}

\bibitem[{{Carnall} {et~al.}(2019){Carnall}, {Leja}, {Johnson}, {McLure},
  {Dunlop}, \& {Conroy}}]{Carnall2019a}
{Carnall}, A.~C., {Leja}, J., {Johnson}, B.~D., {et~al.} 2019, \apj, 873, 44,
  \dodoi{10.3847/1538-4357/ab04a2}

\bibitem[{{Carnall} {et~al.}(2020){Carnall}, {Walker}, {McLure}, {Dunlop},
  {McLeod}, {Cullen}, {Wild}, {Amorin}, {Bolzonella}, {Castellano}, {Cimatti},
  {Cucciati}, {Fontana}, {Gargiulo}, {Garilli}, {Jarvis}, {Pentericci},
  {Pozzetti}, {Zamorani}, {Calabro}, {Hathi}, \& {Koekemoer}}]{Carnall2020a}
{Carnall}, A.~C., {Walker}, S., {McLure}, R.~J., {et~al.} 2020, \mnras, 496,
  695, \dodoi{10.1093/mnras/staa1535}

\bibitem[{{Carnall} {et~al.}(2023{\natexlab{a}}){Carnall}, {McLeod}, {McLure},
  {Dunlop}, {Begley}, {Cullen}, {Donnan}, {Hamadouche}, {Jewell}, {Jones},
  {Pollock}, \& {Wild}}]{Carnall2023a}
{Carnall}, A.~C., {McLeod}, D.~J., {McLure}, R.~J., {et~al.}
  2023{\natexlab{a}}, \mnras, 520, 3974, \dodoi{10.1093/mnras/stad369}

\bibitem[{{Carnall} {et~al.}(2023{\natexlab{b}}){Carnall}, {McLure}, {Dunlop},
  {McLeod}, {Wild}, {Cullen}, {Magee}, {Begley}, {Cimatti}, {Donnan},
  {Hamadouche}, {Jewell}, \& {Walker}}]{Carnall2023b}
{Carnall}, A.~C., {McLure}, R.~J., {Dunlop}, J.~S., {et~al.}
  2023{\natexlab{b}}, \nat, 619, 716, \dodoi{10.1038/s41586-023-06158-6}

\bibitem[{{Carnall} {et~al.}(2024){Carnall}, {Cullen}, {McLure}, {McLeod},
  {Begley}, {Donnan}, {Dunlop}, {Shapley}, {Rowlands}, {Almaini},
  {Arellano-C{\'o}rdova}, {Barrufet}, {Cimatti}, {Ellis}, {Grogin},
  {Hamadouche}, {Illingworth}, {Koekemoer}, {Leung}, {Lovell},
  {P{\'e}rez-Gonz{\'a}lez}, {Santini}, {Stanton}, \& {Wild}}]{Carnall2024a}
{Carnall}, A.~C., {Cullen}, F., {McLure}, R.~J., {et~al.} 2024, \mnras, 534,
  325, \dodoi{10.1093/mnras/stae2092}

\bibitem[{Chabrier(2003)}]{Chabrier2003}
Chabrier, G. 2003, Publications of the Astronomical Society of the Pacific,
  115, pp. 763.
\newblock \url{http://www.jstor.org/stable/10.1086/376392}

\bibitem[{{Chittenden} \& {Tojeiro}(2023)}]{Chittenden2023a}
{Chittenden}, H.~G., \& {Tojeiro}, R. 2023, \mnras, 518, 5670,
  \dodoi{10.1093/mnras/stac3498}

\bibitem[{{Choi} {et~al.}(2016){Choi}, {Dotter}, {Conroy}, {Cantiello},
  {Paxton}, \& {Johnson}}]{Choi2016}
{Choi}, J., {Dotter}, A., {Conroy}, C., {et~al.} 2016, \apj, 823, 102,
  \dodoi{10.3847/0004-637X/823/2/102}

\bibitem[{{Ciesla} {et~al.}(2024){Ciesla}, {Elbaz}, {Ilbert}, {Buat},
  {Magnelli}, {Narayanan}, {Daddi}, {G{\'o}mez-Guijarro}, \&
  {Arango-Toro}}]{Ciesla2024a}
{Ciesla}, L., {Elbaz}, D., {Ilbert}, O., {et~al.} 2024, \aap, 686, A128,
  \dodoi{10.1051/0004-6361/202348091}

\bibitem[{{Conroy}(2013)}]{Conroy2013}
{Conroy}, C. 2013, \araa, 51, 393, \dodoi{10.1146/annurev-astro-082812-141017}

\bibitem[{{Conroy} {et~al.}(2009){Conroy}, {Gunn}, \& {White}}]{Conroy2009}
{Conroy}, C., {Gunn}, J.~E., \& {White}, M. 2009, \apj, 699, 486,
  \dodoi{10.1088/0004-637X/699/1/486}

\bibitem[{{Conroy} \& {van Dokkum}(2012)}]{Conroy2012}
{Conroy}, C., \& {van Dokkum}, P.~G. 2012, \apj, 760, 71,
  \dodoi{10.1088/0004-637X/760/1/71}

\bibitem[{{Conroy} {et~al.}(2018){Conroy}, {Villaume}, {van Dokkum}, \&
  {Lind}}]{Conroy2018a}
{Conroy}, C., {Villaume}, A., {van Dokkum}, P.~G., \& {Lind}, K. 2018, \apj,
  854, 139, \dodoi{10.3847/1538-4357/aaab49}

\bibitem[{{Crain} \& {van de Voort}(2023)}]{Crain2023a}
{Crain}, R.~A., \& {van de Voort}, F. 2023, \araa, 61, 473,
  \dodoi{10.1146/annurev-astro-041923-043618}

\bibitem[{{de Graaff} {et~al.}(2024){de Graaff}, {Setton}, {Brammer}, {Cutler},
  {Suess}, {Labb{\'e}}, {Leja}, {Weibel}, {Maseda}, {Whitaker}, {Bezanson},
  {Boogaard}, {Cleri}, {De Lucia}, {Franx}, {Greene}, {Hirschmann}, {Matthee},
  {McConachie}, {Naidu}, {Oesch}, {Price}, {Rix}, {Valentino}, {Wang}, \&
  {Williams}}]{deGraaff2024a}
{de Graaff}, A., {Setton}, D.~J., {Brammer}, G., {et~al.} 2024, Nature
  Astronomy, \dodoi{10.1038/s41550-024-02424-3}

\bibitem[{{De Lucia} {et~al.}(2024){De Lucia}, {Fontanot}, {Xie}, \&
  {Hirschmann}}]{De-Lucia2024a}
{De Lucia}, G., {Fontanot}, F., {Xie}, L., \& {Hirschmann}, M. 2024, arXiv
  e-prints, arXiv:2401.06211, \dodoi{10.48550/arXiv.2401.06211}

\bibitem[{{Dekel} {et~al.}(2013){Dekel}, {Zolotov}, {Tweed}, {Cacciato},
  {Ceverino}, \& {Primack}}]{Dekel2013a}
{Dekel}, A., {Zolotov}, A., {Tweed}, D., {et~al.} 2013, \mnras, 435, 999,
  \dodoi{10.1093/mnras/stt1338}

\bibitem[{{D'Eugenio} {et~al.}(2021){D'Eugenio}, {Daddi}, {Gobat},
  {Strazzullo}, {Lustig}, {Delvecchio}, {Jin}, {Cimatti}, \&
  {Onodera}}]{DEugenio2021a}
{D'Eugenio}, C., {Daddi}, E., {Gobat}, R., {et~al.} 2021, \aap, 653, A32,
  \dodoi{10.1051/0004-6361/202040067}

\bibitem[{{D'Eugenio} {et~al.}(2024){D'Eugenio}, {Cameron}, {Scholtz},
  {Carniani}, {Willott}, {Curtis-Lake}, {Bunker}, {Parlanti}, {Maiolino},
  {Willmer}, {Jakobsen}, {Robertson}, {Johnson}, {Tacchella}, {Cargile},
  {Rawle}, {Arribas}, {Chevallard}, {Curti}, {Egami}, {Eisenstein}, {Kumari},
  {Looser}, {Rieke}, {Rodr{\'\i}guez Del Pino}, {Saxena}, {{\"U}bler},
  {Venturi}, {Witstok}, {Baker}, {Bhatawdekar}, {Bonaventura}, {Boyett},
  {Charlot}, {Danhaive}, {Hainline}, {Hausen}, {Helton}, {Ji}, {Ji}, {Jones},
  {Joud{\v{z}}balis}, {Maseda}, {P{\'e}rez-Gonz{\'a}lez}, {Perna},
  {Pusk{\'a}s}, {Shivaei}, {Silcock}, {Simmonds}, {Smit}, {Sun}, {Villanueva},
  {Williams}, \& {Zhu}}]{DEugenio2024a}
{D'Eugenio}, F., {Cameron}, A.~J., {Scholtz}, J., {et~al.} 2024, arXiv
  e-prints, arXiv:2404.06531, \dodoi{10.48550/arXiv.2404.06531}

\bibitem[{{Dotter}(2016)}]{Dotter2016a}
{Dotter}, A. 2016, \apjs, 222, 8, \dodoi{10.3847/0067-0049/222/1/8}

\bibitem[{{Eldridge} {et~al.}(2017){Eldridge}, {Stanway}, {Xiao}, {McClelland},
  {Taylor}, {Ng}, {Greis}, \& {Bray}}]{Eldridge2017}
{Eldridge}, J.~J., {Stanway}, E.~R., {Xiao}, L., {et~al.} 2017, PASA, 34, e058,
  \dodoi{10.1017/pasa.2017.51}

\bibitem[{{Endsley} {et~al.}(2024){Endsley}, {Stark}, {Whitler}, {Topping},
  {Johnson}, {Robertson}, {Tacchella}, {Alberts}, {Baker}, {Bhatawdekar},
  {Boyett}, {Bunker}, {Cameron}, {Carniani}, {Charlot}, {Chen}, {Chevallard},
  {Curtis-Lake}, {Danhaive}, {Egami}, {Eisenstein}, {Hainline}, {Helton}, {Ji},
  {Looser}, {Maiolino}, {Nelson}, {Pusk{\'a}s}, {Rieke}, {Rieke}, {Rix},
  {Sandles}, {Saxena}, {Simmonds}, {Smit}, {Sun}, {Williams}, {Willmer},
  {Willott}, \& {Witstok}}]{Endsley2024a}
{Endsley}, R., {Stark}, D.~P., {Whitler}, L., {et~al.} 2024, \mnras, 533, 1111,
  \dodoi{10.1093/mnras/stae1857}

\bibitem[{{Esdaile} {et~al.}(2021{\natexlab{a}}){Esdaile}, {Glazebrook},
  {Labb{\'e}}, {Taylor}, {Schreiber}, {Nanayakkara}, {Kacprzak}, {Oesch},
  {Tran}, {Papovich}, {Spitler}, \& {Straatman}}]{Esdaile2021a}
{Esdaile}, J., {Glazebrook}, K., {Labb{\'e}}, I., {et~al.} 2021{\natexlab{a}},
  \apjl, 908, L35, \dodoi{10.3847/2041-8213/abe11e}

\bibitem[{{Esdaile} {et~al.}(2021{\natexlab{b}}){Esdaile}, {Labb{\'e}},
  {Glazebrook}, {Antwi-Danso}, {Papovich}, {Taylor}, {Marsan}, {Muzzin},
  {Straatman}, {Marchesini}, {Diaz}, {Spitler}, {Tran}, \&
  {Goodsell}}]{Esdaile2021b}
{Esdaile}, J., {Labb{\'e}}, I., {Glazebrook}, K., {et~al.} 2021{\natexlab{b}},
  \aj, 162, 225, \dodoi{10.3847/1538-3881/ac2148}

\bibitem[{{Falc{\'o}n-Barroso} {et~al.}(2011){Falc{\'o}n-Barroso},
  {S{\'a}nchez-Bl{\'a}zquez}, {Vazdekis}, {Ricciardelli}, {Cardiel}, {Cenarro},
  {Gorgas}, \& {Peletier}}]{Falcon-Barroso2011a}
{Falc{\'o}n-Barroso}, J., {S{\'a}nchez-Bl{\'a}zquez}, P., {Vazdekis}, A.,
  {et~al.} 2011, \aap, 532, A95, \dodoi{10.1051/0004-6361/201116842}

\bibitem[{{Ferland} {et~al.}(2017){Ferland}, {Chatzikos}, {Guzm{\'a}n},
  {Lykins}, {van Hoof}, {Williams}, {Abel}, {Badnell}, {Keenan}, {Porter}, \&
  {Stancil}}]{Ferland2017}
{Ferland}, G.~J., {Chatzikos}, M., {Guzm{\'a}n}, F., {et~al.} 2017, \rmxaa, 53,
  385.
\newblock \doarXiv{1705.10877}

\bibitem[{{Finkelstein} {et~al.}(2023){Finkelstein}, {Bagley}, {Ferguson},
  {Wilkins}, {Kartaltepe}, {Papovich}, {Yung}, {Arrabal Haro}, {Behroozi},
  {Dickinson}, {Kocevski}, {Koekemoer}, {Larson}, {Le Bail}, {Morales},
  {P{\'e}rez-Gonz{\'a}lez}, {Burgarella}, {Dav{\'e}}, {Hirschmann},
  {Somerville}, {Wuyts}, {Bromm}, {Casey}, {Fontana}, {Fujimoto}, {Gardner},
  {Giavalisco}, {Grazian}, {Grogin}, {Hathi}, {Hutchison}, {Jha}, {Jogee},
  {Kewley}, {Kirkpatrick}, {Long}, {Lotz}, {Pentericci}, {Pierel}, {Pirzkal},
  {Ravindranath}, {Ryan}, {Trump}, {Yang}, {Bhatawdekar}, {Bisigello}, {Buat},
  {Calabr{\`o}}, {Castellano}, {Cleri}, {Cooper}, {Croton}, {Daddi}, {Dekel},
  {Elbaz}, {Franco}, {Gawiser}, {Holwerda}, {Huertas-Company}, {Jaskot},
  {Leung}, {Lucas}, {Mobasher}, {Pandya}, {Tacchella}, {Weiner}, \&
  {Zavala}}]{Finkelstein2023a}
{Finkelstein}, S.~L., {Bagley}, M.~B., {Ferguson}, H.~C., {et~al.} 2023, \apjl,
  946, L13, \dodoi{10.3847/2041-8213/acade4}

\bibitem[{{Foreman-Mackey} {et~al.}(2013){Foreman-Mackey}, {Hogg}, {Lang}, \&
  {Goodman}}]{Foreman-Mackey2013}
{Foreman-Mackey}, D., {Hogg}, D.~W., {Lang}, D., \& {Goodman}, J. 2013, \pasp,
  125, 306, \dodoi{10.1086/670067}

\bibitem[{{Forrest} {et~al.}(2020{\natexlab{a}}){Forrest}, {Marsan},
  {Annunziatella}, {Wilson}, {Muzzin}, {Marchesini}, {Cooper}, {Chan},
  {McConachie}, {Gomez}, {Kado-Fong}, {Barbera}, {Lange-Vagle}, {Nantais},
  {Nonino}, {Saracco}, {Stefanon}, \& {van der Burg}}]{Forrest2020b}
{Forrest}, B., {Marsan}, Z.~C., {Annunziatella}, M., {et~al.}
  2020{\natexlab{a}}, \apj, 903, 47, \dodoi{10.3847/1538-4357/abb819}

\bibitem[{{Forrest} {et~al.}(2020{\natexlab{b}}){Forrest}, {Annunziatella},
  {Wilson}, {Marchesini}, {Muzzin}, {Cooper}, {Marsan}, {McConachie}, {Chan},
  {Gomez}, {Kado-Fong}, {L Barbera}, {Labb{\'e}}, {Lange-Vagle}, {Nantais},
  {Nonino}, {Pe{\~n}a}, {Saracco}, {Stefanon}, \& {van der
  Burg}}]{Forrest2020a}
{Forrest}, B., {Annunziatella}, M., {Wilson}, G., {et~al.} 2020{\natexlab{b}},
  \apjl, 890, L1, \dodoi{10.3847/2041-8213/ab5b9f}

\bibitem[{{Forrest} {et~al.}(2022){Forrest}, {Wilson}, {Muzzin}, {Marchesini},
  {Cooper}, {Marsan}, {Annunziatella}, {McConachie}, {Zaidi}, {Gomez}, {Urbano
  Stawinski}, {Chang}, {de Lucia}, {La Barbera}, {Lubin}, {Nantais},
  {Pe{\~n}a}, {Saracco}, {Surace}, \& {Stefanon}}]{Forrest2022a}
{Forrest}, B., {Wilson}, G., {Muzzin}, A., {et~al.} 2022, \apj, 938, 109,
  \dodoi{10.3847/1538-4357/ac8747}

\bibitem[{{Glazebrook} {et~al.}(2017){Glazebrook}, {Schreiber}, {Labb{\'e}},
  {Nanayakkara}, {Kacprzak}, {Oesch}, {Papovich}, {Spitler}, {Straatman},
  {Tran}, \& {Yuan}}]{Glazebrook2017}
{Glazebrook}, K., {Schreiber}, C., {Labb{\'e}}, I., {et~al.} 2017, \nat, 544,
  71, \dodoi{10.1038/nature21680}

\bibitem[{{Glazebrook} {et~al.}(2024){Glazebrook}, {Nanayakkara}, {Schreiber},
  {Lagos}, {Kawinwanichakij}, {Jacobs}, {Chittenden}, {Brammer}, {Kacprzak},
  {Labbe}, {Marchesini}, {Marsan}, {Oesch}, {Papovich}, {Remus}, {Tran},
  {Esdaile}, \& {Chandro-Gomez}}]{Glazebrook2024a}
{Glazebrook}, K., {Nanayakkara}, T., {Schreiber}, C., {et~al.} 2024, \nat, 628,
  277, \dodoi{10.1038/s41586-024-07191-9}

\bibitem[{{Gobat} {et~al.}(2012){Gobat}, {Strazzullo}, {Daddi}, {Onodera},
  {Renzini}, {B{\'e}thermin}, {Dickinson}, {Carollo}, \&
  {Cimatti}}]{Gobat2012a}
{Gobat}, R., {Strazzullo}, V., {Daddi}, E., {et~al.} 2012, \apjl, 759, L44,
  \dodoi{10.1088/2041-8205/759/2/L44}

\bibitem[{{Greene} {et~al.}(2024){Greene}, {Labbe}, {Goulding}, {Furtak},
  {Chemerynska}, {Kokorev}, {Dayal}, {Volonteri}, {Williams}, {Wang}, {Setton},
  {Burgasser}, {Bezanson}, {Atek}, {Brammer}, {Cutler}, {Feldmann}, {Fujimoto},
  {Glazebrook}, {de Graaff}, {Khullar}, {Leja}, {Marchesini}, {Maseda},
  {Matthee}, {Miller}, {Naidu}, {Nanayakkara}, {Oesch}, {Pan}, {Papovich},
  {Price}, {van Dokkum}, {Weaver}, {Whitaker}, \& {Zitrin}}]{Greene2024a}
{Greene}, J.~E., {Labbe}, I., {Goulding}, A.~D., {et~al.} 2024, \apj, 964, 39,
  \dodoi{10.3847/1538-4357/ad1e5f}

\bibitem[{{Grogin} {et~al.}(2011){Grogin}, {Kocevski}, {Faber}, {Ferguson},
  {Koekemoer}, {Riess}, {Acquaviva}, {Alexander}, {Almaini}, {Ashby}, {Barden},
  {Bell}, {Bournaud}, {Brown}, {Caputi}, {Casertano}, {Cassata}, {Castellano},
  {Challis}, {Chary}, {Cheung}, {Cirasuolo}, {Conselice}, {Roshan Cooray},
  {Croton}, {Daddi}, {Dahlen}, {Dav{\'e}}, {de Mello}, {Dekel}, {Dickinson},
  {Dolch}, {Donley}, {Dunlop}, {Dutton}, {Elbaz}, {Fazio}, {Filippenko},
  {Finkelstein}, {Fontana}, {Gardner}, {Garnavich}, {Gawiser}, {Giavalisco},
  {Grazian}, {Guo}, {Hathi}, {H{\"a}ussler}, {Hopkins}, {Huang}, {Huang},
  {Jha}, {Kartaltepe}, {Kirshner}, {Koo}, {Lai}, {Lee}, {Li}, {Lotz}, {Lucas},
  {Madau}, {McCarthy}, {McGrath}, {McIntosh}, {McLure}, {Mobasher},
  {Moustakas}, {Mozena}, {Nandra}, {Newman}, {Niemi}, {Noeske}, {Papovich},
  {Pentericci}, {Pope}, {Primack}, {Rajan}, {Ravindranath}, {Reddy}, {Renzini},
  {Rix}, {Robaina}, {Rodney}, {Rosario}, {Rosati}, {Salimbeni}, {Scarlata},
  {Siana}, {Simard}, {Smidt}, {Somerville}, {Spinrad}, {Straughn}, {Strolger},
  {Telford}, {Teplitz}, {Trump}, {van der Wel}, {Villforth}, {Wechsler},
  {Weiner}, {Wiklind}, {Wild}, {Wilson}, {Wuyts}, {Yan}, \& {Yun}}]{Grogin2011}
{Grogin}, N.~A., {Kocevski}, D.~D., {Faber}, S.~M., {et~al.} 2011, \apjs, 197,
  35, \dodoi{10.1088/0067-0049/197/2/35}

\bibitem[{{Hartley} {et~al.}(2023){Hartley}, {Nelson}, {Suess}, {Garcia},
  {Park}, {Hernquist}, {Bezanson}, {Nevin}, {Pillepich}, {Schechter},
  {Terrazas}, {Torrey}, {Wellons}, {Whitaker}, \& {Williams}}]{Hartley2023a}
{Hartley}, A.~I., {Nelson}, E.~J., {Suess}, K.~A., {et~al.} 2023, \mnras, 522,
  3138, \dodoi{10.1093/mnras/stad1162}

\bibitem[{Hunter(2007)}]{Hunter2007}
Hunter, J.~D. 2007, Computing In Science \& Engineering, 9, 90

\bibitem[{{Johnson} {et~al.}(2021){Johnson}, {Leja}, {Conroy}, \&
  {Speagle}}]{Johnson2021a}
{Johnson}, B.~D., {Leja}, J., {Conroy}, C., \& {Speagle}, J.~S. 2021, \apjs,
  254, 22, \dodoi{10.3847/1538-4365/abef67}

\bibitem[{{Kimmig} {et~al.}(2025){Kimmig}, {Remus}, {Seidel}, {Valenzuela},
  {Dolag}, \& {Burkert}}]{Kimmig2025a}
{Kimmig}, L.~C., {Remus}, R.-S., {Seidel}, B., {et~al.} 2025, \apj, 979, 15,
  \dodoi{10.3847/1538-4357/ad9472}

\bibitem[{{Kobayashi} {et~al.}(2020){Kobayashi}, {Karakas}, \&
  {Lugaro}}]{Kobayashi2020a}
{Kobayashi}, C., {Karakas}, A.~I., \& {Lugaro}, M. 2020, \apj, 900, 179,
  \dodoi{10.3847/1538-4357/abae65}

\bibitem[{{Koekemoer} {et~al.}(2011){Koekemoer}, {Faber}, {Ferguson}, {Grogin},
  {Kocevski}, {Koo}, {Lai}, {Lotz}, {Lucas}, {McGrath}, {Ogaz}, {Rajan},
  {Riess}, {Rodney}, {Strolger}, {Casertano}, {Castellano}, {Dahlen},
  {Dickinson}, {Dolch}, {Fontana}, {Giavalisco}, {Grazian}, {Guo}, {Hathi},
  {Huang}, {van der Wel}, {Yan}, {Acquaviva}, {Alexander}, {Almaini}, {Ashby},
  {Barden}, {Bell}, {Bournaud}, {Brown}, {Caputi}, {Cassata}, {Challis},
  {Chary}, {Cheung}, {Cirasuolo}, {Conselice}, {Roshan Cooray}, {Croton},
  {Daddi}, {Dav{\'e}}, {de Mello}, {de Ravel}, {Dekel}, {Donley}, {Dunlop},
  {Dutton}, {Elbaz}, {Fazio}, {Filippenko}, {Finkelstein}, {Frazer}, {Gardner},
  {Garnavich}, {Gawiser}, {Gruetzbauch}, {Hartley}, {H{\"a}ussler},
  {Herrington}, {Hopkins}, {Huang}, {Jha}, {Johnson}, {Kartaltepe},
  {Khostovan}, {Kirshner}, {Lani}, {Lee}, {Li}, {Madau}, {McCarthy},
  {McIntosh}, {McLure}, {McPartland}, {Mobasher}, {Moreira}, {Mortlock},
  {Moustakas}, {Mozena}, {Nandra}, {Newman}, {Nielsen}, {Niemi}, {Noeske},
  {Papovich}, {Pentericci}, {Pope}, {Primack}, {Ravindranath}, {Reddy},
  {Renzini}, {Rix}, {Robaina}, {Rosario}, {Rosati}, {Salimbeni}, {Scarlata},
  {Siana}, {Simard}, {Smidt}, {Snyder}, {Somerville}, {Spinrad}, {Straughn},
  {Telford}, {Teplitz}, {Trump}, {Vargas}, {Villforth}, {Wagner}, {Wandro},
  {Wechsler}, {Weiner}, {Wiklind}, {Wild}, {Wilson}, {Wuyts}, \&
  {Yun}}]{Koekemoer2011}
{Koekemoer}, A.~M., {Faber}, S.~M., {Ferguson}, H.~C., {et~al.} 2011, \apjs,
  197, 36, \dodoi{10.1088/0067-0049/197/2/36}

\bibitem[{{Kriek} \& {Conroy}(2013)}]{Kriek2013}
{Kriek}, M., \& {Conroy}, C. 2013, \apjl, 775, L16,
  \dodoi{10.1088/2041-8205/775/1/L16}

\bibitem[{{Kriek} {et~al.}(2009){Kriek}, {van Dokkum}, {Labb{\'e}}, {Franx},
  {Illingworth}, {Marchesini}, \& {Quadri}}]{Kriek2009}
{Kriek}, M., {van Dokkum}, P.~G., {Labb{\'e}}, I., {et~al.} 2009, \apj, 700,
  221, \dodoi{10.1088/0004-637X/700/1/221}

\bibitem[{{Kroupa}(2001)}]{Kroupa2001_conf}
{Kroupa}, P. 2001, in Astronomical Society of the Pacific Conference Series,
  Vol. 228, Dynamics of Star Clusters and the Milky Way, ed. S.~{Deiters},
  B.~{Fuchs}, A.~{Just}, R.~{Spurzem}, \& R.~{Wielen}, 187

\bibitem[{{Labbe} {et~al.}(2025){Labbe}, {Greene}, {Bezanson}, {Fujimoto},
  {Furtak}, {Goulding}, {Matthee}, {Naidu}, {Oesch}, {Atek}, {Brammer},
  {Chemerynska}, {Coe}, {Cutler}, {Dayal}, {Feldmann}, {Franx}, {Glazebrook},
  {Leja}, {Maseda}, {Marchesini}, {Nanayakkara}, {Nelson}, {Pan}, {Papovich},
  {Price}, {Suess}, {Wang}, {Weaver}, {Whitaker}, {Williams}, \&
  {Zitrin}}]{Labbe2025a}
{Labbe}, I., {Greene}, J.~E., {Bezanson}, R., {et~al.} 2025, \apj, 978, 92,
  \dodoi{10.3847/1538-4357/ad3551}

\bibitem[{{Lagos} {et~al.}(2024){Lagos}, {Bravo}, {Tobar}, {Obreschkow},
  {Power}, {Robotham}, {Proctor}, {Hansen}, {Chandro-G{\'o}mez}, \&
  {Carrivick}}]{Lagos2024a}
{Lagos}, C. d.~P., {Bravo}, M., {Tobar}, R., {et~al.} 2024, \mnras, 531, 3551,
  \dodoi{10.1093/mnras/stae1024}

\bibitem[{{Lagos} {et~al.}(2025){Lagos}, {Valentino}, {Wright}, {de Graaff},
  {Glazebrook}, {De Lucia}, {Robotham}, {Nanayakkara}, {Chandro-Gomez},
  {Bravo}, {Baugh}, {Harborne}, {Hirschmann}, {Fontanot}, {Xie}, \&
  {Chittenden}}]{Lagos2025a}
{Lagos}, C. d.~P., {Valentino}, F., {Wright}, R.~J., {et~al.} 2025, \mnras,
  536, 2324, \dodoi{10.1093/mnras/stae2626}

\bibitem[{{Le Borgne} {et~al.}(2003){Le Borgne}, {Bruzual}, {Pell{\'o}},
  {Lan{\c c}on}, {Rocca-Volmerange}, {Sanahuja}, {Schaerer}, {Soubiran}, \&
  {V{\'{\i}}lchez-G{\'o}mez}}]{LeBorgne2003}
{Le Borgne}, J.-F., {Bruzual}, G., {Pell{\'o}}, R., {et~al.} 2003, \aap, 402,
  433, \dodoi{10.1051/0004-6361:20030243}

\bibitem[{{Leja} {et~al.}(2019{\natexlab{a}}){Leja}, {Carnall}, {Johnson},
  {Conroy}, \& {Speagle}}]{Leja2019a}
{Leja}, J., {Carnall}, A.~C., {Johnson}, B.~D., {Conroy}, C., \& {Speagle},
  J.~S. 2019{\natexlab{a}}, \apj, 876, 3, \dodoi{10.3847/1538-4357/ab133c}

\bibitem[{{Leja} {et~al.}(2020){Leja}, {Speagle}, {Johnson}, {Conroy}, {van
  Dokkum}, \& {Franx}}]{Leja2020a}
{Leja}, J., {Speagle}, J.~S., {Johnson}, B.~D., {et~al.} 2020, \apj, 893, 111,
  \dodoi{10.3847/1538-4357/ab7e27}

\bibitem[{{Leja} {et~al.}(2019{\natexlab{b}}){Leja}, {Johnson}, {Conroy}, {van
  Dokkum}, {Speagle}, {Brammer}, {Momcheva}, {Skelton}, {Whitaker}, {Franx}, \&
  {Nelson}}]{Leja2019b}
{Leja}, J., {Johnson}, B.~D., {Conroy}, C., {et~al.} 2019{\natexlab{b}}, \apj,
  877, 140, \dodoi{10.3847/1538-4357/ab1d5a}

\bibitem[{{Li} {et~al.}(2024){Li}, {Leja}, {Johnson}, {Tacchella}, {Davies},
  {Belli}, {Park}, \& {Emami}}]{Li2024a}
{Li}, Y., {Leja}, J., {Johnson}, B.~D., {et~al.} 2024, arXiv e-prints,
  arXiv:2405.04598, \dodoi{10.48550/arXiv.2405.04598}

\bibitem[{{Long} {et~al.}(2023){Long}, {Casey}, {del P. Lagos}, {Lambrides},
  {Zavala}, {Champagne}, {Cooper}, \& {Cooray}}]{Long2023a}
{Long}, A.~S., {Casey}, C.~M., {del P. Lagos}, C., {et~al.} 2023, \apj, 953,
  11, \dodoi{10.3847/1538-4357/acddde}

\bibitem[{{Looser} {et~al.}(2024){Looser}, {D'Eugenio}, {Maiolino}, {Witstok},
  {Sandles}, {Curtis-Lake}, {Chevallard}, {Tacchella}, {Johnson}, {Baker},
  {Suess}, {Carniani}, {Ferruit}, {Arribas}, {Bonaventura}, {Bunker},
  {Cameron}, {Charlot}, {Curti}, {de Graaff}, {Maseda}, {Rawle}, {Rix}, {Del
  Pino}, {Smit}, {{\"U}bler}, {Willott}, {Alberts}, {Egami}, {Eisenstein},
  {Endsley}, {Hausen}, {Rieke}, {Robertson}, {Shivaei}, {Williams}, {Boyett},
  {Chen}, {Ji}, {Jones}, {Kumari}, {Nelson}, {Perna}, {Saxena}, \&
  {Scholtz}}]{Looser2024a}
{Looser}, T.~J., {D'Eugenio}, F., {Maiolino}, R., {et~al.} 2024, \nat, 629, 53,
  \dodoi{10.1038/s41586-024-07227-0}

\bibitem[{{Lustig} {et~al.}(2023){Lustig}, {Strazzullo}, {Remus}, {D'Eugenio},
  {Daddi}, {Burkert}, {De Lucia}, {Delvecchio}, {Dolag}, {Fontanot}, {Gobat},
  {Mohr}, {Onodera}, {Pannella}, \& {Pillepich}}]{Lustig2023a}
{Lustig}, P., {Strazzullo}, V., {Remus}, R.-S., {et~al.} 2023, \mnras, 518,
  5953, \dodoi{10.1093/mnras/stac3450}

\bibitem[{{Marsan} {et~al.}(2017){Marsan}, {Marchesini}, {Brammer}, {Geier},
  {Kado-Fong}, {Labb{\'e}}, {Muzzin}, \& {Stefanon}}]{Marsan2017}
{Marsan}, Z.~C., {Marchesini}, D., {Brammer}, G.~B., {et~al.} 2017, \apj, 842,
  21, \dodoi{10.3847/1538-4357/aa7206}

\bibitem[{{Mart{\'\i}nez-Mar{\'\i}n} {et~al.}(2024){Mart{\'\i}nez-Mar{\'\i}n},
  {Glazebrook}, {Nanayakkara}, {Jacobs}, {Labb{\'e}}, {Kacprzak}, {Papovich},
  \& {Schreiber}}]{Martinez-Marin2024a}
{Mart{\'\i}nez-Mar{\'\i}n}, M., {Glazebrook}, K., {Nanayakkara}, T., {et~al.}
  2024, \mnras, 531, 3187, \dodoi{10.1093/mnras/stae1335}

\bibitem[{{McCracken} {et~al.}(2012){McCracken}, {Milvang-Jensen}, {Dunlop},
  {Franx}, {Fynbo}, {Le F{\`e}vre}, {Holt}, {Caputi}, {Goranova}, {Buitrago},
  {Emerson}, {Freudling}, {Hudelot}, {L{\'o}pez-Sanjuan}, {Magnard}, {Mellier},
  {M{\o}ller}, {Nilsson}, {Sutherland}, {Tasca}, \& {Zabl}}]{McCracken2012}
{McCracken}, H.~J., {Milvang-Jensen}, B., {Dunlop}, J., {et~al.} 2012, \aap,
  544, A156, \dodoi{10.1051/0004-6361/201219507}

\bibitem[{McLean {et~al.}(2012)McLean, Steidel, Epps, Konidaris, Matthews,
  Adkins, Aliado, Brims, Canfield, Cromer, Fucik, Kulas, Mace, Magnone,
  Rodriguez, Rudie, Trainor, Wang, Weber, \& Weiss}]{McLean2012}
McLean, I.~S., Steidel, C.~C., Epps, H.~W., {et~al.} 2012, in Ground-based and
  Airborne Instrumentation for Astronomy {IV}, ed. I.~S. McLean, S.~K. Ramsay,
  \& H.~Takami, Vol. 8446 ({SPIE}-Intl Soc Optical Eng), 84460J--84460J--15,
  \dodoi{10.1117/12.924794}

\bibitem[{{Nanayakkara} {et~al.}(2022){Nanayakkara}, {Esdaile}, {Glazebrook},
  {Espejo Salcedo}, {Durre}, \& {Jacobs}}]{Nanayakkara2022b}
{Nanayakkara}, T., {Esdaile}, J., {Glazebrook}, K., {et~al.} 2022, \pasa, 39,
  e002, \dodoi{10.1017/pasa.2021.61}

\bibitem[{{Nanayakkara} {et~al.}(2024){Nanayakkara}, {Glazebrook}, {Jacobs},
  {Kawinwanichakij}, {Schreiber}, {Brammer}, {Esdaile}, {Kacprzak}, {Labbe},
  {Lagos}, {Marchesini}, {Marsan}, {Oesch}, {Papovich}, {Remus}, \&
  {Tran}}]{Nanayakkara2024a}
{Nanayakkara}, T., {Glazebrook}, K., {Jacobs}, C., {et~al.} 2024, Scientific
  Reports, 14, 3724, \dodoi{10.1038/s41598-024-52585-4}

\bibitem[{{Nelson} {et~al.}(2018){Nelson}, {Pillepich}, {Springel},
  {Weinberger}, {Hernquist}, {Pakmor}, {Genel}, {Torrey}, {Vogelsberger},
  {Kauffmann}, {Marinacci}, \& {Naiman}}]{Nelson2018a}
{Nelson}, D., {Pillepich}, A., {Springel}, V., {et~al.} 2018, \mnras, 475, 624,
  \dodoi{10.1093/mnras/stx3040}

\bibitem[{{Nenkova} {et~al.}(2008){Nenkova}, {Sirocky}, {Ivezi{\'c}}, \&
  {Elitzur}}]{Nenkova2008a}
{Nenkova}, M., {Sirocky}, M.~M., {Ivezi{\'c}}, {\v{Z}}., \& {Elitzur}, M. 2008,
  \apj, 685, 147, \dodoi{10.1086/590482}

\bibitem[{{Noeske} {et~al.}(2007){Noeske}, {Weiner}, {Faber}, {Papovich},
  {Koo}, {Somerville}, {Bundy}, {Conselice}, {Newman}, {Schiminovich}, {Le
  Floc'h}, {Coil}, {Rieke}, {Lotz}, {Primack}, {Barmby}, {Cooper}, {Davis},
  {Ellis}, {Fazio}, {Guhathakurta}, {Huang}, {Kassin}, {Martin}, {Phillips},
  {Rich}, {Small}, {Willmer}, \& {Wilson}}]{Noeske2007}
{Noeske}, K.~G., {Weiner}, B.~J., {Faber}, S.~M., {et~al.} 2007, \apjl, 660,
  L43, \dodoi{10.1086/517926}

\bibitem[{{Oke} \& {Gunn}(1983)}]{Oke1983}
{Oke}, J.~B., \& {Gunn}, J.~E. 1983, \apj, 266, 713, \dodoi{10.1086/160817}

\bibitem[{{Oser} {et~al.}(2012){Oser}, {Naab}, {Ostriker}, \&
  {Johansson}}]{Oser2012a}
{Oser}, L., {Naab}, T., {Ostriker}, J.~P., \& {Johansson}, P.~H. 2012, \apj,
  744, 63, \dodoi{10.1088/0004-637X/744/1/63}

\bibitem[{{Pallottini} \& {Ferrara}(2023)}]{Pallottini2023a}
{Pallottini}, A., \& {Ferrara}, A. 2023, \aap, 677, L4,
  \dodoi{10.1051/0004-6361/202347384}

\bibitem[{pandas~development team(2020)}]{Pandas2020}
pandas~development team, T. 2020, pandas-dev/pandas: Pandas, latest,  Zenodo,
  \dodoi{10.5281/zenodo.3509134}

\bibitem[{{Park} {et~al.}(2024){Park}, {Belli}, {Conroy}, {Johnson}, {Davies},
  {Leja}, {Tacchella}, {Mendel}, {Benton}, {Bugiani}, {Emami}, {Khoram}, {Li},
  {Maheson}, {Mathews}, {Naidu}, {Nelson}, {Terrazas}, \&
  {Weinberger}}]{Park2024a}
{Park}, M., {Belli}, S., {Conroy}, C., {et~al.} 2024, arXiv e-prints,
  arXiv:2404.17945, \dodoi{10.48550/arXiv.2404.17945}

\bibitem[{{Paxton} {et~al.}(2015){Paxton}, {Marchant}, {Schwab}, {Bauer},
  {Bildsten}, {Cantiello}, {Dessart}, {Farmer}, {Hu}, {Langer}, {Townsend},
  {Townsley}, \& {Timmes}}]{Paxton2015a}
{Paxton}, B., {Marchant}, P., {Schwab}, J., {et~al.} 2015, \apjs, 220, 15,
  \dodoi{10.1088/0067-0049/220/1/15}

\bibitem[{{P{\'e}rez-Gonz{\'a}lez} {et~al.}(2024){P{\'e}rez-Gonz{\'a}lez},
  {D`Eugenio}, {Rodr{\'\i}guez del Pino}, {{\"U}bler}, {Maiolino}, {Arribas},
  {Cresci}, {Lamperti}, {Bunker}, {Carniani}, {Charlot}, {Willott},
  {B{\"o}ker}, {Parlanti}, {Scholtz}, {Venturi}, {Barro}, {Costantin},
  {Mart{\'\i}n-Navarro}, {Dunlop}, \& {Magee}}]{Perez-Gonzalez2024a}
{P{\'e}rez-Gonz{\'a}lez}, P.~G., {D`Eugenio}, F., {Rodr{\'\i}guez del Pino},
  B., {et~al.} 2024, arXiv e-prints, arXiv:2405.03744,
  \dodoi{10.48550/arXiv.2405.03744}

\bibitem[{{Pillepich} {et~al.}(2018{\natexlab{a}}){Pillepich}, {Springel},
  {Nelson}, {Genel}, {Naiman}, {Pakmor}, {Hernquist}, {Torrey}, {Vogelsberger},
  {Weinberger}, \& {Marinacci}}]{Pillepich2018a}
{Pillepich}, A., {Springel}, V., {Nelson}, D., {et~al.} 2018{\natexlab{a}},
  \mnras, 473, 4077, \dodoi{10.1093/mnras/stx2656}

\bibitem[{{Pillepich} {et~al.}(2018{\natexlab{b}}){Pillepich}, {Nelson},
  {Hernquist}, {Springel}, {Pakmor}, {Torrey}, {Weinberger}, {Genel}, {Naiman},
  {Marinacci}, \& {Vogelsberger}}]{Pillepich2018b}
{Pillepich}, A., {Nelson}, D., {Hernquist}, L., {et~al.} 2018{\natexlab{b}},
  \mnras, 475, 648, \dodoi{10.1093/mnras/stx3112}

\bibitem[{{Remus} \& {Kimmig}(2023)}]{Remus2023a}
{Remus}, R.-S., \& {Kimmig}, L.~C. 2023, arXiv e-prints, arXiv:2310.16089,
  \dodoi{10.48550/arXiv.2310.16089}

\bibitem[{{Salmon} {et~al.}(2016){Salmon}, {Papovich}, {Long}, {Willner},
  {Finkelstein}, {Ferguson}, {Dickinson}, {Duncan}, {Faber}, {Hathi},
  {Koekemoer}, {Kurczynski}, {Newman}, {Pacifici}, {P{\'e}rez-Gonz{\'a}lez}, \&
  {Pforr}}]{Salmon2016}
{Salmon}, B., {Papovich}, C., {Long}, J., {et~al.} 2016, \apj, 827, 20,
  \dodoi{10.3847/0004-637X/827/1/20}

\bibitem[{{Schreiber} {et~al.}(2018{\natexlab{a}}){Schreiber}, {Glazebrook},
  {Nanayakkara}, {Kacprzak}, {Labb{\'e}}, {Oesch}, {Yuan}, {Tran}, {Papovich},
  {Spitler}, \& {Straatman}}]{Schreiber2018}
{Schreiber}, C., {Glazebrook}, K., {Nanayakkara}, T., {et~al.}
  2018{\natexlab{a}}, \aap, 618, A85, \dodoi{10.1051/0004-6361/201833070}

\bibitem[{{Schreiber} {et~al.}(2018{\natexlab{b}}){Schreiber}, {Labb{\'e}},
  {Glazebrook}, {Bekiaris}, {Papovich}, {Costa}, {Elbaz}, {Kacprzak},
  {Nanayakkara}, \& {Oesch}}]{Schreiber2018b}
{Schreiber}, C., {Labb{\'e}}, I., {Glazebrook}, K., {et~al.}
  2018{\natexlab{b}}, \aap, 611, A22, \dodoi{10.1051/0004-6361/201731917}

\bibitem[{{Setton} {et~al.}(2024){Setton}, {Khullar}, {Miller}, {Bezanson},
  {Greene}, {Suess}, {Whitaker}, {Antwi-Danso}, {Atek}, {Brammer}, {Cutler},
  {Dayal}, {Feldmann}, {Fujimoto}, {Furtak}, {Glazebrook}, {Goulding},
  {Kokorev}, {Labbe}, {Leja}, {Ma}, {Marchesini}, {Nanayakkara}, {Pan},
  {Price}, {Siegel}, {Shipley}, {Weaver}, {van Dokkum}, {Wang}, \&
  {Williams}}]{Setton2024a}
{Setton}, D.~J., {Khullar}, G., {Miller}, T.~B., {et~al.} 2024, \apj, 974, 145,
  \dodoi{10.3847/1538-4357/ad6a18}

\bibitem[{{Skelton} {et~al.}(2014){Skelton}, {Whitaker}, {Momcheva}, {Brammer},
  {van Dokkum}, {Labb{\'e}}, {Franx}, {van der Wel}, {Bezanson}, {Da Cunha},
  {Fumagalli}, {F{\"o}rster Schreiber}, {Kriek}, {Leja}, {Lundgren}, {Magee},
  {Marchesini}, {Maseda}, {Nelson}, {Oesch}, {Pacifici}, {Patel}, {Price},
  {Rix}, {Tal}, {Wake}, \& {Wuyts}}]{Skelton2014}
{Skelton}, R.~E., {Whitaker}, K.~E., {Momcheva}, I.~G., {et~al.} 2014, \apjs,
  214, 24, \dodoi{10.1088/0067-0049/214/2/24}

\bibitem[{{Smit} {et~al.}(2012){Smit}, {Bouwens}, {Franx}, {Illingworth},
  {Labb{\'e}}, {Oesch}, \& {van Dokkum}}]{Smit2012a}
{Smit}, R., {Bouwens}, R.~J., {Franx}, M., {et~al.} 2012, \apj, 756, 14,
  \dodoi{10.1088/0004-637X/756/1/14}

\bibitem[{{Spitler} {et~al.}(2014){Spitler}, {Straatman}, {Labb{\'e}},
  {Glazebrook}, {Tran}, {Kacprzak}, {Quadri}, {Papovich}, {Persson}, {van
  Dokkum}, {Allen}, {Kawinwanichakij}, {Kelson}, {McCarthy}, {Mehrtens},
  {Monson}, {Nanayakkara}, {Rees}, {Tilvi}, \& {Tomczak}}]{Spitler2014}
{Spitler}, L.~R., {Straatman}, C.~M.~S., {Labb{\'e}}, I., {et~al.} 2014, \apjl,
  787, L36, \dodoi{10.1088/2041-8205/787/2/L36}

\bibitem[{{Springel} {et~al.}(2018){Springel}, {Pakmor}, {Pillepich},
  {Weinberger}, {Nelson}, {Hernquist}, {Vogelsberger}, {Genel}, {Torrey},
  {Marinacci}, \& {Naiman}}]{Springel2018a}
{Springel}, V., {Pakmor}, R., {Pillepich}, A., {et~al.} 2018, \mnras, 475, 676,
  \dodoi{10.1093/mnras/stx3304}

\bibitem[{{Steinborn} {et~al.}(2015){Steinborn}, {Dolag}, {Hirschmann},
  {Prieto}, \& {Remus}}]{Steinborn2015a}
{Steinborn}, L.~K., {Dolag}, K., {Hirschmann}, M., {Prieto}, M.~A., \& {Remus},
  R.-S. 2015, \mnras, 448, 1504, \dodoi{10.1093/mnras/stv072}

\bibitem[{{Straatman} {et~al.}(2014){Straatman}, {Labb{\'e}}, {Spitler},
  {Allen}, {Altieri}, {Brammer}, {Dickinson}, {van Dokkum}, {Inami},
  {Glazebrook}, {Kacprzak}, {Kawinwanichakij}, {Kelson}, {McCarthy},
  {Mehrtens}, {Monson}, {Murphy}, {Papovich}, {Persson}, {Quadri}, {Rees},
  {Tomczak}, {Tran}, \& {Tilvi}}]{Straatman2014}
{Straatman}, C.~M.~S., {Labb{\'e}}, I., {Spitler}, L.~R., {et~al.} 2014, \apjl,
  783, L14, \dodoi{10.1088/2041-8205/783/1/L14}

\bibitem[{{Straatman} {et~al.}(2016){Straatman}, {Spitler}, {Quadri},
  {Labb{\'e}}, {Glazebrook}, {Persson}, {Papovich}, {Tran}, {Brammer},
  {Cowley}, {Tomczak}, {Nanayakkara}, {Alcorn}, {Allen}, {Broussard}, {van
  Dokkum}, {Forrest}, {van Houdt}, {Kacprzak}, {Kawinwanichakij}, {Kelson},
  {Lee}, {McCarthy}, {Mehrtens}, {Monson}, {Murphy}, {Rees}, {Tilvi}, \&
  {Whitaker}}]{Straatman2016}
{Straatman}, C. M.~S., {Spitler}, L.~R., {Quadri}, R.~F., {et~al.} 2016, \apj,
  830, 51, \dodoi{10.3847/0004-637X/830/1/51}

\bibitem[{{Strait} {et~al.}(2023){Strait}, {Brammer}, {Muzzin}, {Desprez},
  {Asada}, {Abraham}, {Brada{\v{c}}}, {Iyer}, {Martis}, {Mowla}, {Noirot},
  {Sarrouh}, {Sawicki}, {Willott}, {Gould}, {Grindlay}, {Matharu}, \&
  {Rihtar{\v{s}}i{\v{c}}}}]{Strait2023a}
{Strait}, V., {Brammer}, G., {Muzzin}, A., {et~al.} 2023, \apjl, 949, L23,
  \dodoi{10.3847/2041-8213/acd457}

\bibitem[{{Turner} {et~al.}(2025){Turner}, {Tacchella}, {D'Eugenio},
  {Carniani}, {Curti}, {Glazebrook}, {Johnson}, {Lim}, {Looser}, {Maiolino},
  {Nanayakkara}, \& {Wan}}]{Turner2025a}
{Turner}, C., {Tacchella}, S., {D'Eugenio}, F., {et~al.} 2025, \mnras,
  \dodoi{10.1093/mnras/staf128}

\bibitem[{{Valentino} {et~al.}(2020){Valentino}, {Tanaka}, {Davidzon}, {Toft},
  {G{\'o}mez-Guijarro}, {Stockmann}, {Onodera}, {Brammer}, {Ceverino},
  {Faisst}, {Gallazzi}, {Hayward}, {Ilbert}, {Kubo}, {Magdis}, {Selsing},
  {Shimakawa}, {Sparre}, {Steinhardt}, {Yabe}, \& {Zabl}}]{Valentino2020}
{Valentino}, F., {Tanaka}, M., {Davidzon}, I., {et~al.} 2020, \apj, 889, 93,
  \dodoi{10.3847/1538-4357/ab64dc}

\bibitem[{{Valentino} {et~al.}(2023){Valentino}, {Brammer}, {Gould}, {Kokorev},
  {Fujimoto}, {Jespersen}, {Vijayan}, {Weaver}, {Ito}, {Tanaka}, {Ilbert},
  {Magdis}, {Whitaker}, {Faisst}, {Gallazzi}, {Gillman}, {Gim{\'e}nez-Arteaga},
  {G{\'o}mez-Guijarro}, {Kubo}, {Heintz}, {Hirschmann}, {Oesch}, {Onodera},
  {Rizzo}, {Lee}, {Strait}, \& {Toft}}]{Valentino2023a}
{Valentino}, F., {Brammer}, G., {Gould}, K. M.~L., {et~al.} 2023, \apj, 947,
  20, \dodoi{10.3847/1538-4357/acbefa}

\bibitem[{{Wang} {et~al.}(2024){Wang}, {Leja}, {de Graaff}, {Brammer},
  {Weibel}, {van Dokkum}, {Baggen}, {Suess}, {Greene}, {Bezanson}, {Cleri},
  {Hirschmann}, {Labb{\'e}}, {Matthee}, {McConachie}, {Naidu}, {Nelson},
  {Oesch}, {Setton}, \& {Williams}}]{Wang2024a}
{Wang}, B., {Leja}, J., {de Graaff}, A., {et~al.} 2024, \apjl, 969, L13,
  \dodoi{10.3847/2041-8213/ad55f7}

\bibitem[{{Weibel} {et~al.}(2024){Weibel}, {de Graaff}, {Setton}, {Miller},
  {Oesch}, {Brammer}, {Lagos}, {Whitaker}, {Williams}, {Baggen}, {Bezanson},
  {Boogaard}, {Cleri}, {Greene}, {Hirschmann}, {Hviding}, {Kuruvanthodi},
  {Labb{\'e}}, {Leja}, {Maseda}, {Matthee}, {McConachie}, {Naidu},
  {Roberts-Borsani}, {Schaerer}, {Suess}, {Valentino}, {van Dokkum}, \&
  {Wang}}]{Weibel2024a}
{Weibel}, A., {de Graaff}, A., {Setton}, D.~J., {et~al.} 2024, arXiv e-prints,
  arXiv:2409.03829, \dodoi{10.48550/arXiv.2409.03829}

\bibitem[{{Weinberger} {et~al.}(2018){Weinberger}, {Springel}, {Pakmor},
  {Nelson}, {Genel}, {Pillepich}, {Vogelsberger}, {Marinacci}, {Naiman},
  {Torrey}, \& {Hernquist}}]{Weinberger2018a}
{Weinberger}, R., {Springel}, V., {Pakmor}, R., {et~al.} 2018, \mnras, 479,
  4056, \dodoi{10.1093/mnras/sty1733}

\bibitem[{{Weller} {et~al.}(2025){Weller}, {Pacucci}, {Ni}, {Hernquist}, \&
  {Park}}]{Weller2025a}
{Weller}, E.~J., {Pacucci}, F., {Ni}, Y., {Hernquist}, L., \& {Park}, M. 2025,
  \apj, 979, 181, \dodoi{10.3847/1538-4357/ada360}

\bibitem[{{Westera} {et~al.}(2002){Westera}, {Lejeune}, {Buser}, {Cuisinier},
  \& {Bruzual}}]{Westera2002a}
{Westera}, P., {Lejeune}, T., {Buser}, R., {Cuisinier}, F., \& {Bruzual}, G.
  2002, \aap, 381, 524, \dodoi{10.1051/0004-6361:20011493}

\bibitem[{{Williams} {et~al.}(2009){Williams}, {Quadri}, {Franx}, {van Dokkum},
  \& {Labb{\'e}}}]{Williams2009}
{Williams}, R.~J., {Quadri}, R.~F., {Franx}, M., {van Dokkum}, P., \&
  {Labb{\'e}}, I. 2009, \apj, 691, 1879, \dodoi{10.1088/0004-637X/691/2/1879}

\bibitem[{{Xie} {et~al.}(2024){Xie}, {De Lucia}, {Fontanot}, {Hirschmann},
  {Bah{\'e}}, {Balogh}, {Muzzin}, {Vulcani}, {Baxter}, {Forrest}, {Wilson},
  {Rudnick}, {Cooper}, \& {Rescigno}}]{Xie2024a}
{Xie}, L., {De Lucia}, G., {Fontanot}, F., {et~al.} 2024, \apjl, 966, L2,
  \dodoi{10.3847/2041-8213/ad380a}

\end{thebibliography}

\end{document}